\documentclass{aa}  

\usepackage{graphicx}
\usepackage{txfonts}
\def\kms{\,km\,s$^{-1}$}

\def\msun{{M$_{\odot}$}}
\def\rsun{{R$_{\odot}$}}
\def\lsun{{L$_{\odot}$}}

\def\rstar{{R$_{\star}$}}

\def\vsini{$v\sin i_\star $}

\begin{document}

   \title{Inner disk structure of the classical T Tauri star LkCa 15}


   \author{S.H.P. Alencar\inst{\ref{inst1},\ref{inst2}\thanks{Based on observations obtained 
           at the Canada-France-Hawaii Telescope (CFHT) which is operated by the 
           National Research Council of Canada, the Institut National des Sciences 
           de l'Univers of the Centre National de la Recherche Scientifique of 
           France, and the University of Hawaii.}}
          \and J. Bouvier\inst{\ref{inst2}}
          \and J.-F. Donati\inst{\ref{inst3}}
          \and E. Alecian\inst{\ref{inst2}}
          \and C.P. Folsom\inst{\ref{inst3}}
          \and K. Grankin\inst{\ref{inst4}}
          \and G.A.J. Hussain\inst{\ref{inst5}}
          \and C. Hill\inst{\ref{inst3}}
          \and A.-M. Cody\inst{\ref{inst6}}
          \and A. Carmona\inst{\ref{inst3}}
          \and C. Dougados\inst{\ref{inst2}}
          \and S.G. Gregory\inst{\ref{inst7}}
          \and G. Herczeg\inst{\ref{inst8}}
          \and F. M\'enard\inst{\ref{inst2}}
          \and C. Moutou\inst{\ref{inst9}}
          \and L. Malo\inst{\ref{inst9}}
          \and M. Takami\inst{\ref{inst10}}
          \and the MaTYSSE collaboration
          }

   \institute{Departamento de Fisica – ICEx – UFMG, Av. Ant\^onio Carlos 6627, 
         30270-901 Belo Horizonte, MG, Brazil, \email{silvia@fisica.ufmg.br}\label{inst1}
         \and Universit\'e Grenoble Alpes, CNRS, IPAG, F-38000 Grenoble, France,
         \email{jerome.bouvier@univ-grenoble-alpes.fr} \label{inst2}
         \and IRAP, Universit\'e de Toulouse, CNRS, UPS, CNES, 14 Avenue Edouard Belin, 
         Toulouse, F-31400, France, \email{jean-francois.donati@irap.omp.eu}\label{inst3}
         \and Crimean Astrophysical Observatory, Scientific Research Institute, 
         298409, Nauchny, Crimea, \label{inst4}
         \and ESO, Karl-Schwarzschild-Str. 2, D-85748 Garching, Germany, \label{inst5}
         \and NASA Ames Research Center, Moffett Field, CA 94035, USA, \label{inst6}
         \and SUPA, School of Physics and Astronomy, Univ. of St Andrews, St Andrews, Scotland KY16 9SS, UK, \label{inst7}
         \and Kavli Institute for Astronomy and Astrophysics, Peking University, Yi He Yuan Lu 5, Haidian Qu, Beijing 100871, China, \label{inst8}
         \and CFHT Corporation, 65–1238 Mamalahoa Hwy, Kamuela, Hawaii 96743, USA, \label{inst9} 
         \and Institute of Astronomy and Astrophysics, Academia Sinica, PO Box 23–141, 106, Taipei, Taiwan \label{inst10}
             }

   \date{Received ; accepted }

 
  \abstract
   {Magnetospheric accretion has been thoroughly studied in young 
stellar systems with full non-evolved accretion disks, 
but it is poorly documented for transition disk objects with large inner 
cavities.}
   {We aim at characterizing the star-disk interaction and the 
accretion process onto the central star of LkCa 15, a prototypical 
transition disk system with an inner dust cavity that is 50 au wide.
   }
   {We obtained quasi-simultaneous photometric and 
spectropolarimetric observations of the system over several rotational 
periods. We analyzed the system light curve and associated color 
variations, as well as changes in spectral continuum and line profile
to derive the properties of the accretion flow from the edge of the inner disk to the central star. We also derived magnetic field measurements at 
the stellar surface. 
   }
   {We find that the system exhibits magnetic, photometric, and 
spectroscopic variability with a period of about 5.70 days. The light 
curve reveals a periodic dip, which suggests the presence of an inner 
disk warp that is located at the corotation radius at about 0.06 au from the star. 
Line profile variations and 
veiling variability are consistent with a magnetospheric accretion model 
where the funnel flows reach the star at high latitudes. This leads to the 
development of an accretion shock close to the magnetic poles. All 
diagnostics point to a highly inclined inner disk that interacts with the 
stellar magnetosphere. 
   }
   {The spectroscopic and photometric variability on a 
timescale of days to weeks of LkCa 15 is remarkably similar to that of AA Tau, the 
prototype of periodic dippers. We therefore suggest that the origin of 
the variability is a rotating disk warp that is located at the inner edge of a 
highly inclined disk close to the star. This contrasts with the moderate 
inclination of the outer transition disk seen on the large scale and 
thus provides evidence for a significant misalignment between the inner 
and outer disks of this planet-forming transition disk system.}


   \maketitle

\section{Introduction}
Classical T Tauri stars (CTTSs) are young ($\sim 1-10$ Myr), low-mass (M $< 2$ \msun)
stars surrounded by a circumstellar disk from which they accrete. Their magnetic fields are strong enough to truncate the inner disks and channel
the accreting gas. These systems also produce outflows
in the form of stellar and disk winds that may collimate in a jet \citep{bou07a,har16}.
At about the same ages, T Tauri stars that are no longer accreting are called 
weak-line T Tauri stars (WTTSs).

LkCa 15 is a K5, $\sim$ 1 \msun\, classical T Tauri star with an age of about 3 to 5 Myr \citep{sim00}, 
located at a distance of $159 \pm 1$ pc \citep{gaia18} in the Taurus-Auriga star-forming region. It presents a transition disk, as defined by \citet{str89}.
Its gas disk component, extending out to 900 au,  was directly imaged through $^{12}$CO 2-1 emission with 
the IRAM Plateau de Bure interferometer (PdBI) by \citet{pie07}.
The 50 au dust cavity in the disk was first spatially resolved by 1.4 mm and 2.8 mm PdBI observations
\citep{pie06} and was later confirmed by dust continuum emission 
at 850$\mu$m with Small Millimeter Array (SMA) observations \citep{and11} and at 7mm with the Very Large Array (VLA) \citep{ise14}.
Scattered light from small dust components was observed in the optical and 
infrared through imaging polarimetry with the Zurich Imaging Polarimeter of the Spectro-Polarimetric 
High-contrast Exoplanet REsearch (SPHERE/ZIMPOL) and the Infra-Red Dual Imaging and Spectrograph (IRDIS) 
\citep{tha15,tha16}. The system is composed of a possibly warped 
inner disk, according to polarimetric differential imaging results obtained with the High-Contrast Coronographic Imager for Adaptive Optics (HiCIAO) of the Subaru telescope 
\citep{oh16}, a $\sim$50 au dust gap, and an outer disk \citep{tha15}. Modelling for the outer disk
suggests inclinations of 44$\degr$ to 50$\degr$ \citep{oh16,tha14}
with respect to our line of sight. Inside the dust cavity, gas was detected through CO 4.7 $\mu$m 
emission by \citet{naj03}, revealing the presence of warm gas down to $0.083\pm0.030$ au (assuming $i=52\degr$).
At even smaller scales, the photometric and spectroscopic variability of LkCa 15 provides some 
insight into the interaction between the central star and the inner 
disk. \citet{gra07} reported a six-year-long light curve for the 
system that exhibits a nearly constant brightness level interrupted by 
recurrent dimmings of up to one magnitude, with a periodicity of about six 
days \citep{art12}. Such a light curve is typical of the 
so-called dippers \citep{cod14} and is thought to result from the 
obscuration of the central star by rotating circumstellar dusty clumps 
in highly inclined systems \citep[e.g.,][]{mcg15}. \citet{whe15}
further reported line profile variability in the optical and 
near-infrared spectra of the system, with the transient occurrence of 
inverse P Cygni profiles, that is, the appearance of redshifted absorption 
components up to velocities of 400 \kms. These features are usually 
interpreted as the signature of accretion funnel flows along an
inclined magnetosphere in young stellar systems \citep[e.g.,][]{har94,har16}. 
The photometric behavior and the spectral line profiles 
suggest that the system has an inner disk that interacts with the central 
star, seen at high inclination. LkCa 15 was previously classified by \citet{esp07}
as a pre-transition disk, implying the presence of an inner disk, and 
\citet{tha16} reported shadows in the outer disk that
would indicate some structure located close to the star. 
The high inclination of the inner disk suggested by the
photometric and spectrocopic observations, however, is somewhat unexpected 
compared to the moderate inclination suggested by the modeling of high angular resolution 
imaging observations of the outer disk of LkCa 15. 

These preliminary results provided the motivation for an in-depth study 
of the spectrophotometric variability of this prototypical transition 
disk system in Taurus. We therefore designed a campaign to obtain 
spectropolarimetric observations together with multi-color photometric 
observations on a timescale of weeks with the hope  of clarifying how 
accretion proceeds onto the central star in a transition disk system 
with a large inner hole, thus attempting to relate the immediate 
circumstellar environment of the star to the large-scale properties of 
its circumstellar disk. In the following sections we detail the observations
(Sect. \ref{observations}), analyze the ground-based 
and K2 photometric data (Sect. \ref{photometry}), and discuss the spectroscopic
results (Sect. \ref{spectroscopy}). In Sect. \ref{discussion} we give 
a global interpretation for the observed variability of LkCa 15, 
and we present our conclusions in Sect. \ref{conclusions}.

\section{Observations}\label{observations}
BVRI photometric observations of the LkCa 15 system were obtained at the 
Crimean Astrophysical 
Observatory (CrAO) from November 13, 2015, to March 31, 2017, on the AZT-11 
1.25m telescope with a five-channel photometer or a CCD camera 
with the ProLine PL23042 detector. A nearby control star of 
similar brightness (TYC 1278-166-1) was used to estimate the 
photometric rms error in each filter, 
which amounts to 0.02, 0.014, 0.013, and 0.012 in the BVRI bands, 
respectively. A comparison star, HD284589, was also observed in the BVRI 
filters with a five-channel photometer, thus providing the required 
calibration for the differential light curves derived from CCD 
photometry. Averaged extinction coefficients were assumed for the site. The 
brightness and colors of the comparison star in the Johnson system 
are 10.78 (V), 1.57 (B-V), 1.21 (V-R$_J$), 
and 2.03 (V-I$_J$). Unfortunately, only five measurements were obtained in 
the B band, and they are not considered further here. CCD images were obtained in 
VRI filters and were bias subtracted and flat-field calibrated 
following a standard procedure. 

High-resolution spectroscopy was performed on the LkCa 15 system 
at the 3.6m Canada-France-Hawaii Telescope (CFHT), using the Echelle SpectroPolarimetric Device
for the Observation of Stars (ESPaDOnS) \citep{don03,sil12}.  A first
series of 14 optical spectra, ranging from 370 to 1,050 nm, were 
obtained between November 18 and December 3, 2015 (the 2015B dataset), at a resolving power of about 
68,000 and with a signal-to-noise ratio (S/N) of about 150 at 666 nm. A second series of 11 observations were 
obtained from December 13, 2016, to January 17, 2017 (the 2016B dataset) with the same setting.
Observations were made using the spectropolarimetric mode, which provides simultaneous Stokes V 
(circularly polarized) and I (total intensity) spectra. Observations 
were reduced using the Libre-ESpRIT package \citep{don97}. In 
this paper we mostly investigate the 2015B spectral variability of the LkCa 15 system 
based on the intensity spectra, while a detailed discussion of the polarized 
spectra is deferred to a companion paper \citep{don18}. 

\section{Photometry}\label{photometry}
\begin{figure}
\centering
\includegraphics[width=9cm]{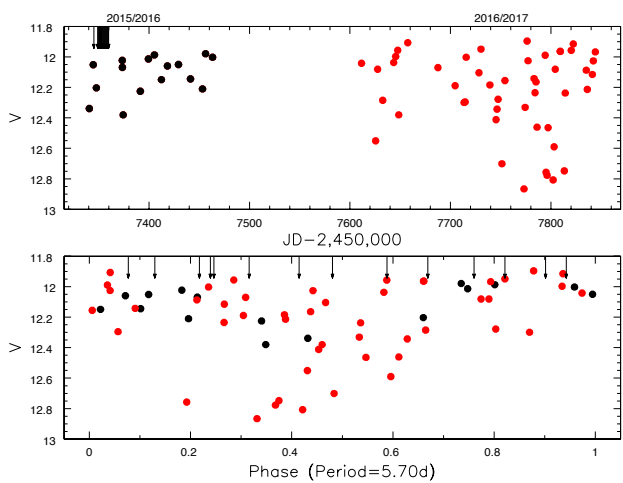}
\caption{Upper panel: V-band light curve of LkCa 15 obtained at CrAO over 
seasons 2015/16 (red) and 2016/17 (black). Lower panel: CrAO light curve folded 
in phase with a period of 5.70 days and JD0=2457343.8. The dates of spectroscopic
observations of the 2015B dataset are indicated by black arrows.}
\label{fig_lc_CrAO}
\end{figure}

\begin{figure}
\centering
\includegraphics[width=9cm]{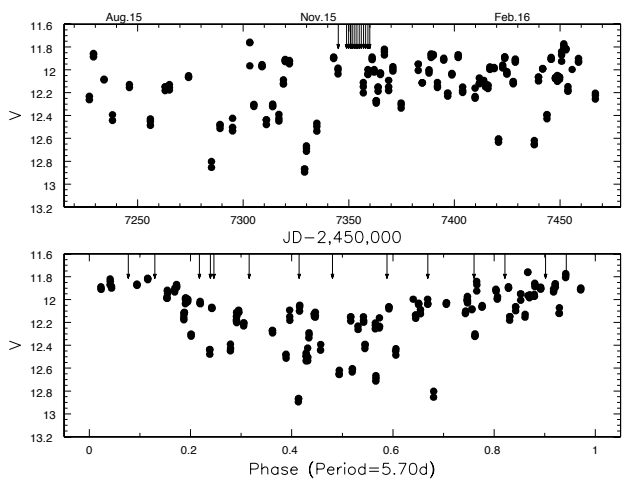}
\caption{Upper panel: V-band light curve obtained from ASAS-SN for the 
season 2015/16. Lower panel: ASAS-SN light curve folded in phase with a
period of 5.70 days and JD0=2457343.8. The dates of spectroscopic
observations are indicated by black arrows.}
\label{fig_lc_ASAS}
\end{figure}

A V-band light curve of LkCa 15, shown in Fig. \ref{fig_lc_CrAO}, was obtained at CrAO over 
two seasons. 
The system exhibits large V-band variability of up to 0.4 mag during 
the 2015/16 season and up to 1 mag during the 2016/17 season. 
Fig. \ref{fig_lc_CrAO} suggests that the variability amplitude was
smaller in 2015 than in 2016. However, the former epoch contains far fewer 
measurements than the second, and it is unclear whether deeper minima might have 
been missed. the better sampled 2015 ASAS-SN light curve shown in Fig. 
\ref{fig_lc_ASAS}  has the same amplitude as the CrAO 2016 light curve, which 
suggests that the photometric behavior did not change drastically between the two epochs.
Only some photometric measurements were obtained simultaneously with the 2015B spectroscopic 
observations, namely on November 17 and 20, 2015. Fortunately, the LkCa 15 photometric 
variability appears to be relatively stable over the years, as demonstrated by 
the V-band light curve obtained by \citet{gra07} for this system over six 
seasons from 1992 to 1997. The average V-band magnitude and the variability amplitude 
 we measure on the 2015-2017 light curve compare well with the 
values they report. Another similarity is the nearly constant maximum brightness 
level interrupted by deep minima that occur on a timescale of days. Even though 
the sampling of the CrAO light curve is not optimal, this type of light curve 
is strongly reminiscent of dippers \citep[e.g.,][]{mcg15,rod17,bod17}. 

We searched for periodicities in the light curve in the full dataset and 
independently for each season. A CLEAN periodogram analysis 
\citep{rob87} and a string-length analysis \citep{dwo83} both resulted 
in periods being detected in the full dataset as well as in each season, ranging 
from 5.62 to 5.70 days. This is slightly shorter than the periods of 5.86 to 6.03 
days that were previously reported for the system by \citet{bou93} and 
\citet{art12}. Figure \ref{fig_lc_CrAO} shows the V-band light curve phased with a period 
of 5.70 days. It concentrates most of the photometric minima around phases 0.4-0.6, 
but with a significant photometric scatter at all phases, which is typical 
of dippers. Additional photometry for the 2015/2016 season is available from 
ASAS-SN \citep{sha14,koc17}, and the corresponding light curve 
is shown in Fig. \ref{fig_lc_ASAS}. A periodogram analysis yields a period of $5.70 \pm 0.05$ 
days, consistent with that derived from the CrAO dataset. We used this 
as the stellar rotation period here. The corresponding phase diagram is shown 
in the lower panel of Fig. \ref{fig_lc_ASAS}. 

\begin{figure}
\centering
\includegraphics[width=9cm]{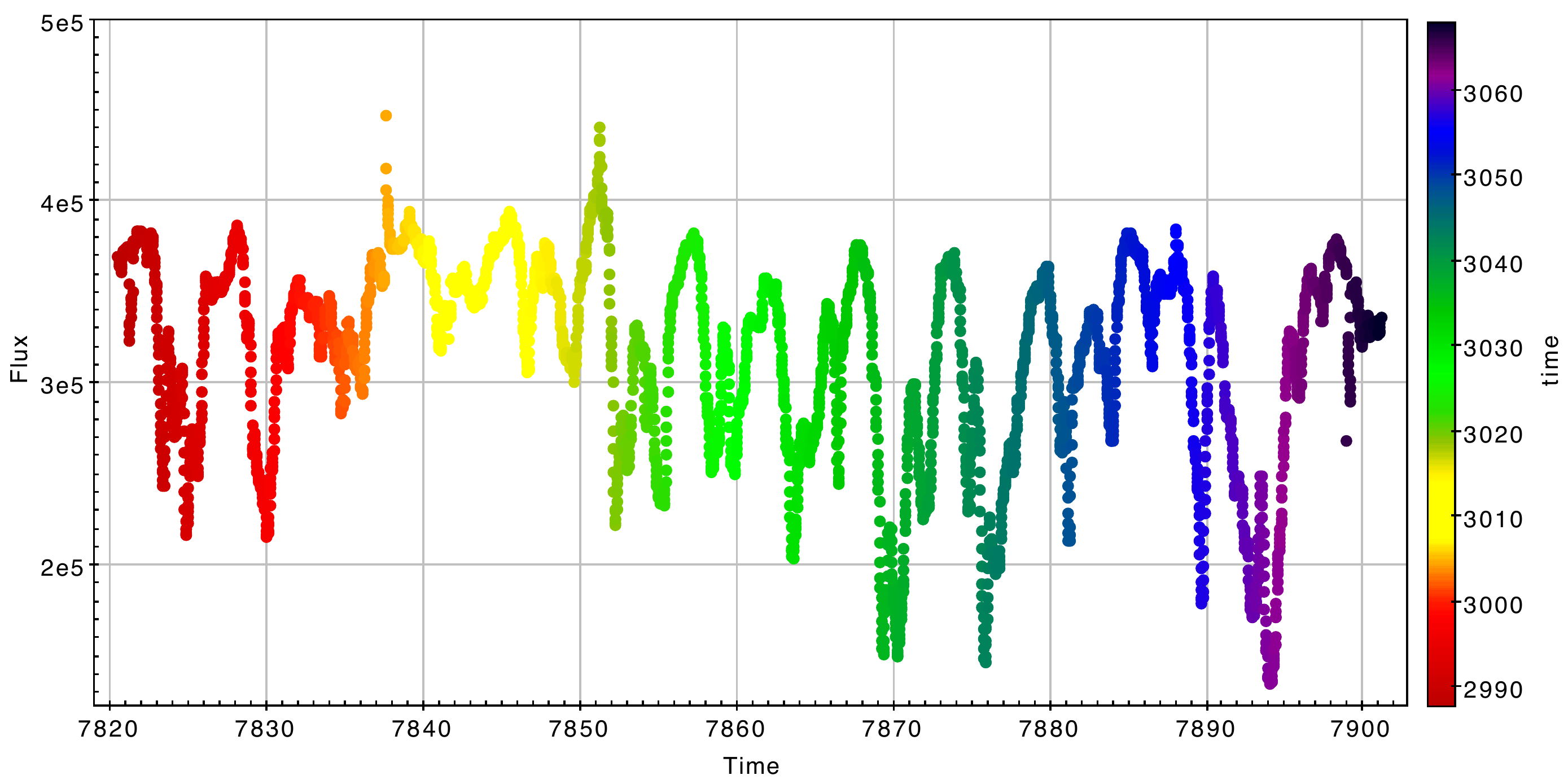}
\includegraphics[width=9cm]{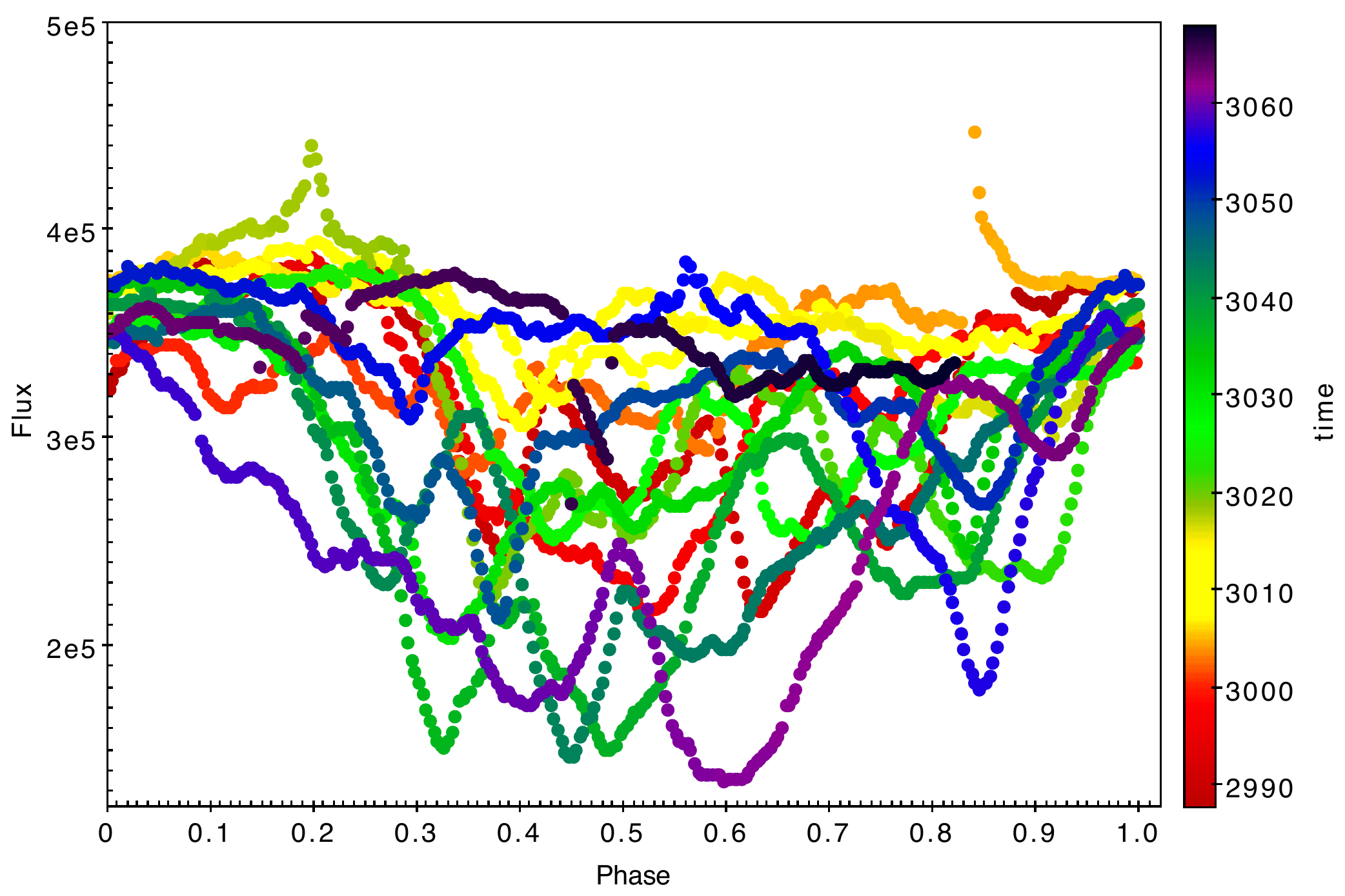}
\caption{Upper panel: Kepler K2 light curve obtained for LkCa 15 from 
March 7 to May 28, 2017. The light-curve morphology is clearly that of a dipper.
Lower panel: K2 light curve folded in phase with a period of 5.78 days. The
color code is the same in both panels and reflects the Julian Date of
observations. Two flare-like events are visible in the light curve (t=7838 and 7851).
}
\label{fig_lc_K2}
\end{figure}

Finally, the Kepler K2 satellite monitored LkCa 15 in a broad bandpass ranging
from 420 to 900 nm during Campaign F13 from March 
8 to May 27, 2017. While these observations occurred more than a year after the 
spectral series discussed in this paper (2015B dataset), the unsurpassed quality of the K2 light 
curve provides a much deeper insight into the photometric variability of the 
system. The light curve, encompassing 80 days uninterrupted, is shown in Fig. 
\ref{fig_lc_K2} and clearly confirms that LkCa 15 is a dipper: a nearly constant flux level 
is periodically interrupted by brightness dips lasting for a few days with an 
amplitude of up to one magnitude. A period analysis reveals a clear signal at 
5.78 days, present over the 13 rotational cycles the K2 light curve samples. 
The phased light curve is shown in the lower panel of Fig. \ref{fig_lc_K2}. The photometric minima often last 
for a significant fraction of the period, which suggests that an extended inner 
disk warp produces the dips \citep{rom13,rom15}. While periodic, the dips vary in shape, and their 
width, depth, and phase evolve from one cycle to the next. This is indicative of a dynamic 
interaction between the inner disk and the stellar magnetosphere. This behavior 
is quite reminiscent of that of the prototype of the dipper class, 
AA Tau \citep[e.g.,][]{bou07}. 

\begin{figure}
\centering
\includegraphics[width=9cm]{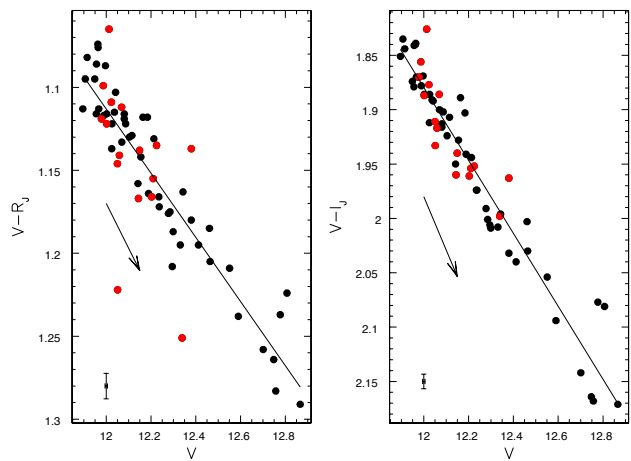}
\caption{(V-R$_J$) vs. V (left panel) and (V-I$_J$) vs. V (right panel)
color-magnitude variations of the system during the 2015/16 (red) and 2016/17
(black) season from CrAO photometry. The straight line in each panel is a linear
regression fit to the color slope. Arrows indicate ISM-like extinction vectors.
Photometric error bars are indicated in the bottom left corner of each panel.
}
\label{fig_CrAO_colors}
\end{figure}

CrAO multi-band photometry provides information on the color changes associated 
with the flux variations. Over the two observing seasons, the system exhibited color 
variations amounting to a few tenths of a magnitude, as shown in Fig. \ref{fig_CrAO_colors}. The 
system becomes redder when fainter, with little scatter around a mean color slope 
of 0.19 in the (V, V-R$_J$) color-magnitude diagram and 0.33 in the (V, V-I$_J$) diagram. 
These slopes are different from those expected for an interstellar medium (ISM)-like 
reddening vector, which would amount to 0.27 and 0.49 in the (V, V-R$_J$) and (V, V-I$_J$) 
diagrams, respectively. The (V-R$_J$) color range and color slope are similar to the 
values reported by \citet{gra07} from the long-term photometric 
monitoring of the system.  

\section{Spectroscopy}\label{spectroscopy}

\subsection{Veiling and radial velocities}\label{veiling_vrad}

Classical T Tauri stars (CTTSs) present hot spots at the stellar 
surface that emit an extra continuum, which
is added to the stellar continuum and veils the photospheric lines, decreasing their
depths. Veiling is measured as the hot-spot continuum flux divided by the stellar     
continuum flux at a given wavelength. Therefore, a veiling of 1 corresponds to an 
equal contribution from the hot spot and the star to the observed continuum.
CTTSs also have cold spots that are due to magnetic field emergence at the stellar 
surface. Hot and cold spots represent regions at different temperatures than the stellar photosphere
and can cause significant distortions in the photospheric line profiles
that mimic radial velocity variations.

Veiling and photospheric radial velocities 
were calculated from the continuum-normalized spectra of LkCa 15 in five spectral 
regions of about 100 \AA\, wide, close to 4800 \AA, 5300 \AA, 5600 \AA, 6000 \AA,\, and 6300 \AA.
We used as a standard the weak-lined T Tauri star V819 Tau (K4, $T_{\rm eff}=4250 \pm 50$ K)
that was observed with ESPaDOnS (CFHT), with the same settings as LkCa 15, as part of the MaTYSSE
program \citep{don15}. V819 Tau rotates slowly (\vsini$=9.5 \pm 0.5$ \kms) and matches the LkCa 15 photospheric lines well. The standard spectrum was rotationally broadened,
veiled, and cross-correlated with each spectrum of LkCa 15 until a best match was found.
This procedure allows the simultaneous determination of veiling and radial velocities.

\begin{figure}
\centering
\includegraphics[width=9cm]{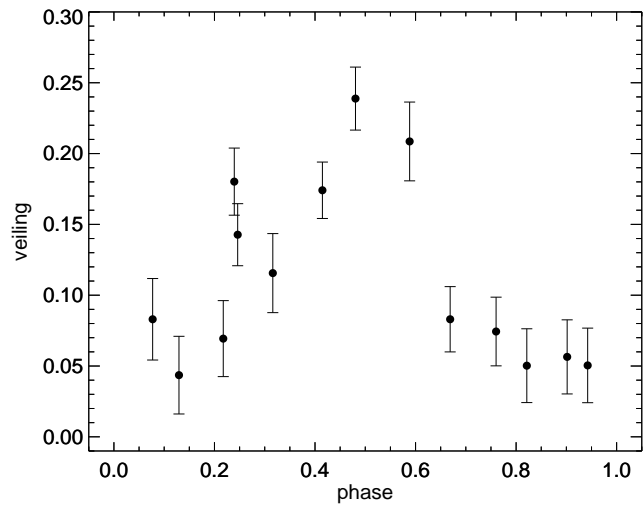}
\includegraphics[width=9cm]{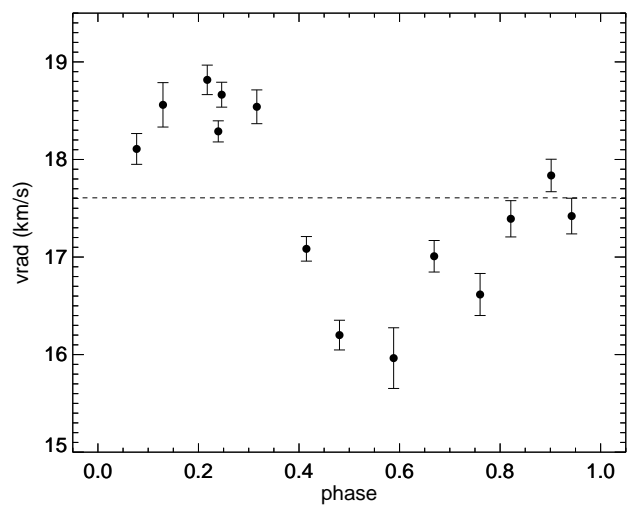}
\caption{Veiling (top) and radial velocities (bottom) of LkCa 15 in phase with the $5.7$-day
ASAS-SN photometric period and JD0=2457343.8. The dashed line corresponds to the mean
v$_{\rm rad}=17.83 \pm 0.86$ \kms}.
\label{fig_veiling_vrad}
\end{figure}

We searched for periodical variations in the veiling and radial velocities
using the Scargle periodogram as modified by \citet{hor86}.
Errors were estimated as the standard deviation of a Gaussian fitted to the main peak of the periodogram power spectrum.
The veiling and radial velocity present periodicities at $5.55 \pm 0.71$ days and $5.55 \pm 0.77$ days, respectively, which 
agree with the photometric period of $5.70 \pm 0.05$ days within the errors.
The veiling and the radial velocity periods should correspond to the stellar rotation period,
since hot and cold spots are located at the stellar surface. The agreement between the veiling and
radial velocity period and the photometric period supports the hypothesis that the inner
disk warp is located close to the disk corotation radius.
In Fig. \ref{fig_veiling_vrad} we show the mean veiling and radial velocity values obtained at each
phase and the corresponding standard deviations. The standard deviations were computed with 
the veiling and radial velocity values calculated at each observing date in the five different spectral 
regions we analyzed.
The phases in Fig. \ref{fig_veiling_vrad} were calculated with a photometric period of 5.70 days 
and an arbitrary JD0=2457343.8, chosen to ensure that the photometric minimum of the ASAS-SN light
curve was around phase 0.5. The maximum veiling is coincident with the photometric minimum, which indicates that the
accretion spot is aligned with the inner disk warp.

We fit synthetic spectra to the observations in the spectral region that extends 
from 6060\, to 6155\AA\, using a version of the ZEEMAN spectrum synthesis code \citep{lan88,wad01,fol16}
that allows for an additional continuum flux to be fit together with the stellar spectrum.
This spectral region was chosen because
it presents many atomic lines with few blends. We made some initial
fits, and it became clear that the surface gravity ($g$) was poorly determined. 
We therefore estimated $\log{g}$ using the stellar mass value determined by \citet{sim00} 
from the circumstellar disk rotation curve, and the stellar radius obtained from the stellar luminosity 
calculated by \citet{her14}, after scaling the values to take into account the LkCa 15 
distance of $159 \pm 1$ pc measured by the Gaia satellite \citep{gaia18}.
With the scaled values of $M_{\star}=1.10 \pm 0.03$ \msun\, and $L_{\star}=0.80 \pm 0.15$ \lsun, 
and using T$_{\rm eff}$ in the range of 4350 K 
to 4500 K, which corresponded to our best initial fits, we obtained $\log{g}=4.10 \pm 0.10$. 
Keeping $\log{g}$ fixed at 4.0, 4.1, or 4.2, we again fit the observed spectra with T$_{\rm eff}$,
the projected rotational velocity (v$\sin{i_\star}$), the radial velocity, and the veiling as free
parameters. We then noted that two of the spectra, from JD=7350.8645 and 7352.8529, converged to much 
higher values of v$\sin{i_\star}$ than the others. At these dates, the Moon
was very close to the star, and we decided to forego using these two spectra in the determination
of stellar parameters to avoid contamination.
We applied our fitting procedure to the 12 remaining spectra with the fixed $\log{g}$ values
and obtained T$_{\rm eff}=4492 \pm 50$ K and v$\sin{i_\star}=13.82 \pm 0.50$ \kms, where the 
values correspond to the average and the errors to the standard deviation of the best-fit 
results of our grid. With our effective temperature 
and the stellar luminosity of $0.80 \pm 0.15$ \lsun\, , we calculated a stellar 
radius of $1.49 \pm 0.15$ \rsun. According to the \citet{sie00} and \citet{mar13} evolutionary models, for 
{a mass estimate in the range between 1.10 \msun\,} \citep{sim00} and 1.25 \msun\, \citep{don18}, 
LkCa 15 is no longer fully convective and has already developed a radiative core.

The mean v$\sin{i_\star}=13.82 \pm 0.50$ \kms\, obtained during the spectral fit together with the adopted 
photometric period of $5.70 \pm 0.05$ days (see Sect. \ref{photometry}) and the stellar radius 
of $1.49 \pm 0.15$ \rsun\, yield $\sin{i_\star}$ larger than 0.90, within the error bars
of the measured values. This corresponds to stellar inclinations $i_\star > 65$\degr, which are higher than the outer disk inclination
of 44$\degr$ to 50$\degr$ measured from optical and infrared polarimetric imaging \citep{tha15,tha16,oh16}. This
suggests a misalignment between the inner and outer disks.

\subsection{Photospheric LSD profiles}

We calculated least-squares deconvolution (LSD) photospheric profiles \citep{don97} 
using a line mask computed from a Vienna Atomic Line Database \citep[VALD;][]{pis95,kup00} line list 
with T$_{\rm eff}=4500$ K and $\log{g}=4.0$. 
Emission, telluric, and broad lines were excluded from the mask, since they do not 
represent the photospheric weak-line profile. We also removed all the lines before 
5000 \AA\, because the S/N in the blue of our spectra was weak. The 
photospheric LSD profiles were normalized with a mean wavelength, Doppler width, line depth,
and Land\'e factor of 670 nm, 1.8 \kms, 0.55, and 1.2, following \citet{don14}.
We corrected the LSD profiles for veiling, using the mean veiling values displayed
in Fig. \ref{fig_veiling_vrad}. The intensity of each point of the normalized
and veiled spectrum ($I$) is related to the mean veiling value ($v$) and the non-veiled
line intensity ($I_0$) by $I=\frac{I_0+v}{1+v}$, which allows the recovery of $I_0$,
knowing $I$ and $v$. The Stokes I LSD profiles show a periodic variability 
at $5.56 \pm 0.74$ days, as we show Fig. \ref{lines_periodogram}.
Stellar radial velocities, computed from the first moment of the Stokes I LSD profiles, also 
presented a periodicity at $5.62 \pm 0.75$ days. These period values agree with the
photometric and veiling periodicities.
 
\begin{figure}
\centering
\includegraphics[width=9cm]{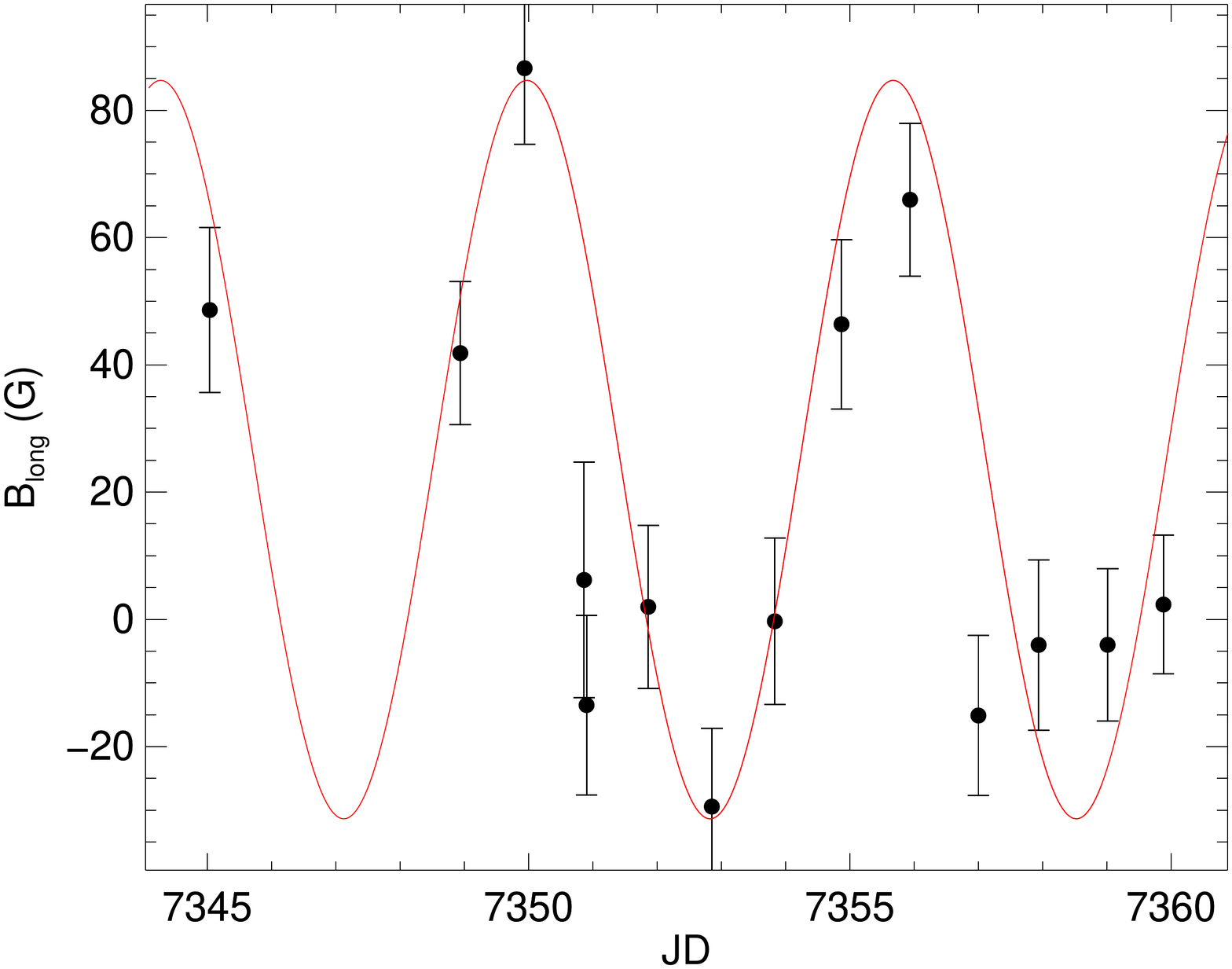}
\includegraphics[width=9cm]{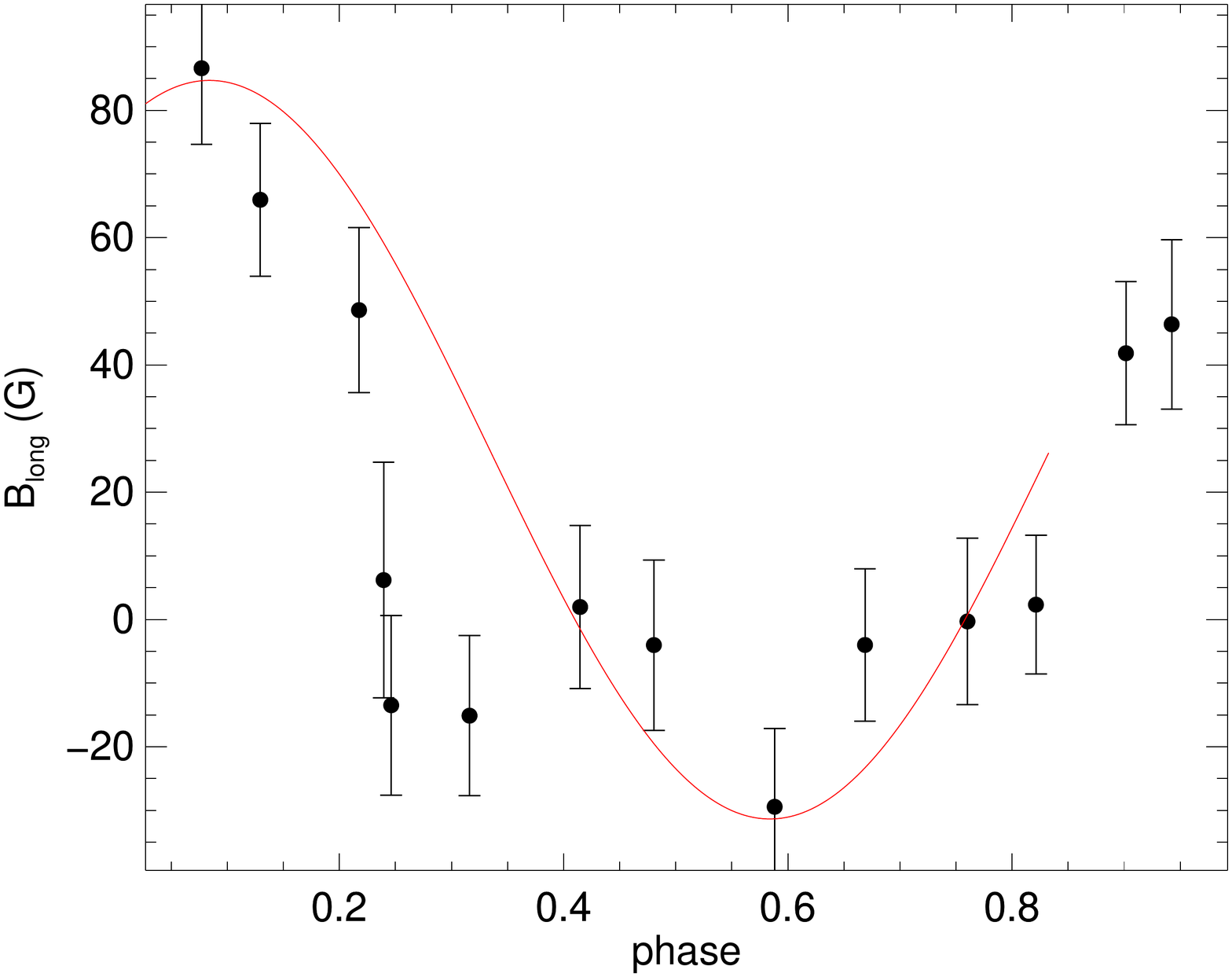}
\caption{Longitudinal component of the magnetic field of LkCa 15 from LSD profiles as a function 
of time (top) and in phase (bottom) with the $5.7$-day ASAS-SN photometric period and JD0=2457343.8.
A sine wave with the $5.7$-day period is overplotted.}
\label{fig_blong}
\end{figure}

The longitudinal component of the magnetic field can be computed from the first moment
of the Stokes V LSD profiles \citep[see][]{don97}. 
The longitudinal magnetic field values computed 
with our veiling-corrected LSD profiles are shown in Fig. \ref{fig_blong}. 
They vary from $-29$ G to $+87$ G and show a periodicity at $5.76 \pm 0.74$ days, 
corresponding to the mean rotation period of the star, and in agreement 
with the veiling, radial velocities, and the photometric periods.

\subsection{Circumstellar line profiles}\label{lines}

LkCa 15 displayed emission line profiles that were strongly variable
during our observations. The high resolution of the ESPaDOnS
spectra enabled us to identify several emission and absorption
components and analyze their variability. We also calculated 2D Scargle periodograms, 
as modified by \citet{hor86}, on the line profile intensities. Errors were estimated as the mean standard
deviation of a Gaussian fit to the main peak of the periodogram power spectrum
over the velocity range of the period detection.
The results are displayed in Fig. \ref{lines_periodogram}.
Our spectroscopic data in 2015A span 15 days, therefore we only considered period
detections shorter than 7.5 days.

\begin{figure}
\centering
\includegraphics[width=4.5cm]{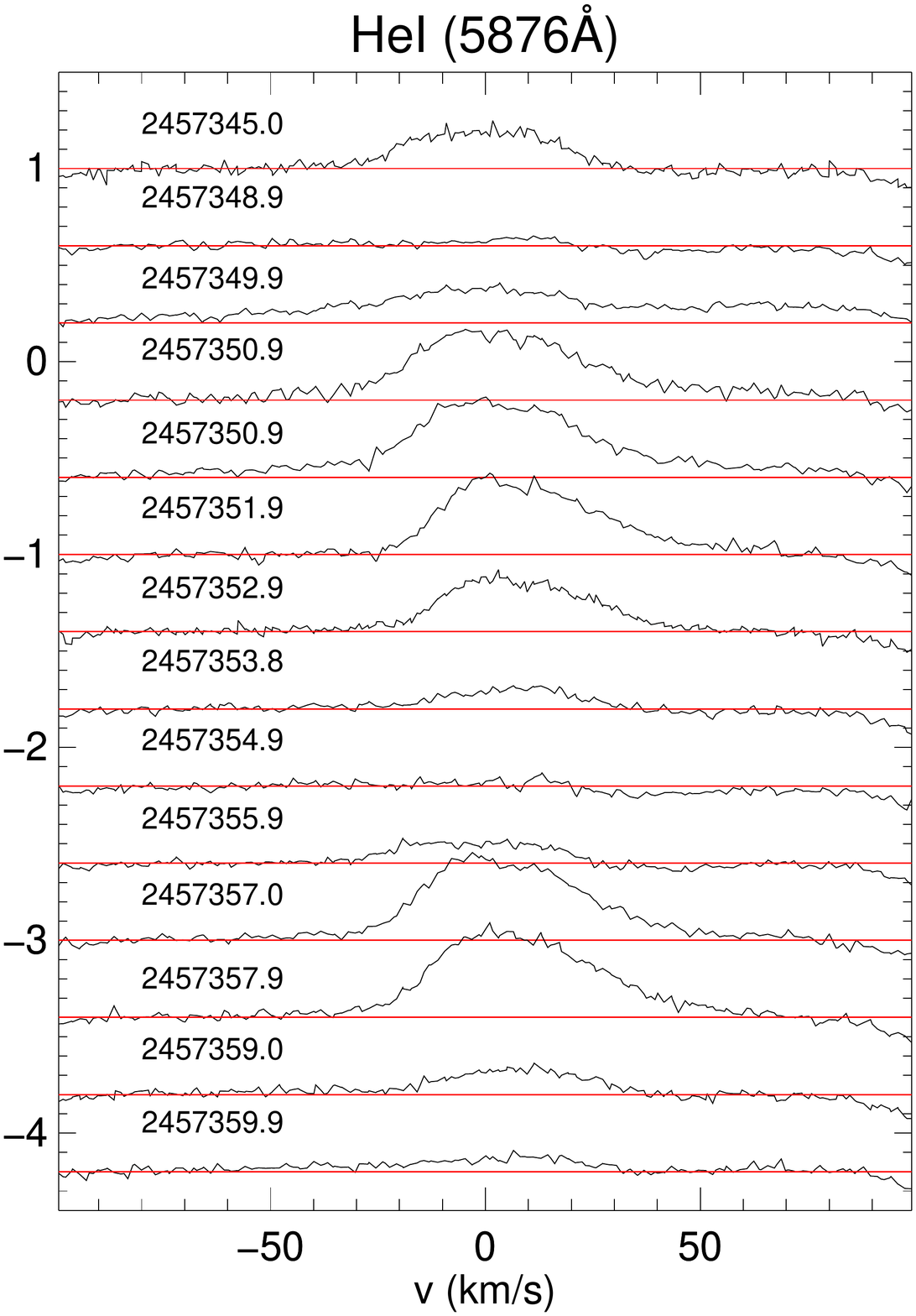}\includegraphics[width=4.5cm]{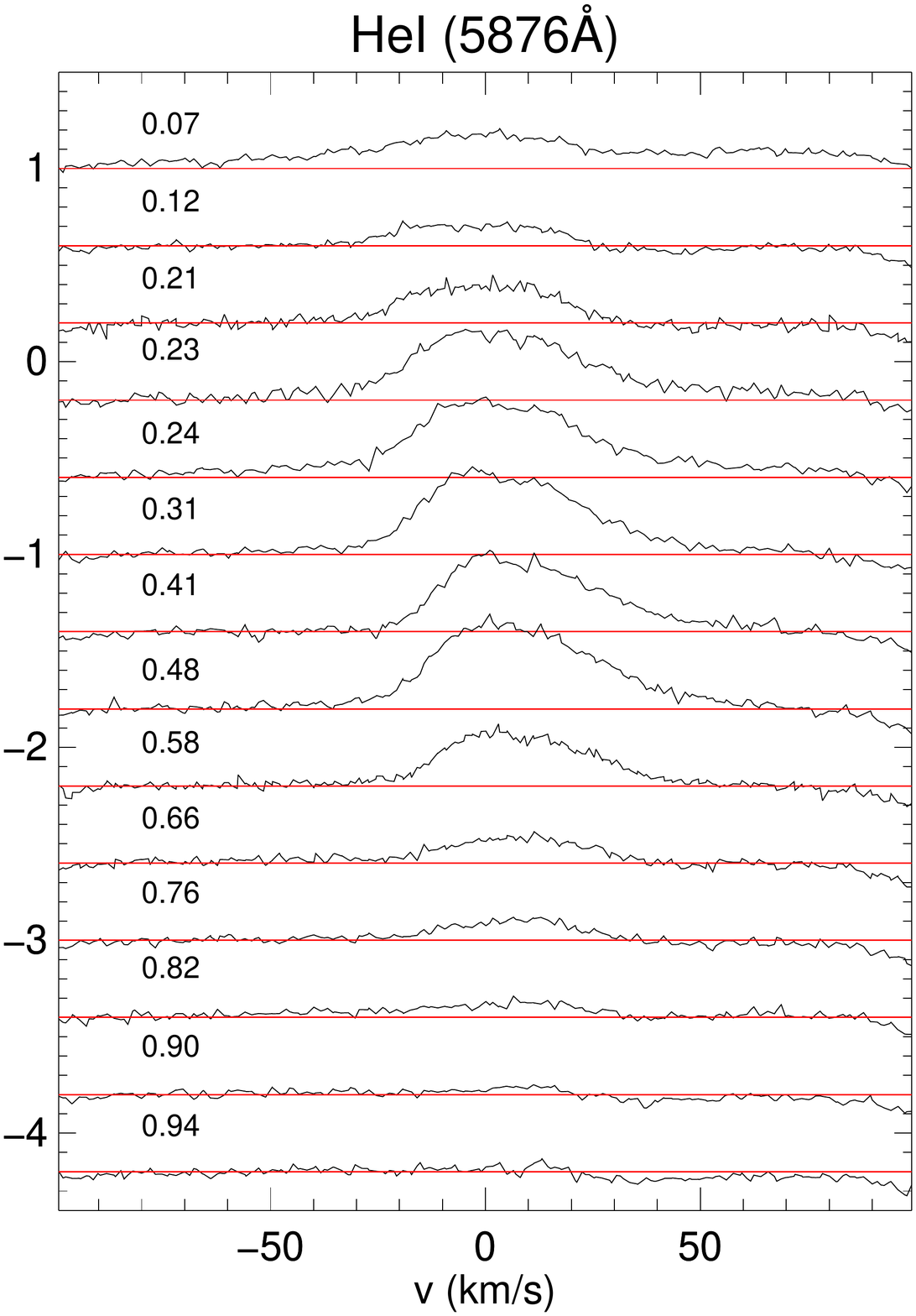}
\caption{Observed HeI 5876\AA\, profiles in time (left) and in phase (right) with the $5.7$-day
ASAS-SN photometric period and JD0=2457343.8. The red lines represent the continuum level.}
\label{he_profiles}
\end{figure}

\begin{figure}
\centering
\includegraphics[width=9cm]{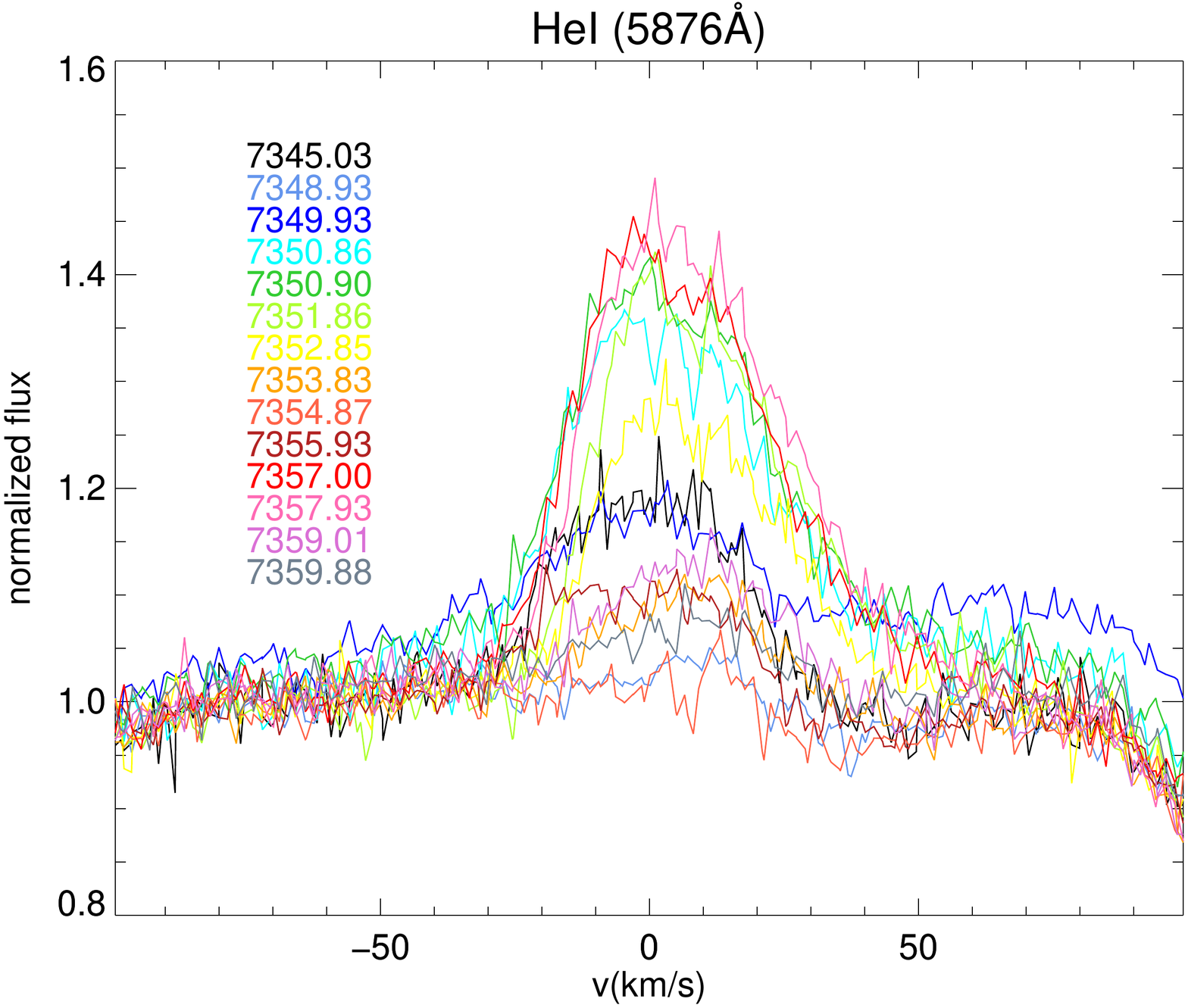}
\caption{Observed HeI 5876\AA\, profiles overplotted. The observation dates are shown
with the same color code as the profiles.}
\label{he_profiles_oplot}
\end{figure}

\begin{figure*}
\centering
\includegraphics[width=9cm]{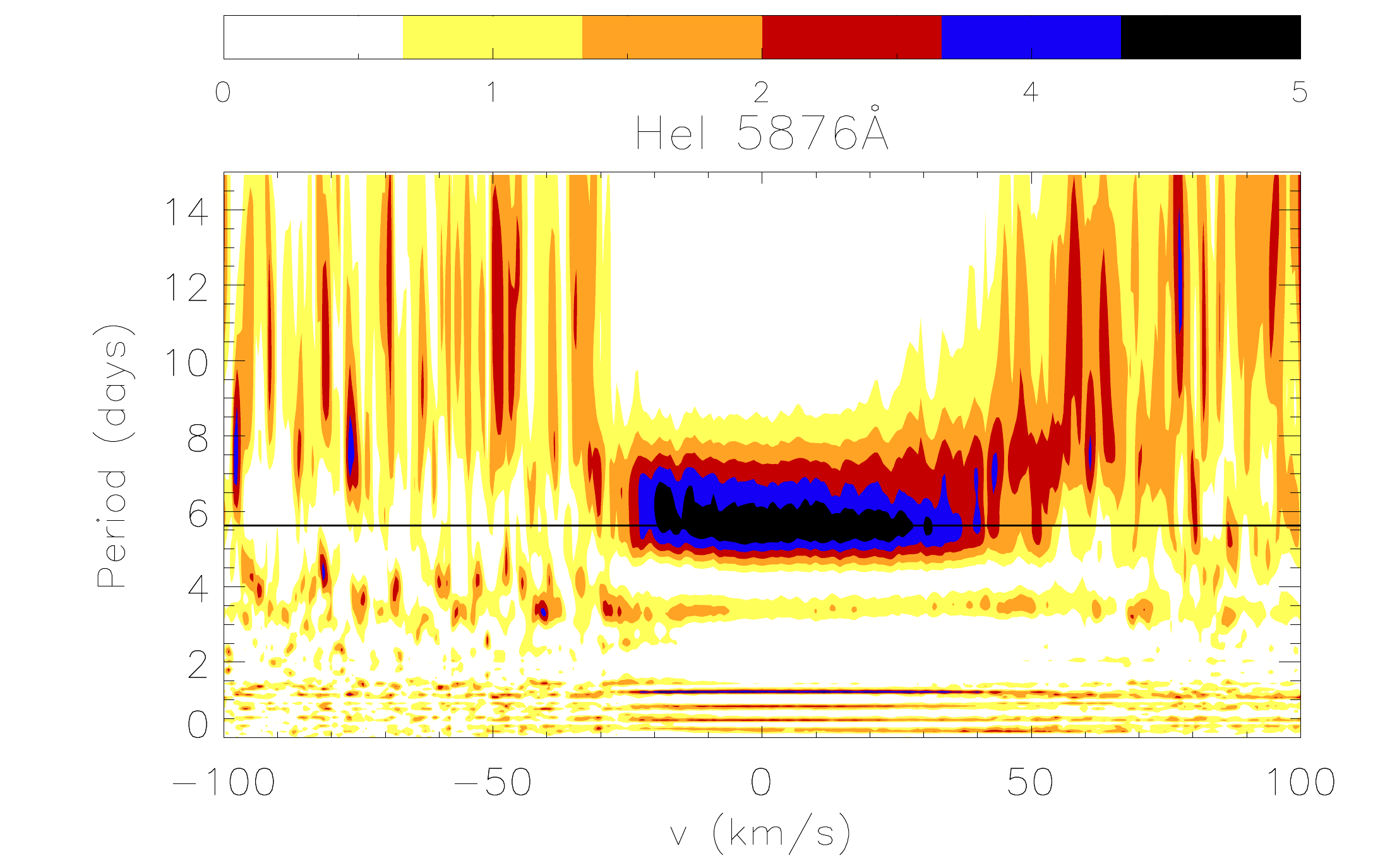}\includegraphics[width=9cm]{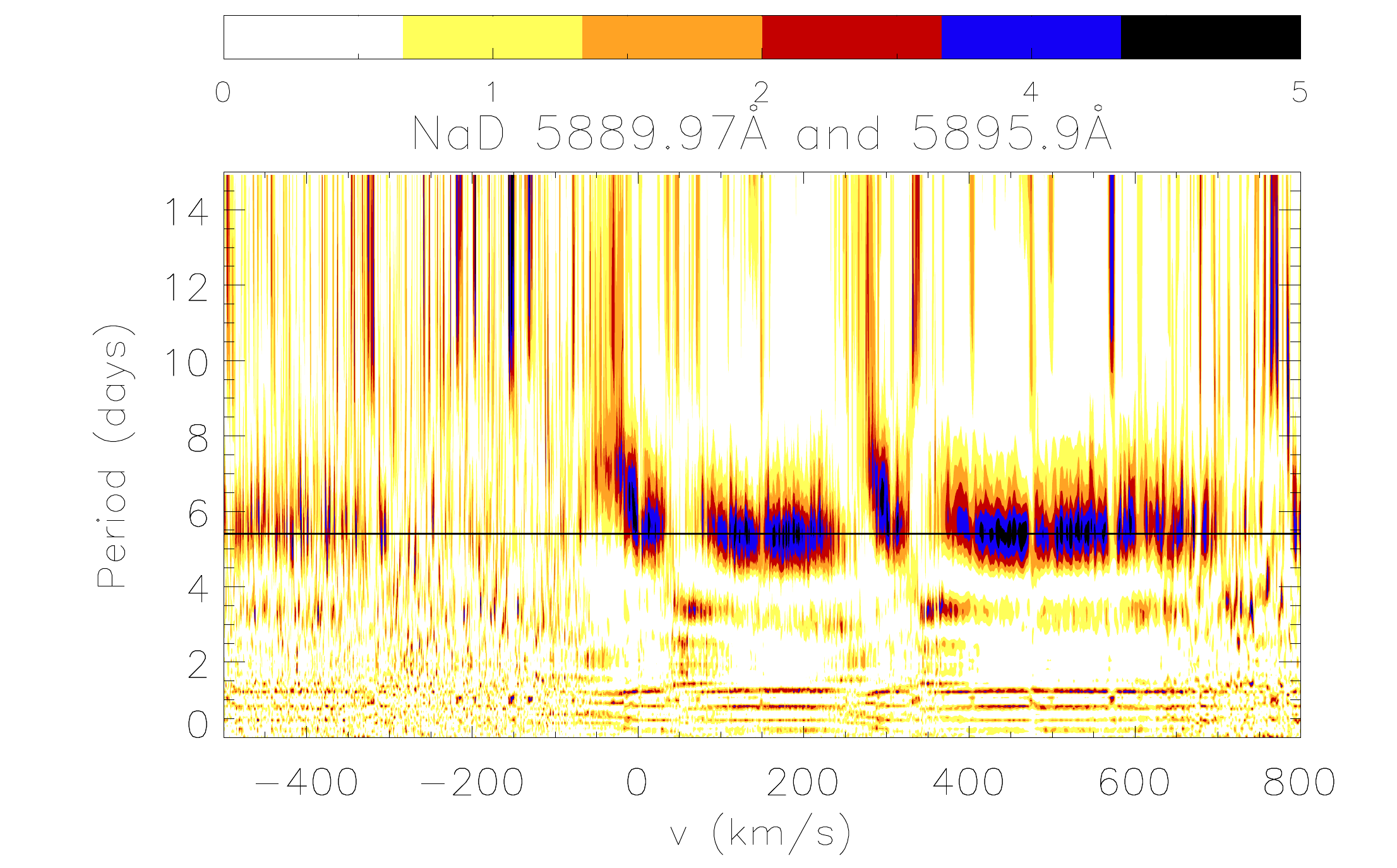}
\includegraphics[width=9cm]{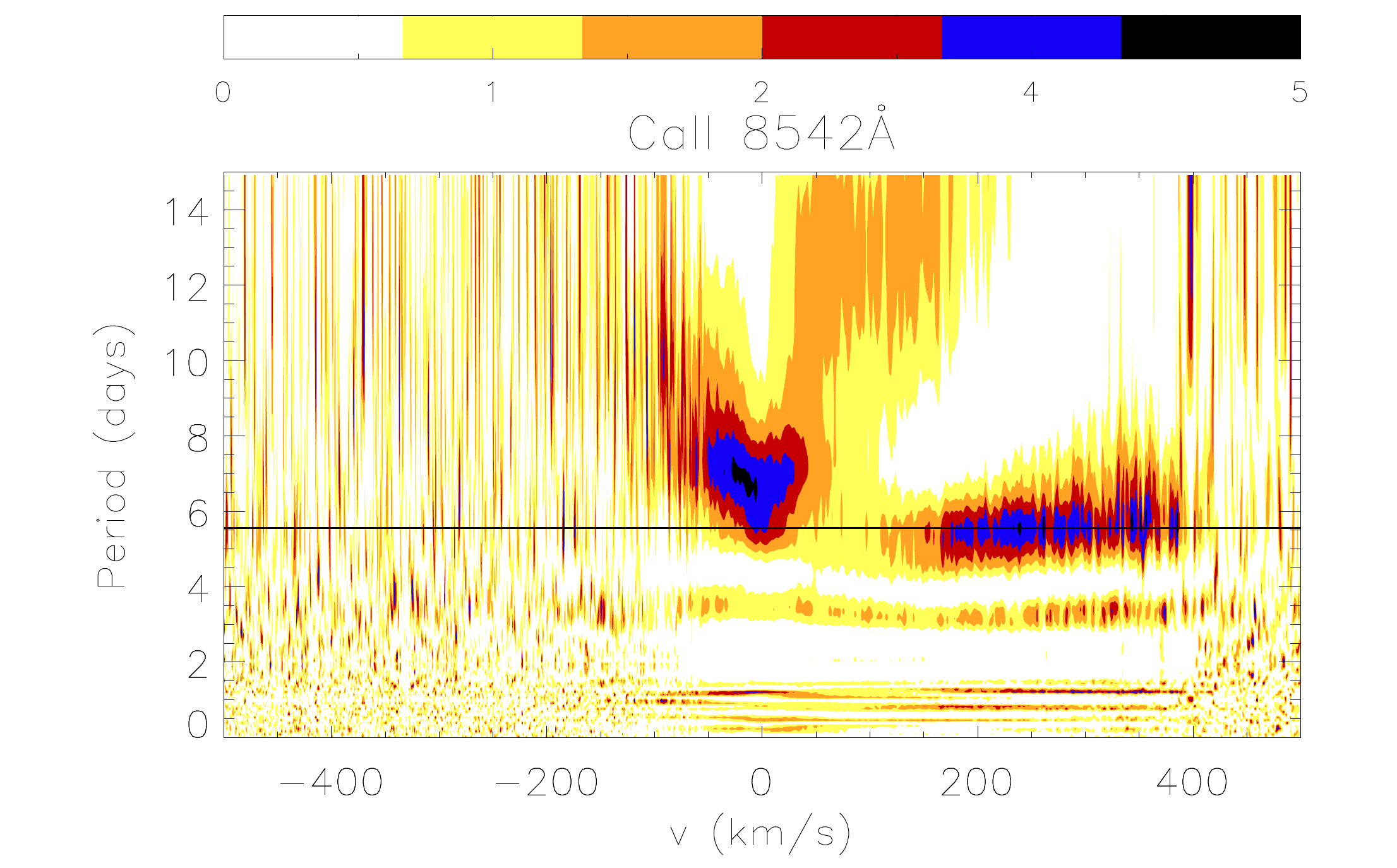}\includegraphics[width=9cm]{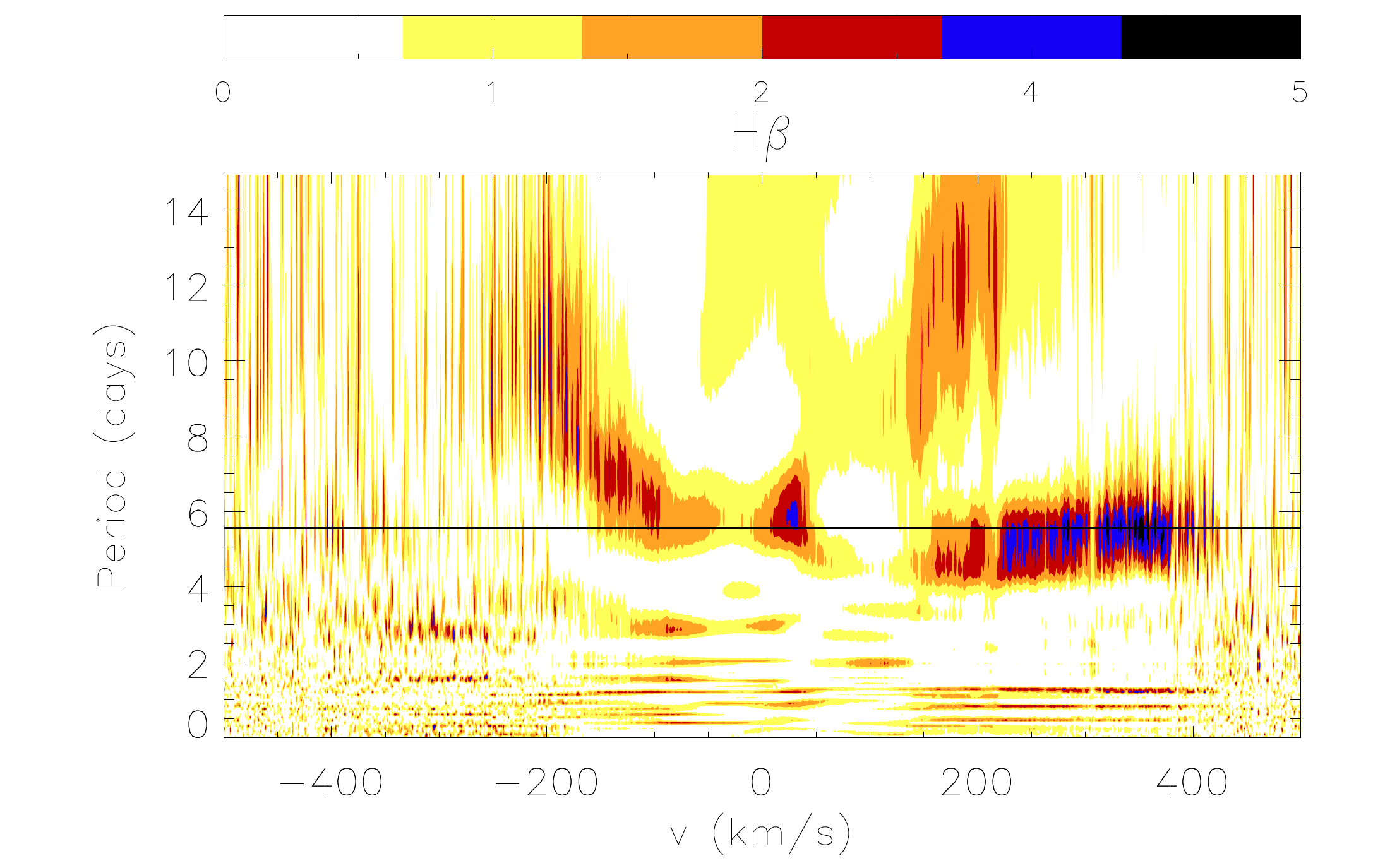}
\includegraphics[width=9cm]{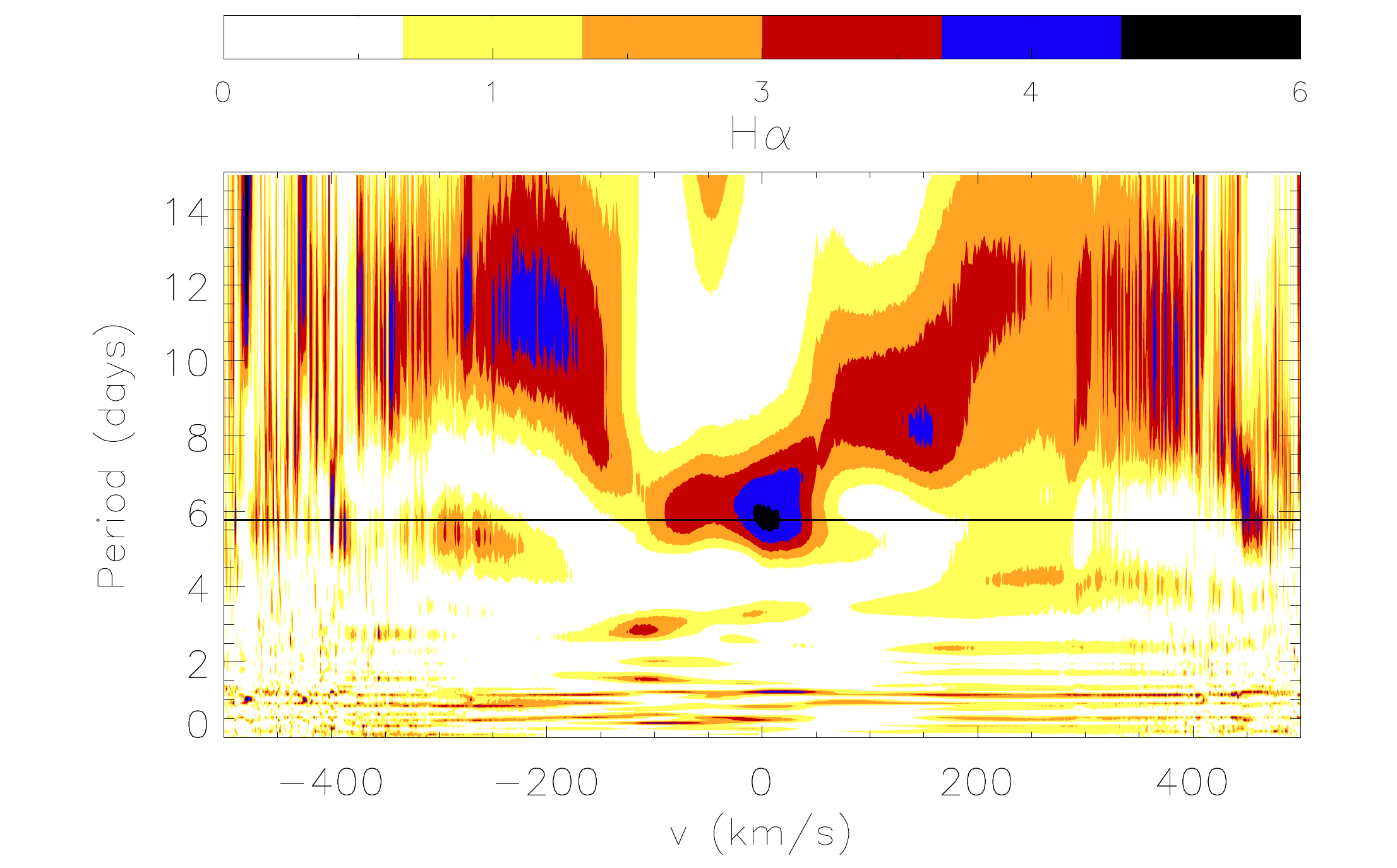}\includegraphics[width=9cm]{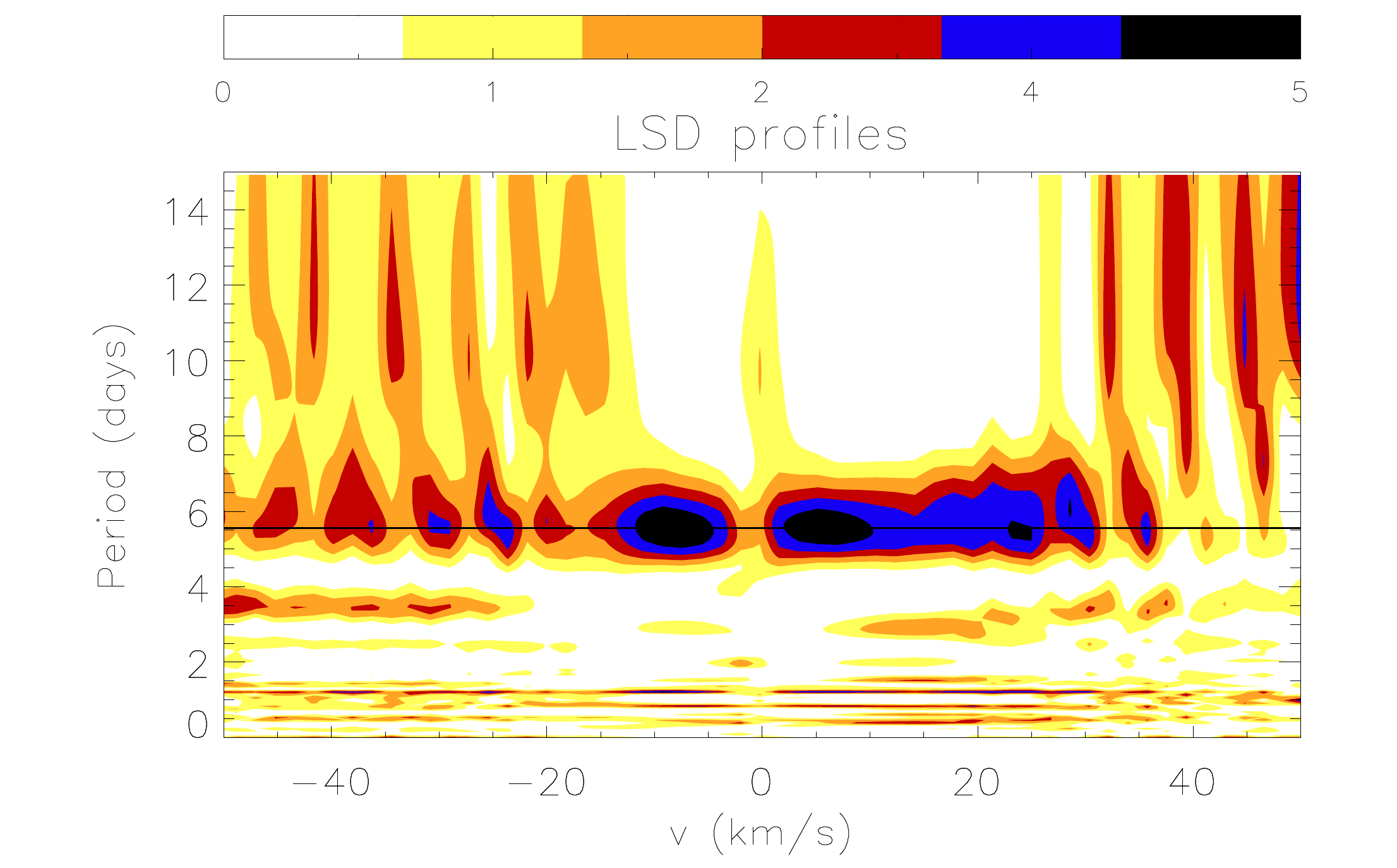}
\caption{HeI 5876\AA\, NaD, CaII 8542\AA\, H$\beta$, H$\alpha,$ and LSD line periodograms.
The top color bars represent the periodogram power, varying from zero (white) to the maximum power (black).
The black horizontal lines correspond to the highest power period of each line, as discussed in the text.}
\label{lines_periodogram}
\end{figure*}

HeI 5876\AA\, presents one of the simplest emission line profiles in our observations,
showing only a narrow emission component \citep[FWHM $< 60$ \kms, according to][]{ber01} 
that periodically varies in intensity, as shown in Figs. \ref{he_profiles}, \ref{he_profiles_oplot}, and \ref{lines_periodogram}. The
strongest emissions are seen close to phase 0.5, when the veiling is maximum
and we expect the accretion spot to be in our line of sight (Fig. \ref{he_profiles}, right). This is in agreement
with a scenario where the HeI 5876\AA\, line is mostly formed in the postshock region,
at the base of the accretion column \citep{ber01}.
The strong modulation of the HeI 5876\AA\, emission, which almost completely disappears at some phases, suggests a high inclination of the inner disk, given the relatively
low obliquity ($20-25$\degr) of the stellar magnetic field 
with respect to the stellar rotation axis \citep[see Sect. \ref{discussion} and][]{don18}.
The HeI 5876\AA\, line shows a well-defined periodic signal at $5.63 \pm 0.66$ days in all velocity bins.
The accretion spot period should correspond to the stellar rotation period, and it indeed
agrees with the period calculated from the variation of the longitudinal magnetic
field of the star. It also agrees within the errors
with the photometric period and supports the assumption that the inner disk warp,
mapped by the photometric variations, is located near the disk corotation radius.

\begin{figure}
\centering
\includegraphics[width=9cm]{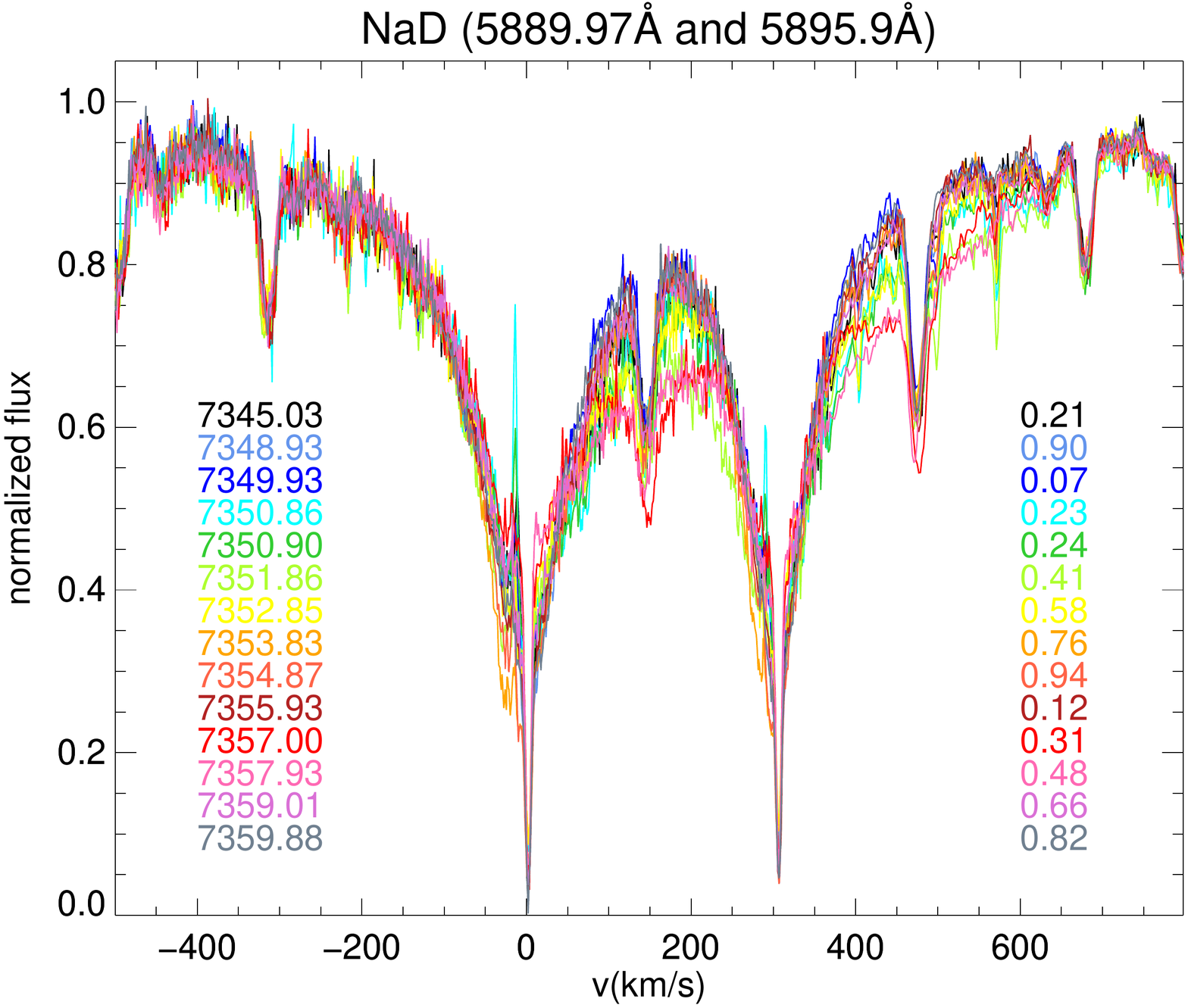}
\caption{Observed  NaD doublet profiles overplotted. The observation dates (left)
and the phases (right) are shown with the same color code as the profiles.}
\label{nad_profiles_oplot}
\end{figure}

\begin{figure}
\centering
\includegraphics[width=4.5cm]{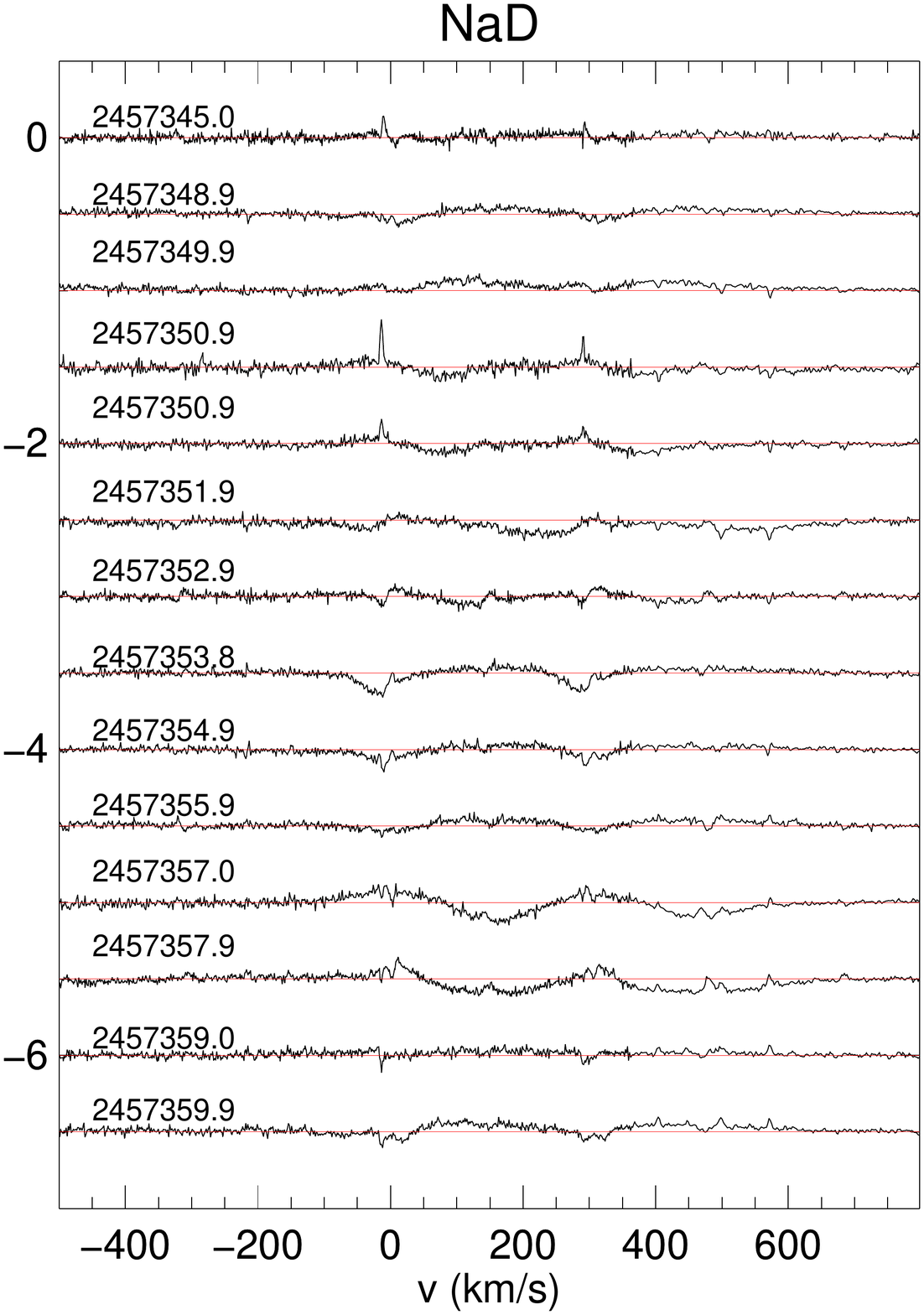}\includegraphics[width=4.5cm]{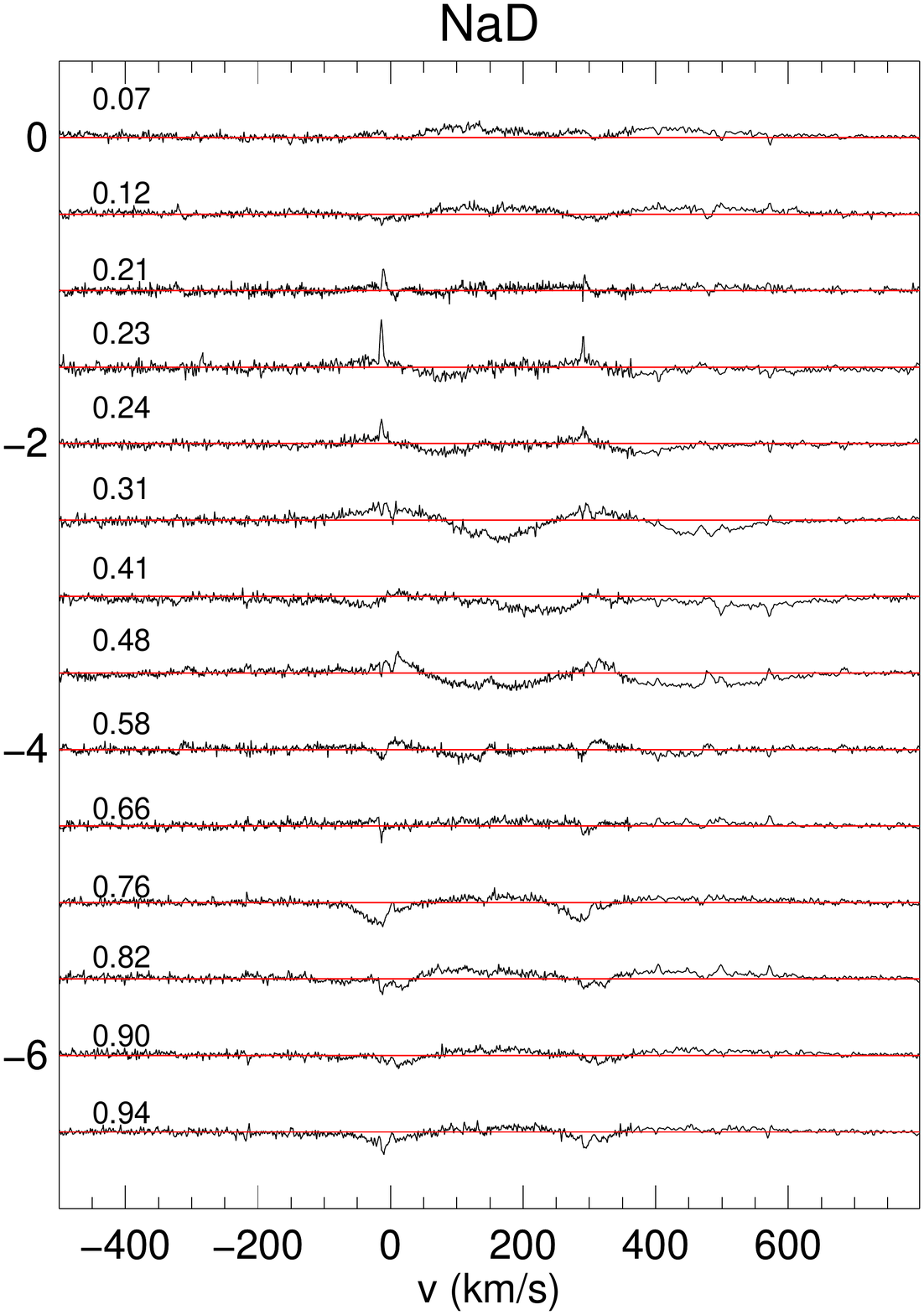}
\caption{Mean profile subtracted from the NaD profiles, in time (left) and in phase (right) with the $5.7$-day
ASAS-SN photometric period and JD0=2457343.8. The red lines represent the continuum level.}
\label{nad_profiles}
\end{figure}

The NaD doublet is mostly photospheric and in absorption, but a strong variability
is present in its redshifted wing, hinting at a possible circumstellar contribution
from high-velocity free-falling gas to this part of the profile, as shown in
Fig. \ref{nad_profiles_oplot}. To enhance the redshifted variability, we subtracted
the mean observed profile from the observations, and we show in Fig. \ref{nad_profiles}
that it extends up to 350 \kms. These redshifted absorptions are periodic
at $5.41 \pm 0.64$ days (Fig. \ref{lines_periodogram}), in agreement with an origin close to
the stellar surface in the high-velocity gas that falls freely toward the star.

\begin{figure}
\centering
\includegraphics[width=4.5cm]{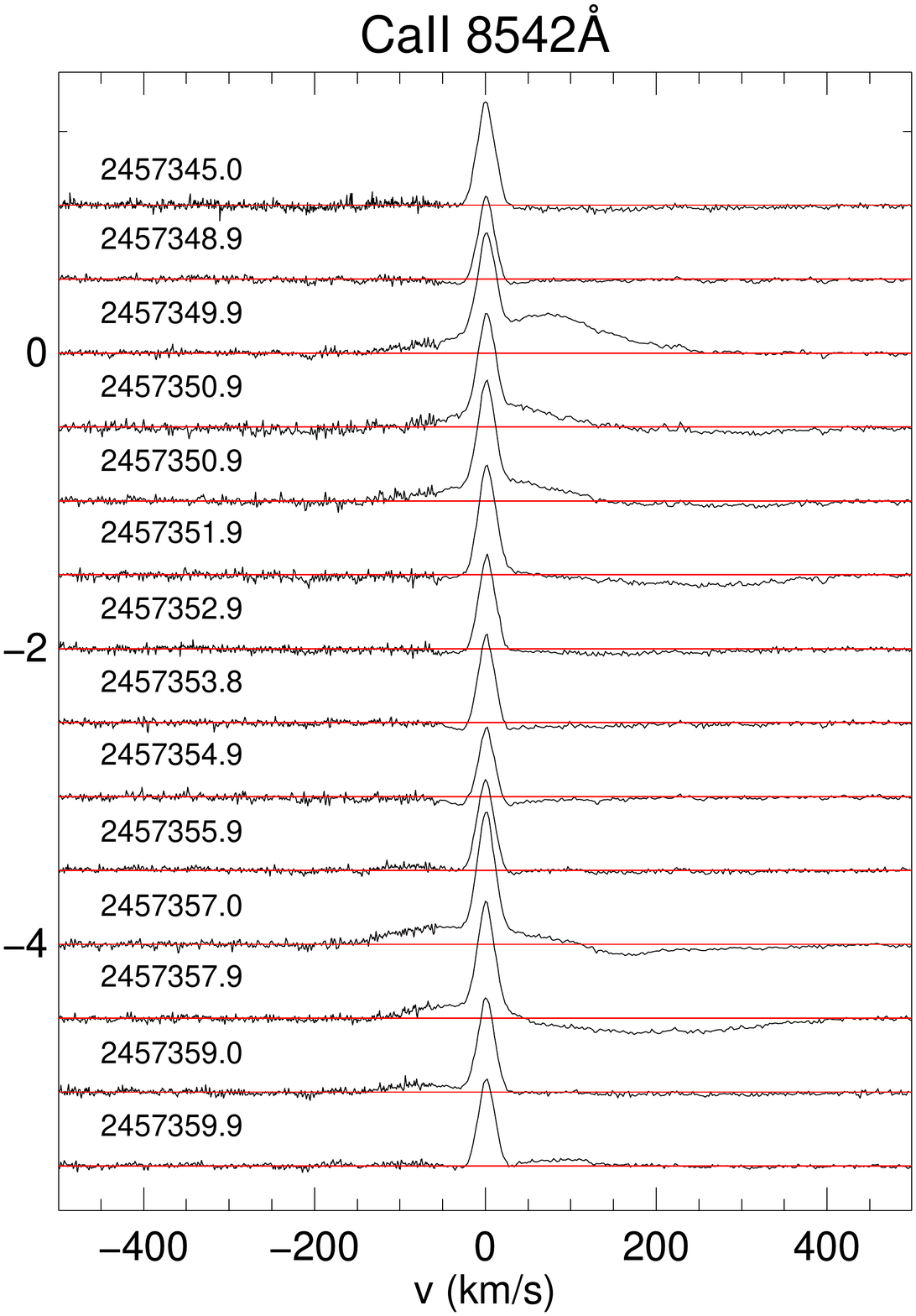}\includegraphics[width=4.5cm]{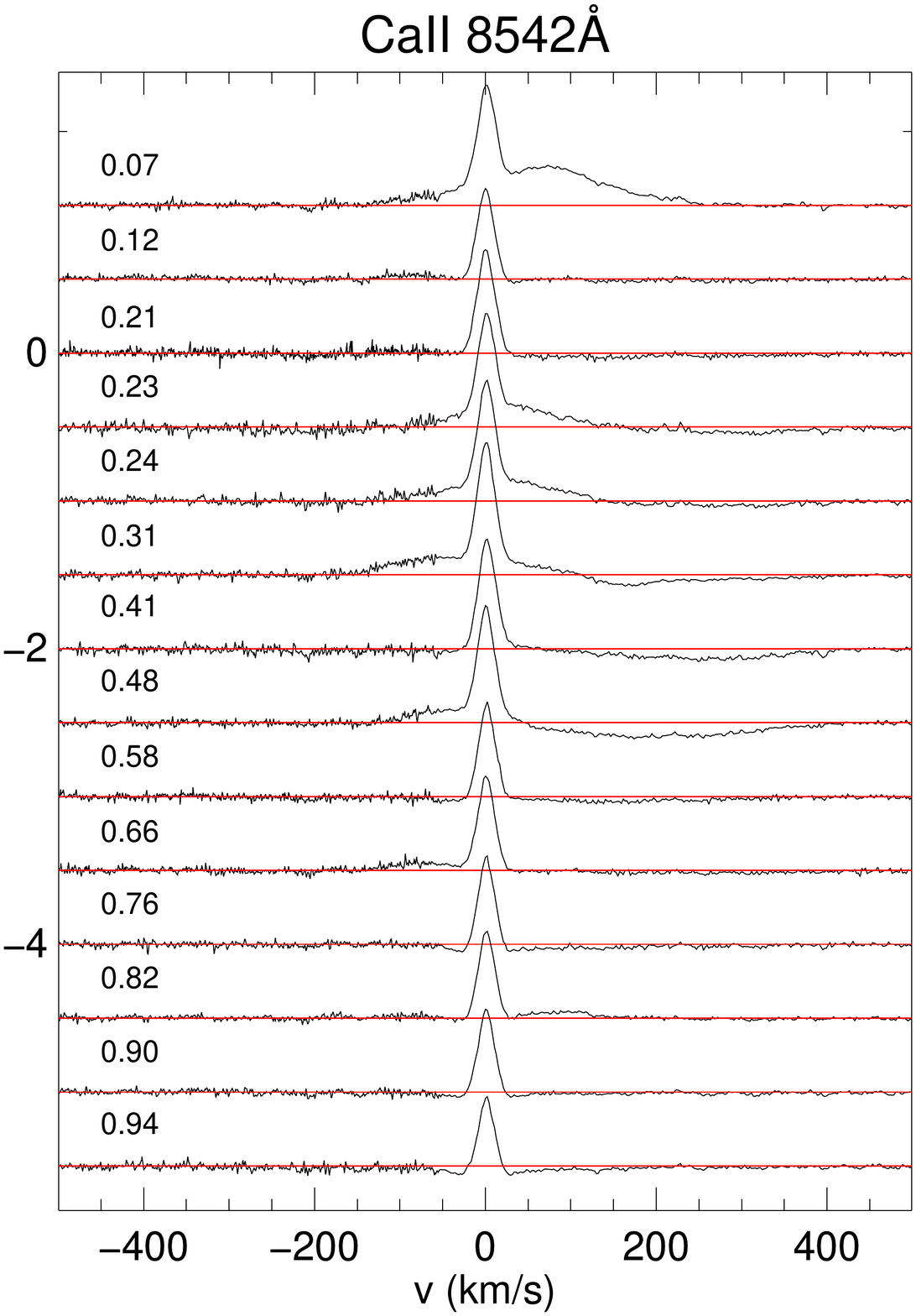}
\caption{Residual CaII 8542\AA\, profiles in time (left) and in phase (right) with the $5.7$-day
ASAS-SN photometric period and JD0=2457343.8. The red lines represent the continuum level.}
\label{ca_profiles}
\end{figure}

\begin{figure}
\centering
\includegraphics[width=9cm]{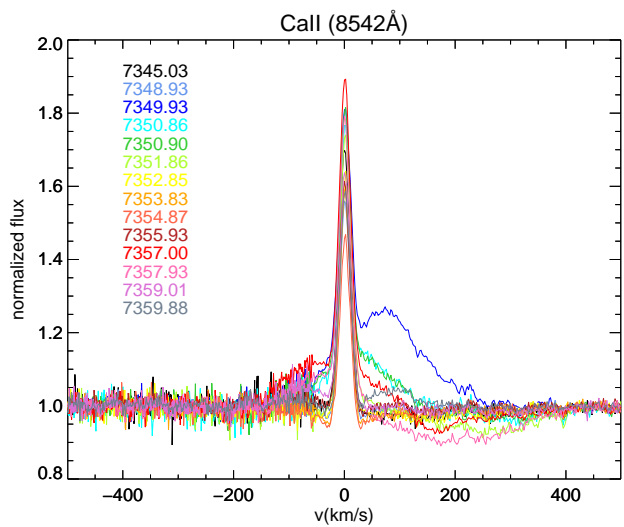}
\caption{Residual CaII 8542\AA\,  profiles overplotted. The observation dates are shown
with the same color code as the profiles.}
\label{ca_profiles_oplot}
\end{figure}

The CaII 8542\AA\, line is part of the CaII infrared triplet. The original spectrum
shows a narrow emission of chromospheric and postshock origin inside a photospheric absorption profile.
To evaluate only the circumstellar contribution to the line profiles, we calculated
residual spectra by subtracting the rotationally broadened and veiled spectrum of
V819 Tau from the observed spectra. The residual line profile (Figs. \ref{ca_profiles} and
\ref{ca_profiles_oplot}) is composed of a narrow emission, a broad emission that can be either blueshifted
or redshifted, and a redshifted
absorption component that preferentially appears around phase 0.5, when the main
accretion column is expected to be in our line of sight.
The redshifted absorption component is indicative of accretion, and is thought to originate
in the accretion funnel, when photons from the accretion spot are absorbed by gas in the funnel at
high velocity, moving away from us. This part of the profile presents a periodicity at
$5.56 \pm 0.65$ days (Fig. \ref{lines_periodogram}),
in agreement with the HeI 5876\AA\, and the photometric period. The narrow emission, however,
apparently shows a slightly longer variability at $6.7 \pm 1.0$ days. To investigate the different
periodicity presented by the narrow emission, we decomposed the CaII 8542\AA\, profiles by
fitting 1, 2, or 3 Gaussians, depending on the number of components present. The decomposition
allows the individual analysis of each component variability. After separating the narrow
emission and the broad component emission, we searched for periodicities in the component
intensities, equivalent widths (EW), and radial velocities. The narrow emission intensity
and EW present periods of $5.76 \pm 0.80$ days and $5.83 \pm 0.80$ days, respectively,
but no significant period was found with the radial velocity variations.
The broad emission component shows strong variability, both in radial velocity and intensity,
but we were unable to detect a significant period. Only ten observations present this component, which can be either blueshifted or redshifted.
The period at $6.7 \pm 1.0$ days detected in the 2D periodogram near the line center therefore probably comes from the mixed variability of the superposed narrow and broad emissions.

\begin{figure}
\centering
\includegraphics[width=9cm]{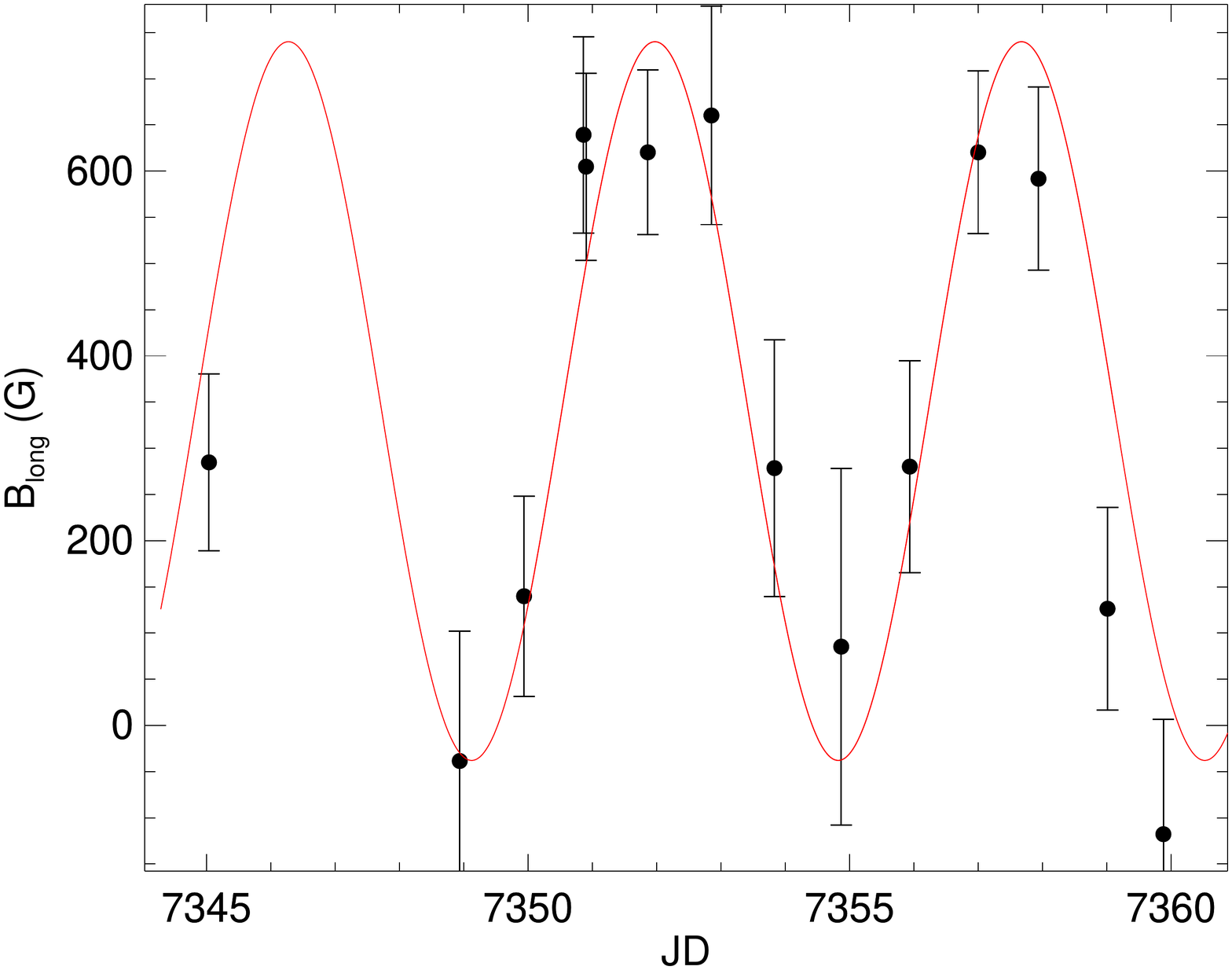}
\includegraphics[width=9cm]{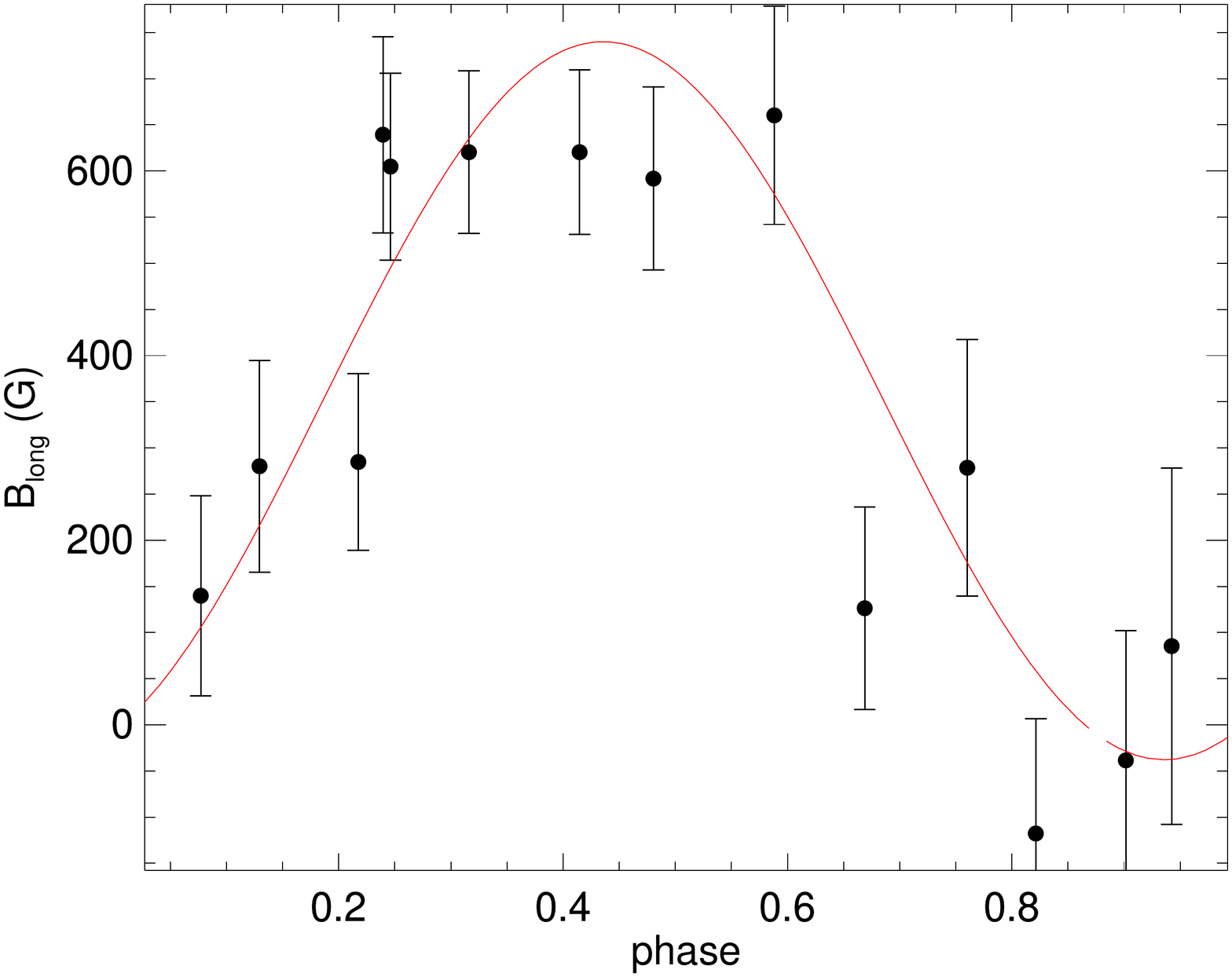}
\caption{Longitudinal component of the magnetic field of LkCa 15 from CaII 8542\AA\, 
profiles as a function of time (top) and in phase (bottom) with the $5.7$-day ASAS-SN 
photometric period and JD0=2457343.8. A sine wave at the $5.7$-day period is overplotted.}
\label{fig_ca_blong}
\end{figure}

The CaII 8542\AA\ line presents a clear signal in Stokes V, which is as broad as the narrow
emission component present in the Stokes I data. We subtracted from the residual CaII 8542\AA\, 
Stokes I profiles a Gaussian fit to the broad
emission component present in some spectra, and we calculated the longitudinal magnetic field
as the first moment of the Stokes V data. The results are shown in Fig. \ref{fig_ca_blong}. 
The longitudinal magnetic field in the region of CaII 8542\AA\, formation varies in
polarity and intensity from $-118$ G 
to $+660$ G, and is stronger near phase 0.5, in agreement with the maximum veiling,
which indicates that this is when the accretion spot faces us. 
The variation is periodical at $5.41 \pm 0.74$ days, which agrees with the stellar rotation period within the errors.
This result strongly suggests that part of the narrow CaII emission is produced in the postshock region.

\begin{figure}
\centering
\includegraphics[width=4.5cm]{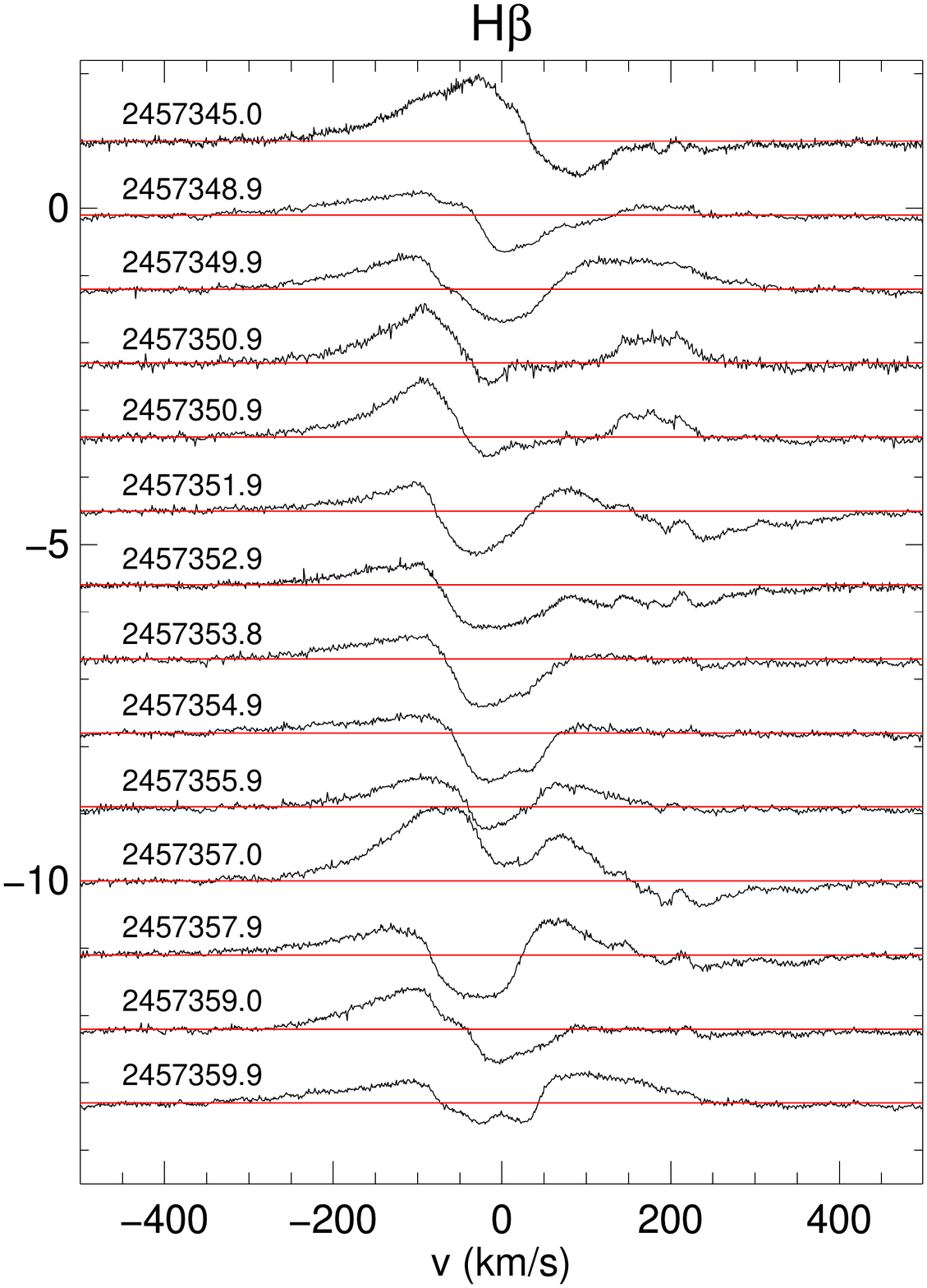}\includegraphics[width=4.5cm]{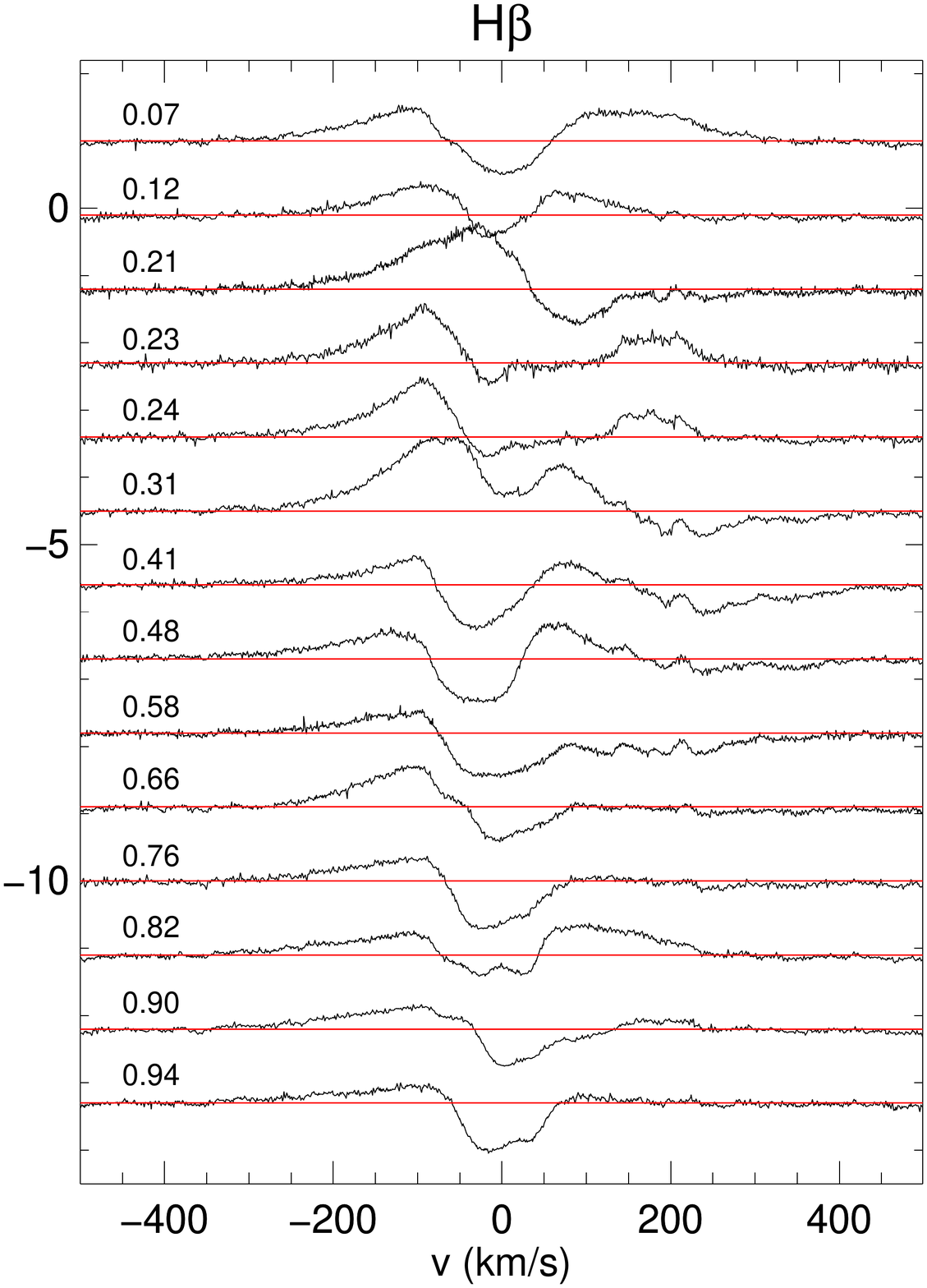}
\caption{Residual H$\beta$  profiles in time (left) and in phase (right) with the $5.7$-day
ASAS-SN photometric period and JD0=2457343.8. The red lines represent the continuum level.}
\label{hb_profiles}
\end{figure}

\begin{figure}
\centering
\includegraphics[width=9cm]{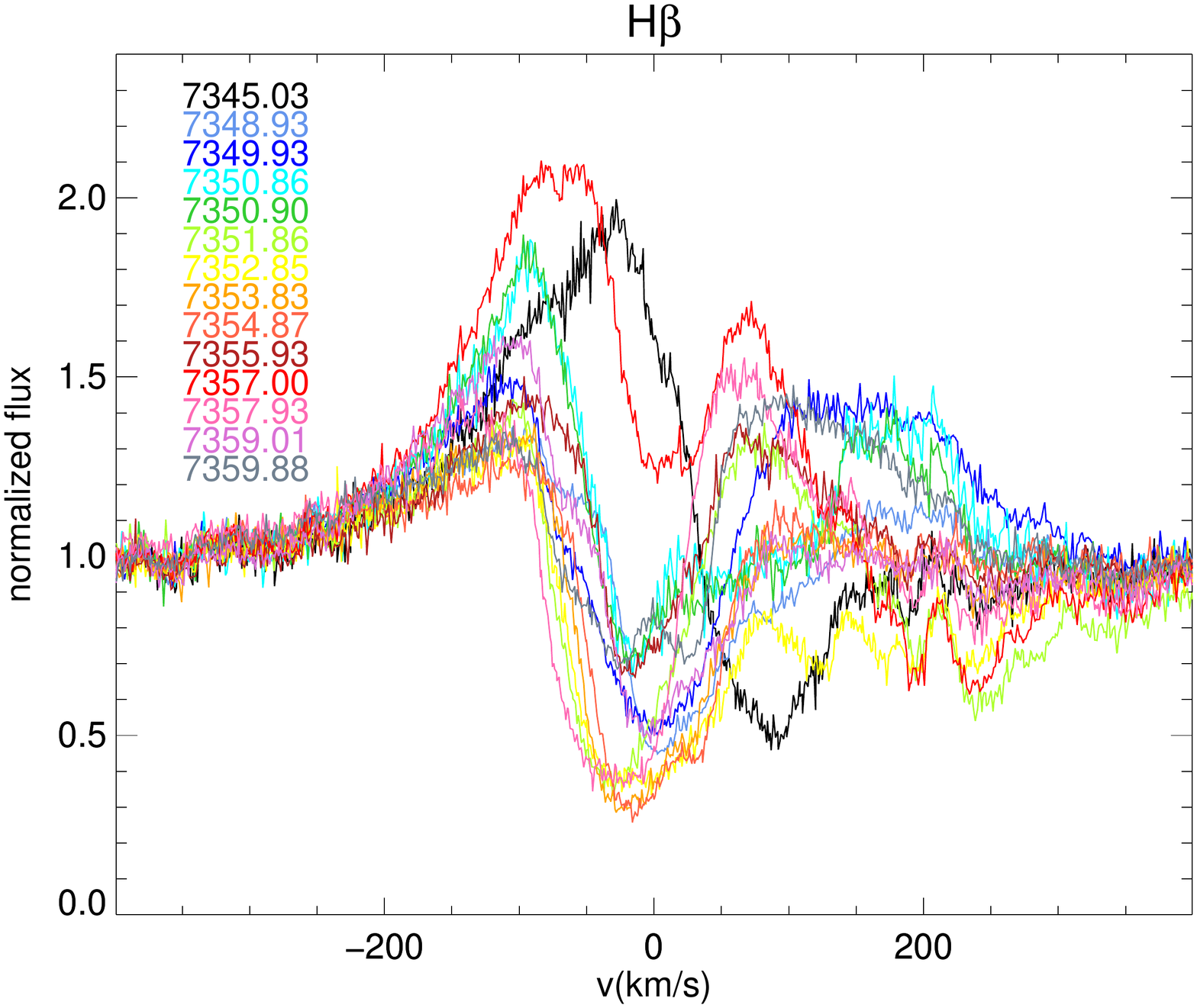}
\caption{Residual H$\beta$ profiles overplotted. The observation dates are shown
with the same color code as the profiles.}
\label{hb_profiles_oplot}
\end{figure}

The H$\beta$ spectral region presents several photospheric absorption lines that
are superposed on the H$\beta$ emission. To remove the photospheric lines, we again calculated
residual profiles by subtracting the rotationally broadened and veiled spectrum
of V819 Tau. The resulting spectra are shown in Figs. \ref{hb_profiles} and \ref{hb_profiles_oplot}.
The residual H$\beta$ line profile is complex and presents an emission with several absorption
components of non-photospheric origin. We can identify a blueshifted absorption component near
50 \kms, a low-velocity redshifted absorption component, and at some phases, a high-velocity
redshifted absorption component. The latter is more visible near phase 0.5, but
cannot be discarded at other phases. The red wing varies with
a period of $5.56 \pm 0.75$ days and the line center at $5.85 \pm 0.76$ days (Fig. \ref{lines_periodogram}),
both agreeing with the variability of the HeI 5876\AA\, line,
the redshifted wing of CaII 8542\AA,\, and with the photometric period, hinting at an origin close to
the stellar surface. The blue wing of the profile shows a more complex variability
without a significant periodicity and may originate in various regions, such as the accretion funnel and a wind,
that do not necessarily vary at the same period.

\begin{figure}
\centering
\includegraphics[width=4.5cm]{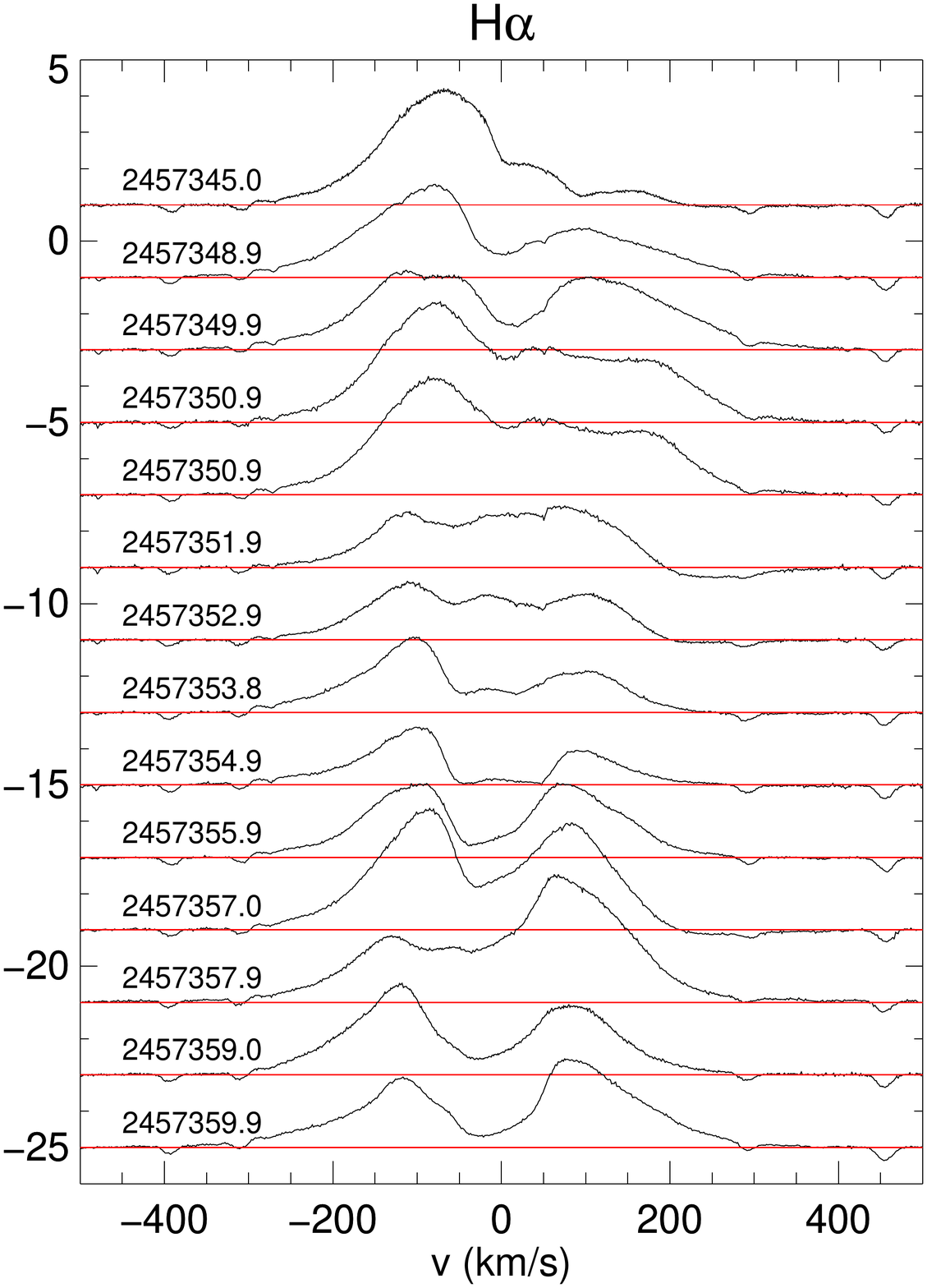}\includegraphics[width=4.5cm]{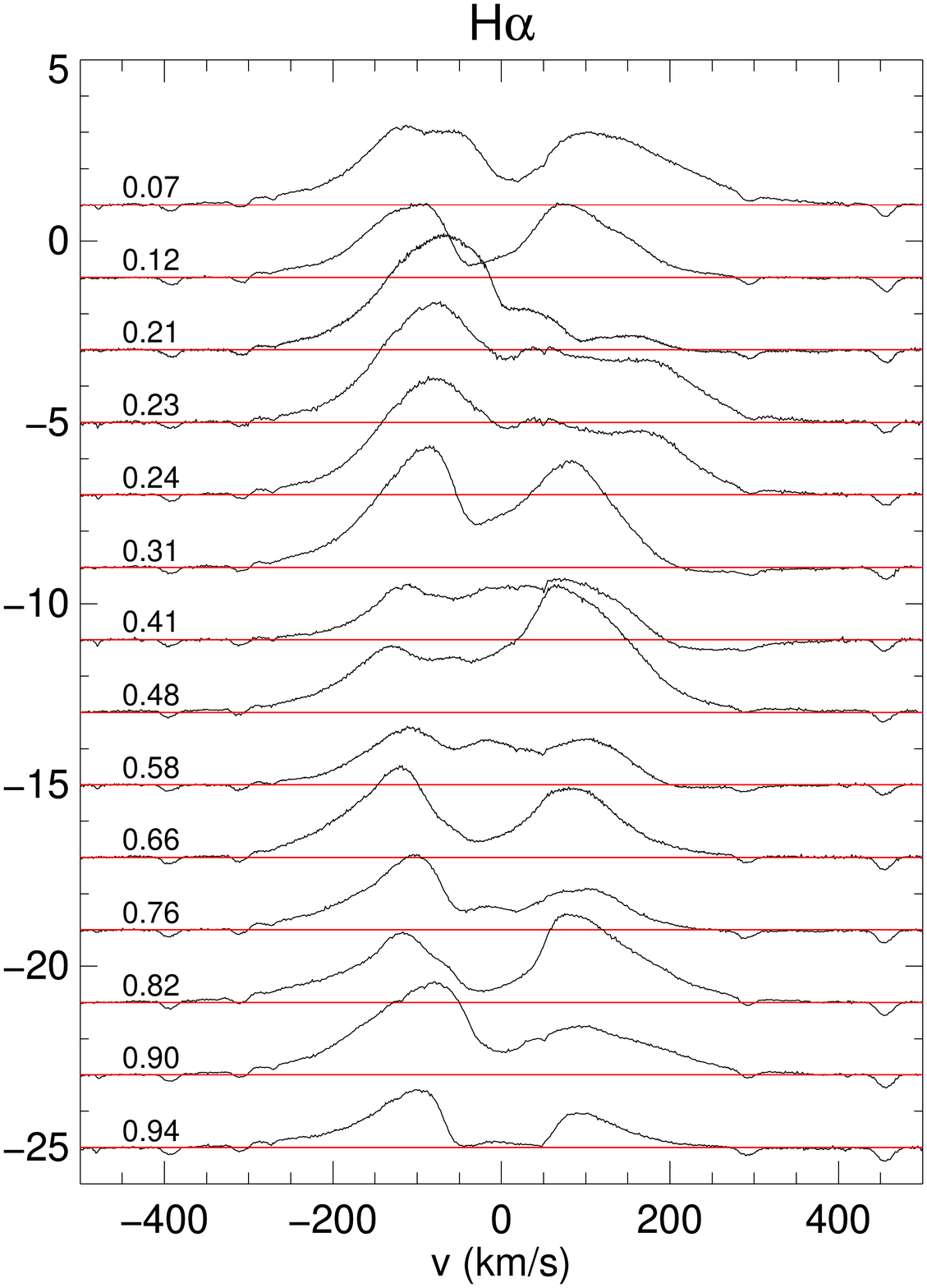}
\caption{H$\alpha$ observed profiles in time (left) and in phase (right) with the $5.7$-day
ASAS-SN photometric period and JD0=2457343.8. The red lines represent the continuum level.}
\label{ha_profiles}
\end{figure}

\begin{figure}
\centering
\includegraphics[width=9cm]{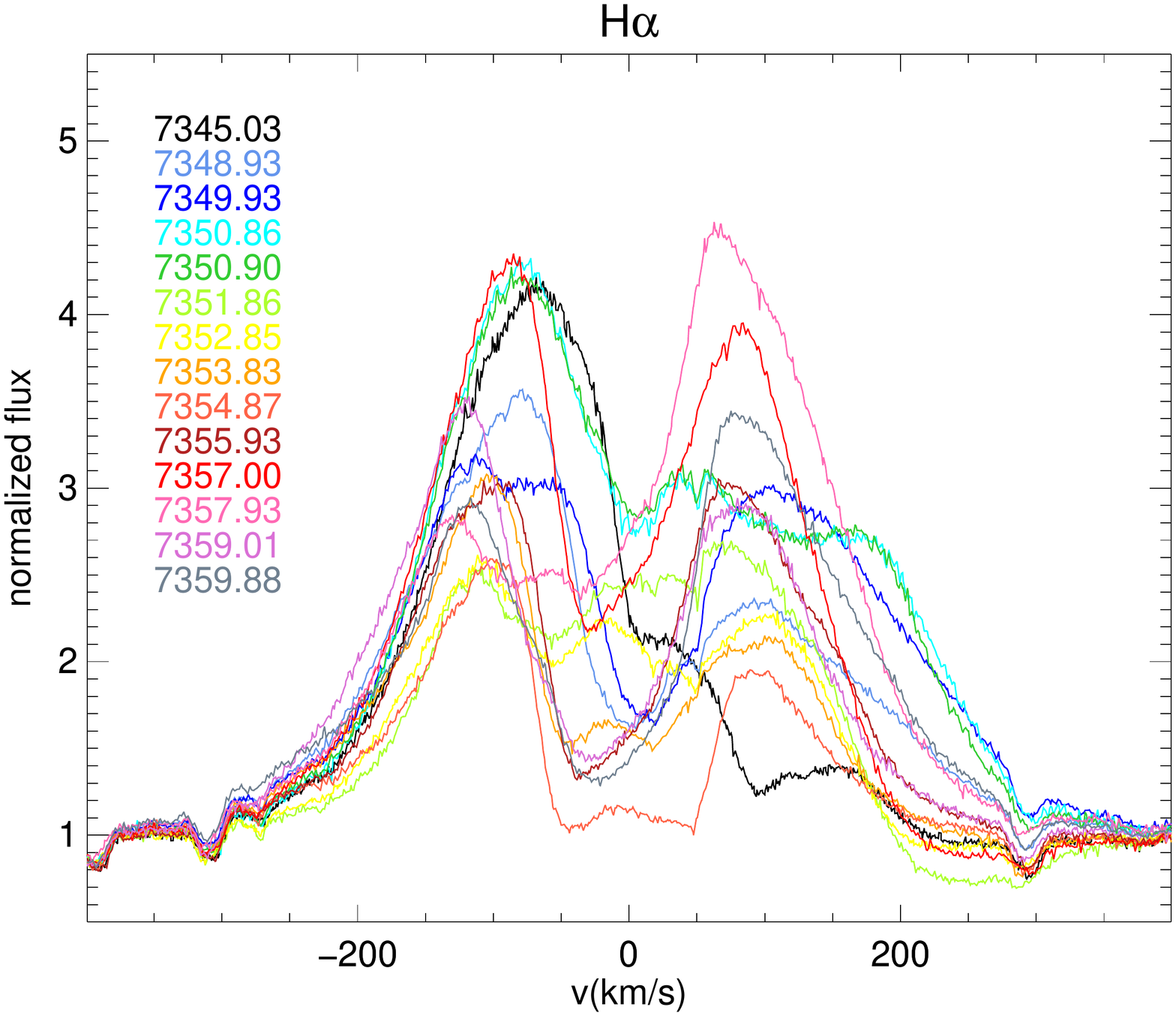}
\caption{Observed H$\alpha$  profiles overplotted. The observation dates are shown
with the same color code as the profiles.}
\label{ha_profiles_oplot}
\end{figure}

The H$\alpha$ emission line, like H$\beta$, presents a rather complex and variable profile,
as we show in Figs. \ref{ha_profiles} and \ref{ha_profiles_oplot}.
A strong emission, a blueshifted absorption, and two
redshifted absorption components can be clearly distinguished. One component lies close to line center and another, which is only
present in a few spectra, is present at high velocity. 
In the Balmer series this is the line with the highest transition probability, and it is expected to form throughout the circumstellar environment of a CTTS: in the disk wind, the accretion funnel,
the accretion spot, and stellar winds. Its variability can therefore be influenced by
variations in all the circumstellar components, and it is not a
surprise that most of the line does not present a single and highly significant period
(Fig. \ref{lines_periodogram}).
The only exception is the region near the line center, which shows a periodicity at $5.78 \pm 0.68$ days,
like the same region of H$\beta$.

\begin{figure}
\centering
\includegraphics[width=9cm]{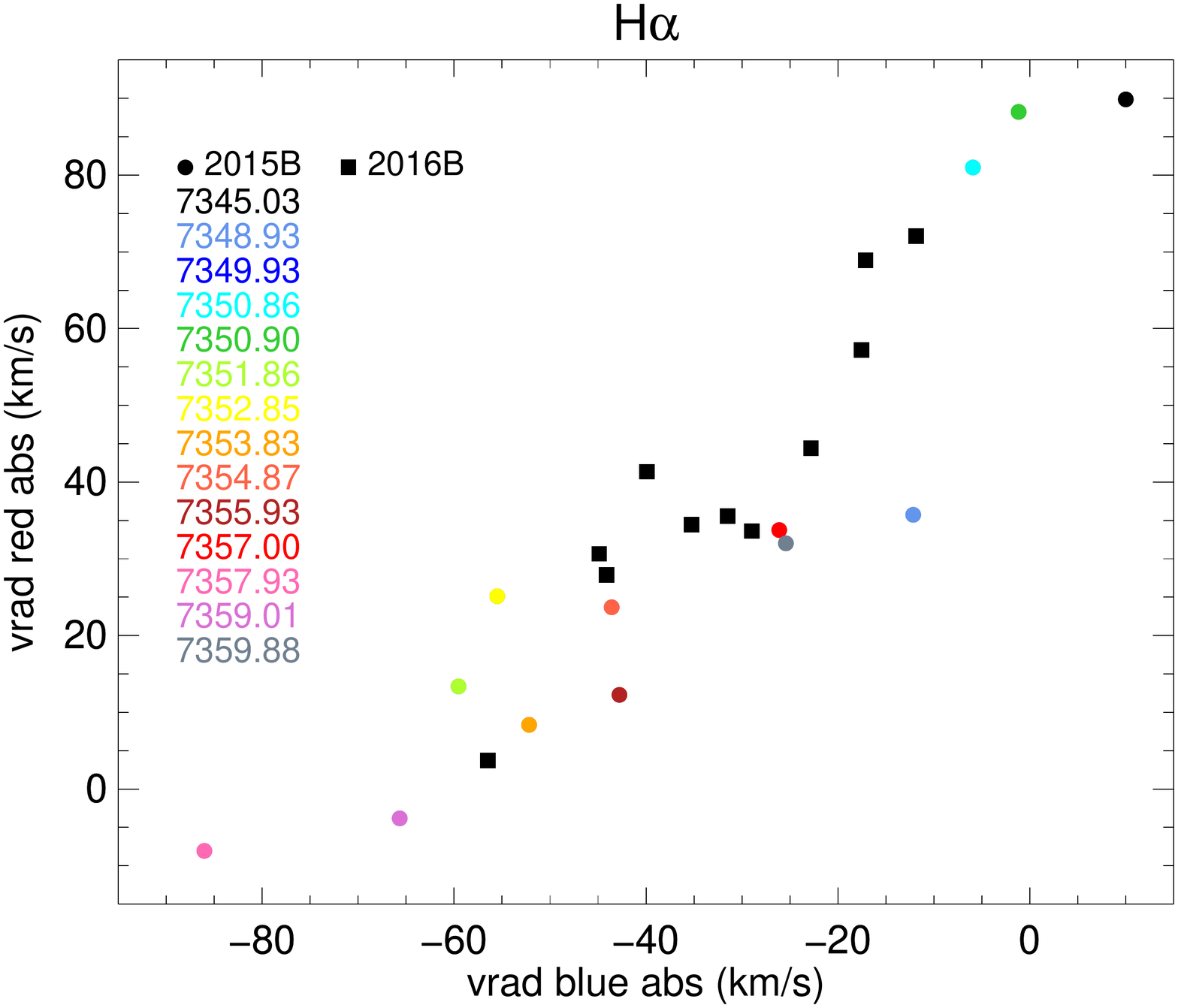}
\caption{H$\alpha$ blueshifted and redshifted radial velocity variability.
Filled circles represent the 2015B data and filled squares the 2016B data.
The observation dates of the 2015B run we analyzed here are shown with
the same color code as the corresponding points.}
\label{line_inflation}
\end{figure}

Although complex, the different components of the H$\alpha$ profile are clearly 
distinguishable, which allowed us to decompose the profiles with
Gaussians in order to measure their parameters. The decomposition is shown in Sect. \ref{appendixA} of the Appendix. 
The star-disk interaction is expected to be very dynamic, as shown by magnetohydrodynamics (MHD) simulations
\citep{zan13,rom09,kur12}. The stellar magnetic field may interact with the inner disk at a 
range of radii, and as a result of differential rotation between the star and the disk, as 
the system rotates, the magnetic field lines inflate, twist, reconnect, and the cycle 
starts again. To probe the inflation of the magnetic field lines, we can analyze the
radial velocity of the H$\alpha$ absorption components, measured as
the central velocity of the corresponding Gaussian components. 
The redshifted component is thought to come from the absorption of photons by the
gas in the accretion column, and the blueshifted component is expected to trace
the inner disk wind. As depicted in Fig. 19 of \citet{bou03}, when the
field lines inflate, the radial velocities of the redshifted and blueshifted components
vary in opposite directions. This behavior was seen in two different
observing campaigns of AA Tau \citep[see Fig. 11 of][]{bou07}, and it is clearly seen in our two observing
seasons of LkCa 15 (Fig. \ref{line_inflation}).

\subsection{Correlation matrices}

Correlation matrices compare two sets of spectral lines that are observed simultaneously and show how
correlated their variabilities are across the line profiles. A strong correlation
among emission components, for example, indicates a common origin
of the components. Correlation matrices can be calculated with datasets
of the same line (autocorrelation), or different lines.

\begin{figure*}
\centering
\includegraphics[width=16cm]{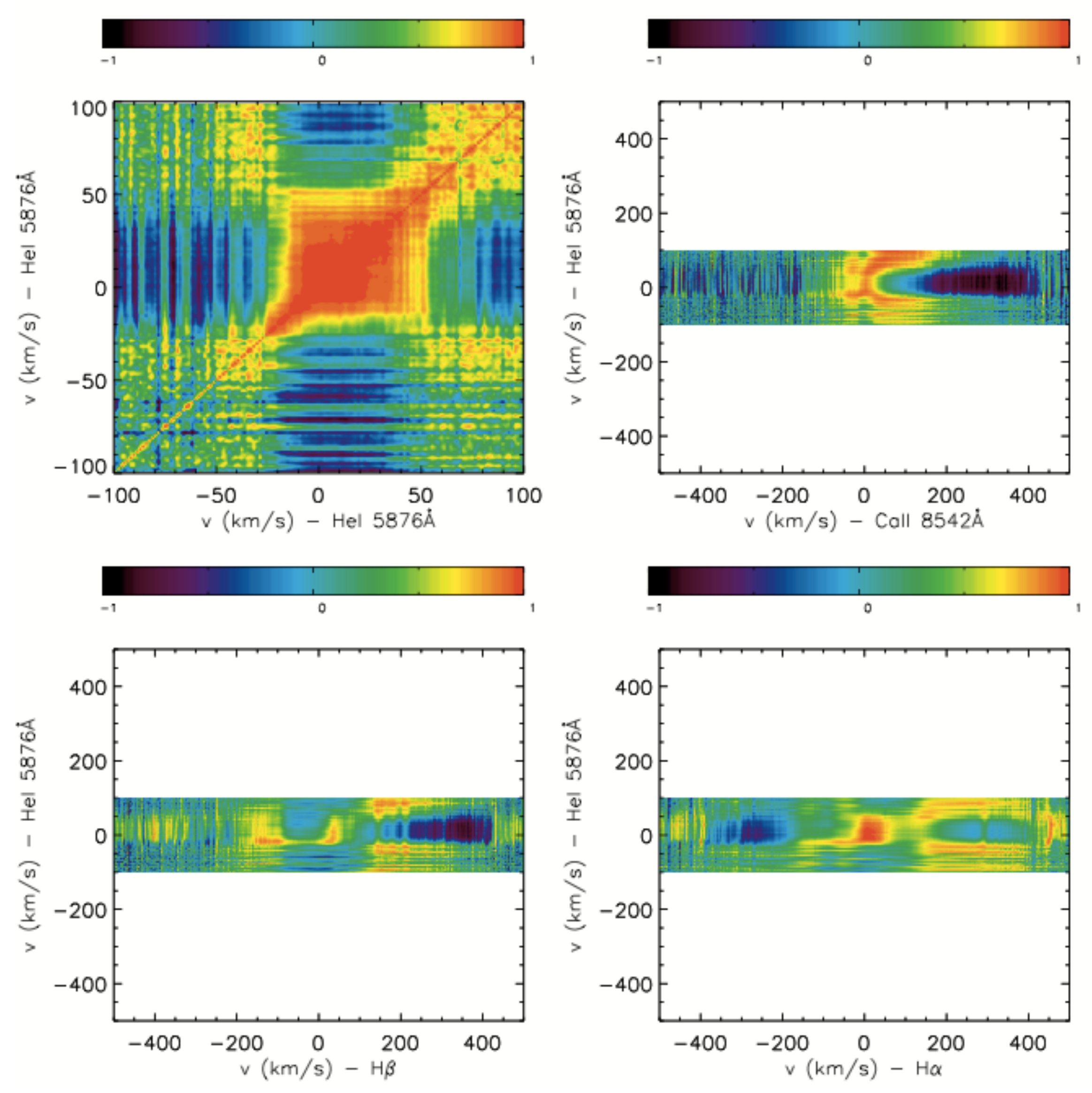}
\caption{HeI 5876\AA\, autocorrelation matrix (top left), CaII 8542\AA\, vs. HeI 5876\AA\,
correlation matrix (top right), H$\beta$ vs. HeI 5876\AA\, correlation matrix (bottom left),
and H$\alpha$ vs. HeI 5876\AA\, correlation matrix (bottom right). The top color bars represent 
the linear correlation coefficient, which varies from $-1$ (black), corresponding to anticorrelated
regions, to $+1$ (red), corresponding to correlated regions.}
\label{line_matrix}
\end{figure*}

The HeI 5876\AA\, line consists of a single narrow component that varies at the stellar
rotation period. Its autocorrelation matrix shows that the narrow component is coherent
with a formation in a single region, as the entire profile is well correlated (Fig. \ref{line_matrix}, top left). 

A comparison of the HeI 5876\AA\, line variations with the line profile variations of other
emission lines can show which parts of the other lines are formed in about the same region
as the HeI 5876\AA\, line.

The HeI 5876\AA\, line and the CaII 8542\AA\, narrow chromospheric emission present a good correlation (Fig.
\ref{line_matrix} top right), which indicates that both
components are related to the accretion shock region. We also see a strong anticorrelation
between the HeI 5876\AA\, line and the CaII 8542\AA\, or the H$\beta$ high-velocity ($v > 150$ \kms) redshifted wings
(Fig. \ref{line_matrix} top right and bottom left).
Since these parts of the CaII 8542\AA\, and H$\beta$ profiles are dominated by the periodic appearance of
a redshifted absorption component, the anticorrelation indicates that when the HeI profile
is stronger, the redshifted component decreases (becomes deeper). This strongly relates
the redshifted absorption component to the appearance of the accretion spot, as expected
from magnetospheric accretion models.

The correlation matrix of H$\alpha$ versus HeI reflects the complexity of H$\alpha$
(Fig. \ref{line_matrix} bottom right).
Slightly redshifted from the H$\alpha$ line center, we see a good correlation with
the HeI line profile. This is a highly variable region in H$\alpha,$ and
a substantial part of it is apparently related to the accretion flow close to the star. The
H$\alpha$ high-velocity blue wing, which does not present a strong variability,
shows some anticorrelation with the HeI line that is difficult to interpret, since
no clear component is associated with this part of the H$\alpha$ line.

\begin{figure*}
\centering
\includegraphics[width=16cm]{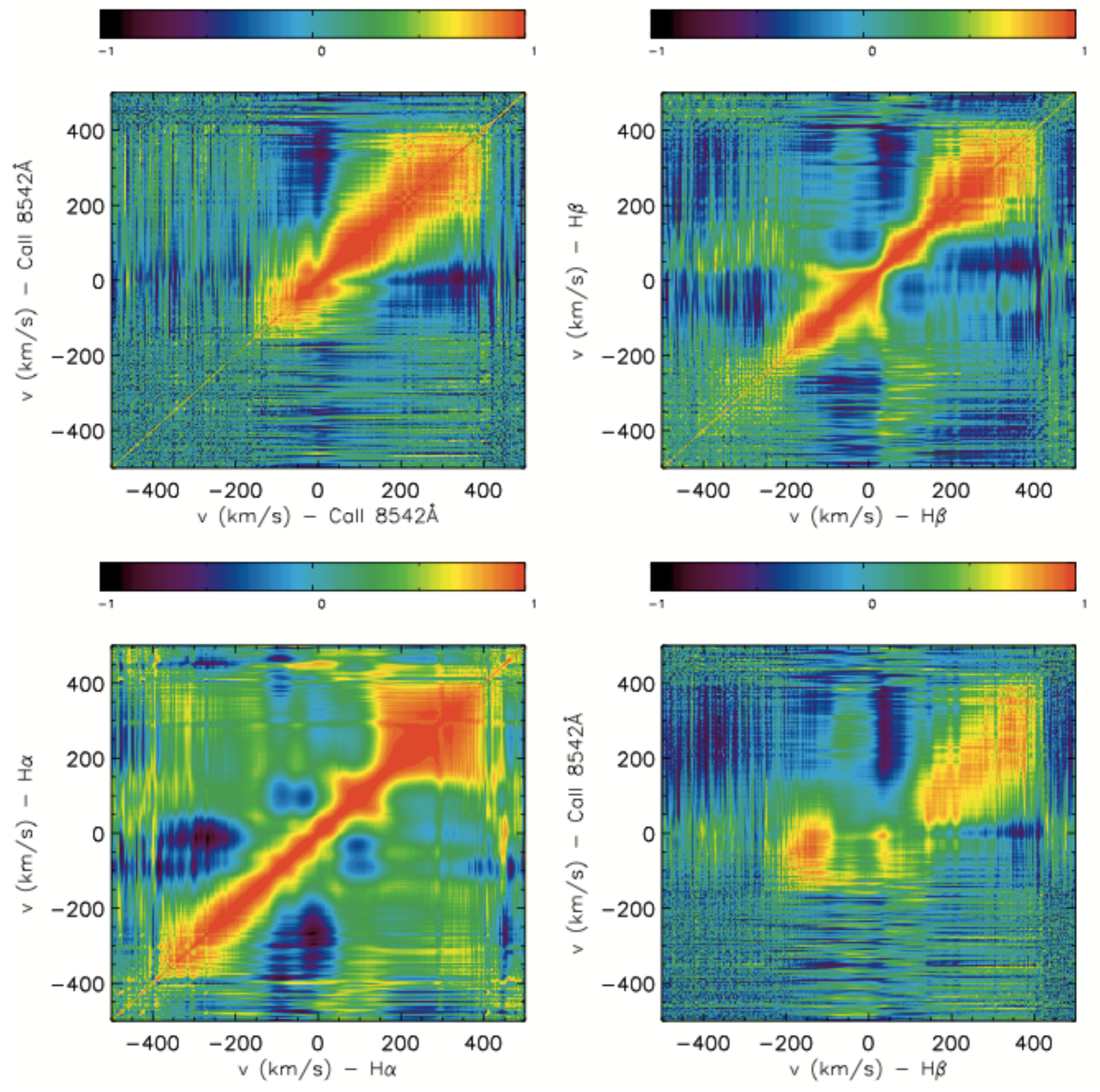}
\caption{CaII 8542\AA\,, H$\beta,$ and H$\alpha$ autocorrelation matrices (top left, right, and bottom left), 
and H$\beta$ vs. CaII 8542\AA\, correlation matrix (bottom right). The top color bars represent 
the linear correlation coefficient, which varies from $-1$ (black), corresponding to anticorrelated 
regions, to $+1$ (red), corresponding to correlated regions.}
\label{line_matrix2}
\end{figure*}

The CaII 8542\AA\, autocorrelation matrix shows that the red high-velocity wing
of the profile correlates well with itself, but different parts of the
line are not strongly correlated with each other (Fig. \ref{line_matrix2} top left). This
indicates a formation in different regions.
The H$\beta$ autocorrelation matrix shows that the red high-velocity wing
of the profile correlates well with itself, but different parts of the
line are not strongly correlated with each other 
(Fig. \ref{line_matrix2} top right).

The H$\alpha$ autocorrelation matrix shows that the red high-velocity wing
of the profile correlates well with itself, as in H$\beta$. We also clearly
see that the line center is anticorrelated with the high-velocity blue wing, which is
difficult to interpret (Fig. \ref{line_matrix2} bottom left).

The correlation matrix of H$\beta$ versus CaII shows that the high-velocity
red wings of both lines correlate well, as expected, since they are dominated by
variations that are due to the redshifted absorption component (Fig. \ref{line_matrix2} bottom right). 
The CaII chromospheric
emission and the blue wing of CaII show some correlation with the blueshifted
H$\beta$ emission. Finally, a strong anticorrelation is seen between the H$\beta$
low-velocity and highly variable redshifted region and the CaII high-velocity
component. This may indicate that this H$\beta$ region originates preferentially
close to the accretion spot.

\section{Discussion}\label{discussion}
A consistent picture seems to emerge from the combined spectroscopic and 
photometric variability of the LkCa 15 system, as most of the observed 
continuum and spectral line variability can be understood in the 
framework of a magnetic interaction between an inclined inner disk with
respect to our line of sight and the
stellar magnetosphere. In the prototypical case AA Tau 
\citep[e.g.,][]{bou99,bou03} and, more generally, in most periodic dippers 
\citep[e.g.,][]{mcg15,rod17,bod17}, recurrent luminosity dimmings of up to one 
magnitude or more are interpreted as resulting from the occultation of 
the central star by a rotating inner disk warp located close to the 
corotation radius. The warp itself stems from the inclination of the 
large-scale stellar magnetosphere onto the stellar rotational axis: the 
magnetic obliquity forces the accreted material out of the disk midplane, 
thus creating a dusty, azimuthally extended inner disk warp \citep{rom13,rom15}. As the warp 
rotates at a Keplerian rate, it occults the central star periodically. 
The light curve and color variations of LkCa15 are fully consistent with this 
interpretation: the luminosity period dips (P=5.70 days), their depth 
of up to one magnitude in the V band, the reddening of the system when 
fainter, and the changing shape of the eclipses from one cycle to the 
next are all characteristics of periodic dippers 
\citep[e.g.,][]{bou07,ale10,cod14,mcg15,rod17,bod17}. For a stellar mass of $1.10 \pm0. 03$ \msun, 
and assuming Keplerian rotation for the inner disk, the occultation period of $5.70 \pm 0.05$ days 
would locate the inner disk warp at a distance of $9.3 \pm 0.1$ \rstar $=0.064 \pm 0.001$ au from the 
central star.

We can compare the warp location with the magnetospheric radius that is due to a dipolar 
magnetic field as derived by \citet{bes08}, $r_{\rm m}(R_\star)=2m_s^{2/7}B^{4/7}_{\star}\dot{M}^{{-2}/7}_{\rm acc}M^{{-1}/7}_{\star}R^{5/7}_{\star}$,
where $m_s\sim 1$ is the sonic Mach number measured at the disk midplane 
and $B_{\star}$, $\dot{M}_{\rm acc}$, $M_\star$ and $R_\star$ are in units of
140 G (at the stellar equator), $10^{-8}$ \msun yr$^{-1}$, 0.8 \msun\, , and 2 \rsun\, , respectively. 
We therefore need to estimate the stellar dipole component 
and the mass accretion rate during our observations. Our ZDI analysis of the 2015B data, which includes
the LSD and CaII spectropolarimetric data, indicates that the dipole component of 
the magnetic field is of about 1.35 kG and tilted by $\sim$20\degr\, with respect to the
stellar rotation axis \citep{don18}. 
To calculate the mass accretion rate, we selected spectra outside the eclipses 
(rotational phases earlier than 0.2 or later than 0.8), and assumed V=11.8 mag and
$(V-R)_J=1.1$ at the time 
of the spectroscopic observations, which corresponds to the brightest state of 
LkCa 15 (see Figs. \ref{fig_lc_CrAO}, \ref{fig_lc_ASAS} and \ref{fig_CrAO_colors}). 
Using the expressions from \citet{fer83}, we obtained $(V-R)_C=0.78$ and $R_C=11.02$.
We measured the H$\alpha$ equivalent widths in the five selected spectra and calculated the H$\alpha$ flux
(Table \ref{macc}) with the out-of-eclipse value of $R_C$. Adopting a distance to 
LkCa 15 of $159 \pm 1$ pc \citep{gaia18}, 
and the empirical relation obtained by \citet{fan09} between accretion and H$\alpha$ line luminosity,
we calculated the accretion luminosity. Mass accretion rates were derived from the
relation $\dot{M}_{\rm acc}=\frac{L_{\rm acc}R_\star}{GM\star(1-R_\star/R_{\rm in})}$ \citep{har98},
with the typical value of $R_{\rm in}=5 R_\star$ that is commonly used in the literature. 
We obtained $\dot{M}_{\rm acc}=(7.4 \pm 2.8)\times 10^{-10}$ \msun${\rm yr}^{-1}$, which corresponds
to the mean and standard deviation of our values presented in Table \ref{macc}.
Changing $R_{\rm in}$ from 5 to 10 $R_\star$ introduces a variation of $\pm1 \times 10^{-10}$ \msun${\rm yr}^{-1}$ 
in our mass accretion rate values, which is smaller than the standard deviation 
of the mass accretion rate from our five out-of-eclipse measurements. 
The dipole component of the magnetic field, together with the mass accretion rate,
the stellar mass ($1.10 \pm0. 03$ \msun), and stellar radius ($1.49 \pm 0.15$ \rsun), yield a magnetospheric radius
of $r_m=8.0 \pm 1.4$  \rstar $= 0.055 \pm 0.015$ au, 
which is consistent within the error bars with the corotation radius at $0.064 \pm 0.001$ au.

When we assume a sublimation temperature of 1500 K, the minimum dust sublimation
radius is located at $r_{\rm sub}=4.48 \pm 0.10$ \rstar$=0.031 \pm 0.001$ au \citep[Eq. 1 with $Q_R=1$ from][]{mon02}, and 
consequently at $r_m=(8.0 \pm 1.4)$ \rstar,\, there is dust left to form a warp.
At the corotation radius, $r_{\rm cor}=(9.3 \pm 0.1)$ \rstar, dust would be present
and the interaction of the stellar magnetic field and
the inner disk could explain the presence of a warp there.

\begin{table}
\scriptsize
\caption{Accretion parameters of LkCa 15 calculated from the H$\alpha$ line flux
of out-of-eclipse observations.}
\label{macc}
\centering
\begin{tabular}{c c c c c c c}
\hline\hline
JD & phase & Flux H$\alpha$ & L H$\alpha$ & $\log{L_{\rm acc}}$ & $\dot{M}_{\rm acc}$\\ 
   &       & [erg cm$^{-2}$ s$^{-1}$]  & [erg cm$^{-2}$] & [\lsun] & [\msun yr$^{-1}$]\\ 
\hline
7348.9390 & 0.902 &  1.22$\times 10^{-12}$ &  3.70$\times 10^{30}$ &  -1.82 &  8.15$\times 10^{-10}$ \\
7349.9391 & 0.077 &  1.55$\times 10^{-12}$ &  4.68$\times 10^{30}$ &  -1.70 &  1.09$\times 10^{-9}$ \\
7354.8715 & 0.942 &  5.58$\times 10^{-13}$ &  1.69$\times 10^{30}$ &  -2.25 &  3.05$\times 10^{-10}$ \\
7355.9376 & 0.129 &  1.09$\times 10^{-12}$ &  3.30$\times 10^{30}$ &  -1.89 &  7.05$\times 10^{-10}$ \\
7359.8815 & 0.821 &  1.20$\times 10^{-12}$ &  3.63$\times 10^{30}$ &  -1.83 &  7.96$\times 10^{-10}$ \\
\hline
\end{tabular}
\end{table}

We can also estimate the magnetospheric radius from the free-fall 
velocity (v$_{\rm ff}$) of the accreting gas.
The redshifted absorption components that are often observed at high 
velocities in NaD, CaII, and H$\beta$, extending up to 350 \kms, correspond to 
the projected velocity of the accreting gas that in a simple dipolar case is
v$_{\rm red}=$v$_{\rm ff}\cos{i_{\rm m}}$, 
where $i_{\rm m}$ is the inclination of the magnetosphere with respect 
to our line of sight. We can then write 

\begin{equation}\label{rmag} 
{\rm v}_{\rm ff}=\frac{{\rm v}_{\rm red}}{\cos(i_{\rm 
m})}=\left[\frac{2GM_{\star}}{R_{\star}}\left(1-\frac{R_{\star}}{r_{\rm 
m}}\right)\right]^{1/2} 
,\end{equation} 

with $i_{\rm m}=i_\star-\beta$, where $i_\star$ is the inclination 
of the stellar axis onto the line of sight, and $\beta$ represents the magnetic 
obliquity, that is, the angle between the magnetic and stellar axes. 
If the accretion spot is on the order of a few percent of the stellar photosphere, 
which is commonly the case in CTTSs, we can estimate the obliquity of 
the magnetic field from the radial velocity amplitude of the HeI 5876\AA\, line as 
$\Delta {\rm v}_{\rm rad}=2{\rm v}\sin{i_\star}\cos({\pi/2 - \beta})$ \citep{bou07}. 
We fit the HeI 5876\AA\, emission line with a single Gaussian and 
measured $\Delta {\rm v}_{\rm rad}=11.6\pm1.0$ \kms. With v$\sin{i_\star}=13.82\pm0.50$ \kms, 
we obtain $\beta=25\degr\pm5\degr$, which is consistent with our value of 
$\beta\sim20\degr$ calculated with ZDI \citep{don18}. 
With $\beta=25\degr=i_\star-i_{\rm m}$, $i_\star > 65\degr$, $R_{\star}=1.49 \pm 0.15$ \rsun, 
$M_{\star}=1.10 \pm 0.03$ \msun\, 
and v$_{\rm red}=350$ \kms, we obtain from Eq. \ref{rmag}, $r_{\rm 
m} > 0.014$ au for $i_\star=65\degr$, and $r_{\rm m} > 0.026$ au for $i_\star=75\degr$, which 
are compatible with the warp location, given our errors.

When the dominant component of the stellar magnetosphere on a scale of a 
few stellar radii is a tilted dipole, the free-falling accretion funnel 
flow reaches the stellar surface at high latitudes, close to the magnetic 
poles. A strong shock forms, which yields continuum excess flux from 
X-ray to optical wavelengths, and produces line emission, in particular, 
the narrow component of the HeI 5876\AA\, feature, which is thought to arise in the 
postshock region \citep[e.g.,][]{ber01}. Since the accretion spot 
rotates with the star, its visibility is modulated at the stellar 
rotation period, as should the veiling and line profile 
variations. This is observed in dippers, as well as in spotted, lower 
inclination systems, where spectroscopic observations have been 
conducted simultaneously with photometric observations 
\citep[e.g.,][]{bou07,ale12,fon14,sou16}.
This is precisely what we observe here in the LkCa 15 system: the 
veiling variations are periodic (P=5.6 days), and reach a maximum at 
the rotational phase corresponding to the maximum longitudinal magnetic field 
measured in the CaII 8542\AA\, line profile, which is thought to form at least partly close 
to the accretion shock. The intensity of the HeI 5876\AA\, line profile also 
varies periodically (P=5.6 days), being maximum at the phase of maximum 
veiling, as expected from the modulation of the accretion shock 
visibility. Like in AA Tau, the magnetospheric radius ($0.055 \pm 0.015$ au) and the corotation 
radius ($0.064 \pm 0.001$ au) are coincident within the errors. 
They are also consistent with the innermost CO emission radius ($0.083 \pm 0.030$ au)
measured by \citet{naj03}. Moreover, the sublimation radius at ($0.031 \pm 0.001$ au) 
indicates that dust can be present at the magnetospheric radius and
fill an inner disk warp that is thought to be created by the dynamical 
interaction of an inclined stellar magnetosphere and the inner disk near corotation.
All these features provide a direct measurement of the 
stellar rotation period, which agrees with the photometric 
period due to the inner disk warp, which must thus be located close to 
the corotation radius. Additional support for the magnetospheric 
accretion scenario to explain the observed variability of the LkCa 15 
system comes from the appearance of high-velocity redshifted absorptions 
occurring in the profiles of the Balmer, NaD I, and CaII 8542\AA\, lines at phases 
ranging from 0.3 to 0.7, that is, corresponding to maximum accretion 
shock visibility. Redshifted absorptions occur when the colder gas in 
the upper part of the accretion flow absorbs the photons originating 
from the hotter gas located deeper down in the accretion column. The 
simultaneous occurrence of stellar occultations, redshifted absorption, 
and continuum excess confirms that the disk warp, the accreting gas, and 
the accretion shock are located at the same azimuth from the disk 
inner edge to the stellar surface, and thus are physically connected. 

However, the magnetospheric accretion model does not straightforwardly 
account for other aspects of the spectral variability of the system 
reported here. For instance, neither the strong redshifted emission 
feature observed in the CaII 8542\AA\, line profile at phase 0.07 nor the 
anticorrelation of intensity variations between the HeI 5876\AA\, line and the 
extreme blue wing of the H$\alpha$ line, possibly related to a fast wind, 
are easily understood. Moreover, the nearly opposite polarities measured in 
the longitudinal component of the magnetic field for the photospheric 
and CaII lines suggest a complex, multipolar field close to the stellar 
surface, even though the dipole may still dominate at the corotation 
radius. This would not be surprising for LkCa 15, whose partly radiative 
interior may trigger a multipolar dynamo field \citep[e.g.,][]{gre12,don13}.

Another interesting feature is the opposite radial velocity 
variations of the blue and red absorption components seen in the H$\alpha$ 
line, from 0 to about 80 \kms. To the best of our knowledge, this 
behavior has been reported before only for AA Tau \citep{bou03,bou07}
and was interpreted as magnetospheric inflation \citep[see Fig.19 of][]{bou03}: 
as magnetic field lines are twisted by the 
accretion flow, they inflate, open, and reconnect on a timescale of 
several rotational periods, that is, a few weeks. During this process, the 
projected velocities of the accretion flow and of the associated 
magnetospheric wind vary, as they follow the evolving curvature of the 
expanding field lines, thus producing the observed correlated radial 
velocities of the red and blue absorption components in the H$\alpha$ line 
profile. In the LkCa 15 system, we monitored these variations over 15 
days, and observed a monotonic decrease of the radial velocities during 
this time span, from about 0 to -80 \kms for the blue component, and 
from about +80 to 0 \kms for the red one, but with a large day-to-day 
scatter. We thus offer the same interpretation, pending additional 
constraints on these longer term variations. 

LkCa 15 appears to be a typical dipper, like AA Tau. 
Not only do the two systems display 
the same type of spectroscopic and photometric day-to-day variability, 
which is stable over a timescale of week to years \citep{gra07}, 
but they also have strong dipole fields,
and both exhibit moderate mass accretion rates of about $10^{-9}$ \msun yr$^{-1}$ \citep{don10,don18}.
Their outer disk structure, although different, also shows similarities.
LkCa 15 exhibits an inner 50 au dust cavity, which makes it a 
{\it \textup{bona fide}} transition disk system, while AA Tau, despite not
being classified as a transition disk, presents a multi-ringed circumstellar 
disk, as seen in ALMA observations \citep{loo17}.

The question then arises how a transition disk object, with a large 
inner dust cavity and a moderately inclined outer disk 
\citep[44$\degr$-50$\degr$,][]{oh16,tha14}, can
exhibit AA Tau-like variability that is usually considered as evidence 
for the presence of an inner disk seen almost edge-on. In order to 
reconcile the small-scale and large-scale properties of the LkCa 15 system, we has 
to assume that the system hosts a highly inclined, compact inner disk, 
in agreement with our estimated value of $i_{\star}> 65\degr$ (Sect.\ref{veiling_vrad}),
and with the value of $i_{\star}=70\degr$ needed to correctly model the 
spectropolarimetric data of LkCa 15 \citep{don18}, hence pointing 
to a significant misalignment with the outer disk 
structure. Recently, evidence for a large-scale warp in the multi-ringed 
circumstellar disk of AA Tau has been reported by \citet{loo17}. 
On a scale of a few 10 au, the HL Tau-like ringed structure of the AA Tau 
disk as seen by ALMA has an inclination of 59$\degr$ \citep{loo17}, 
while from scattered light HST imaging, \citet{cox13} derived an 
inclination of 71$\degr$, and from linear polarimetry, \citet{osu05}
obtained $\sim$75$\degr$ for the inner disk close to the star. 
However, care must be taken to compare these different inclination values, 
since scattered light sees the flared surface of the disk, while the
ALMA continuum observation sees the flat disk midplane. 
Different values may be expected for ALMA and HST inclinations, the ALMA continuum being more 
trustworthy, because it probes the disk midplane. 
\citet{van18} recently summarized the growing observational 
evidence for large-scale warps and inner and outer disk misalignment in the 
circumstellar disks of young stellar objects, particularly in the case of 
transition disk systems. The evidence comes either from the report of moderately inclined
large-scale disks observed for dippers \citep[e.g.,][]{ans16} or from
scattered light high-angular resolution imaging of a few Herbid Ae/Be systems
such as HD100453 \citep{ben17}, HD142527 \citep{mar15}, and HD135344B \citep{min17} 
that were modeled with inner disks misaligned by 72$\degr$, 70$\degr$ , and 22$\degr$ , respectively, 
to explain the observed shadows in the outer disks. The misalignement proposed for 
LkCa 15 is of about 15$\degr$ to 25$\degr$, if we compare the inner (>65$\degr$-70$\degr$) 
and outer disk (44$\degr$-50$\degr$) inclination values, therefore similar to the case of HD135344B.
One possible explanation for an inner disk with a small misalignement,
as observed in LkCa 15, is the presence of a Jupiter-mass planetary companion
inside the disk gap, as proposed by \citet{owe17}. The 
results we report here for LkCa 15 thus seem to support the emerging 
trend for transition disk objects to have misaligned inner and outer disks. 

\section{Conclusions}\label{conclusions}
The remarkable similarity between the spectroscopic and photometric 
variability of LkCa 15 with AA Tau suggests that it is driven by the same 
process, namely the interaction of an inner circumstellar disk with the 
stellar magnetosphere that leads to the development of an inner disk 
warp, responsible for the periodic stellar eclipses, and of accretion 
funnel flows down to the stellar surface, responsible for the line 
profile variability and the rotational modulation of the continuum 
excess flux. 
Different as they are on a scale of a few 10 to 100 au, 
with LkCa 15 having a large dust cavity extending about 50 au inward, 
while AA Tau appears to have an HL Tau-like multi-ringed circumstellar 
disk on the large scale, they appear quite similar on a scale of a few 
0.1 au around the central star. In particular, our results suggest that 
LkCa 15 hosts a compact inner disk that is seen at high inclination, 
that is, significantly misaligned with the outer transition disk. Such a 
misalignment might have various causes, including possibly the presence 
of a Jupiter-mass planet inside the disk gap. 
Linking the small scales to the large scales in planet-forming systems is 
now becoming a reality thanks to high-angular resolution 
imaging studies and time-domain monitoring campaigns. This will 
certainly prove a fruitful approach to investigate the primeval 
architecture of nascent planetary systems in circumstellar disks.

\begin{acknowledgements}
We thank CFHT's QSO Team and especially Nadine Manset for the efficient scheduling 
of service observations at the telescope.
This project was funded in part by INSU/CNRS Programme National de Physique Stellaire 
and Observatoire de Grenoble Labex OSUG\@2020. 
This project has received funding from the European Research Council (ERC) under 
the European Union’s Horizon 2020 research and innovation programme (grant agreements 
No 742095; {\it SPIDI}: Star-Planets-Inner Disk-Interactions, PI: JB, spidi-eu.org; 
and No 740651; NewWorlds, PI: JFD).
SHPA acknowledges financial support from CNPq, CAPES and Fapemig.
FM acknowledges funding from ANR of France under contract number ANR-16-CE31-0013.
This paper includes data collected by the K2 mission. Funding for the K2 mission 
is provided by the NASA Science Mission directorate. 
\end{acknowledgements}

%
%

\begin{appendix}
\section{H$\alpha$ profile decomposition}\label{appendixA}
\begin{figure*}
\centering
\includegraphics[width=3.2cm]{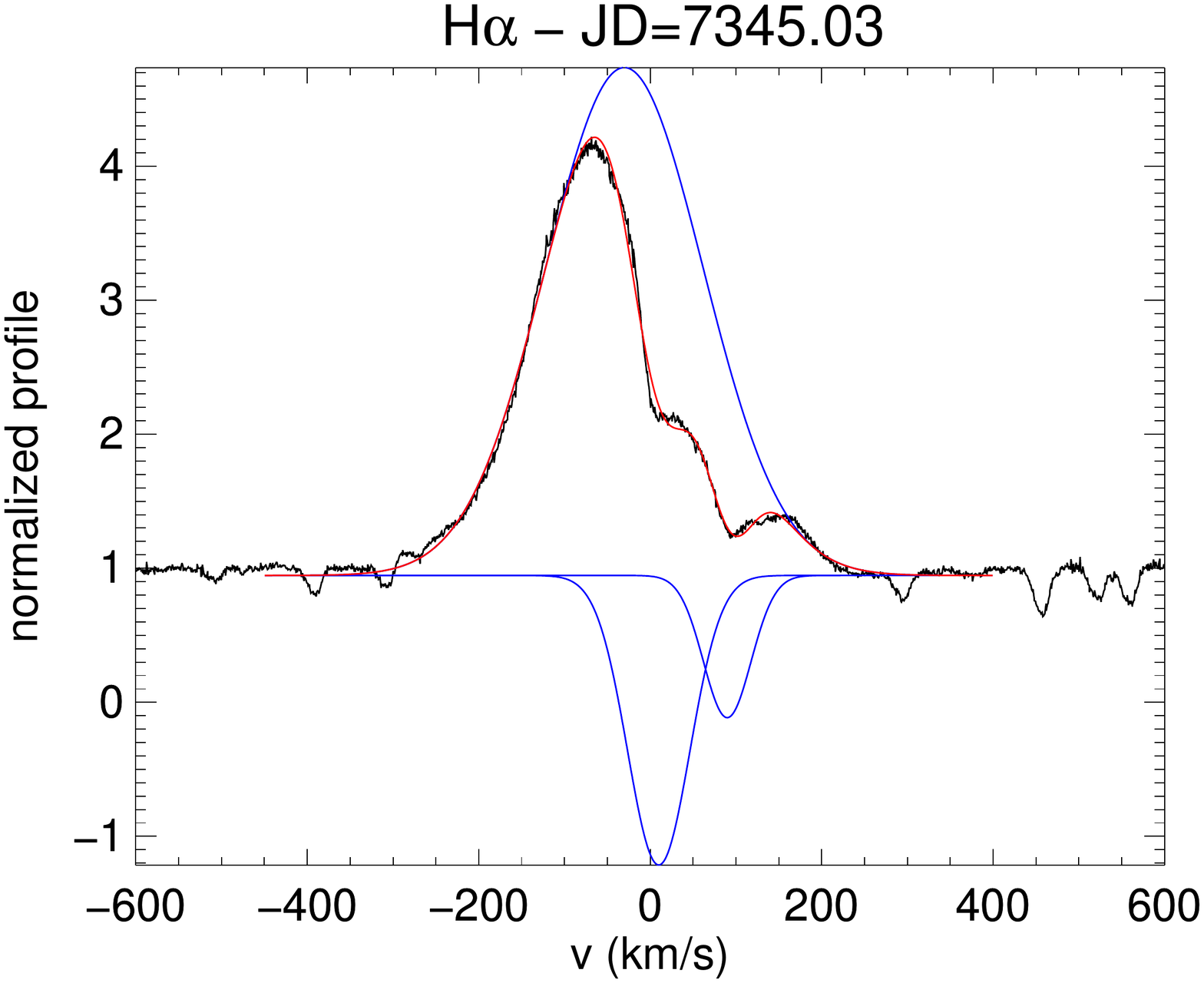}\includegraphics[width=3.2cm]{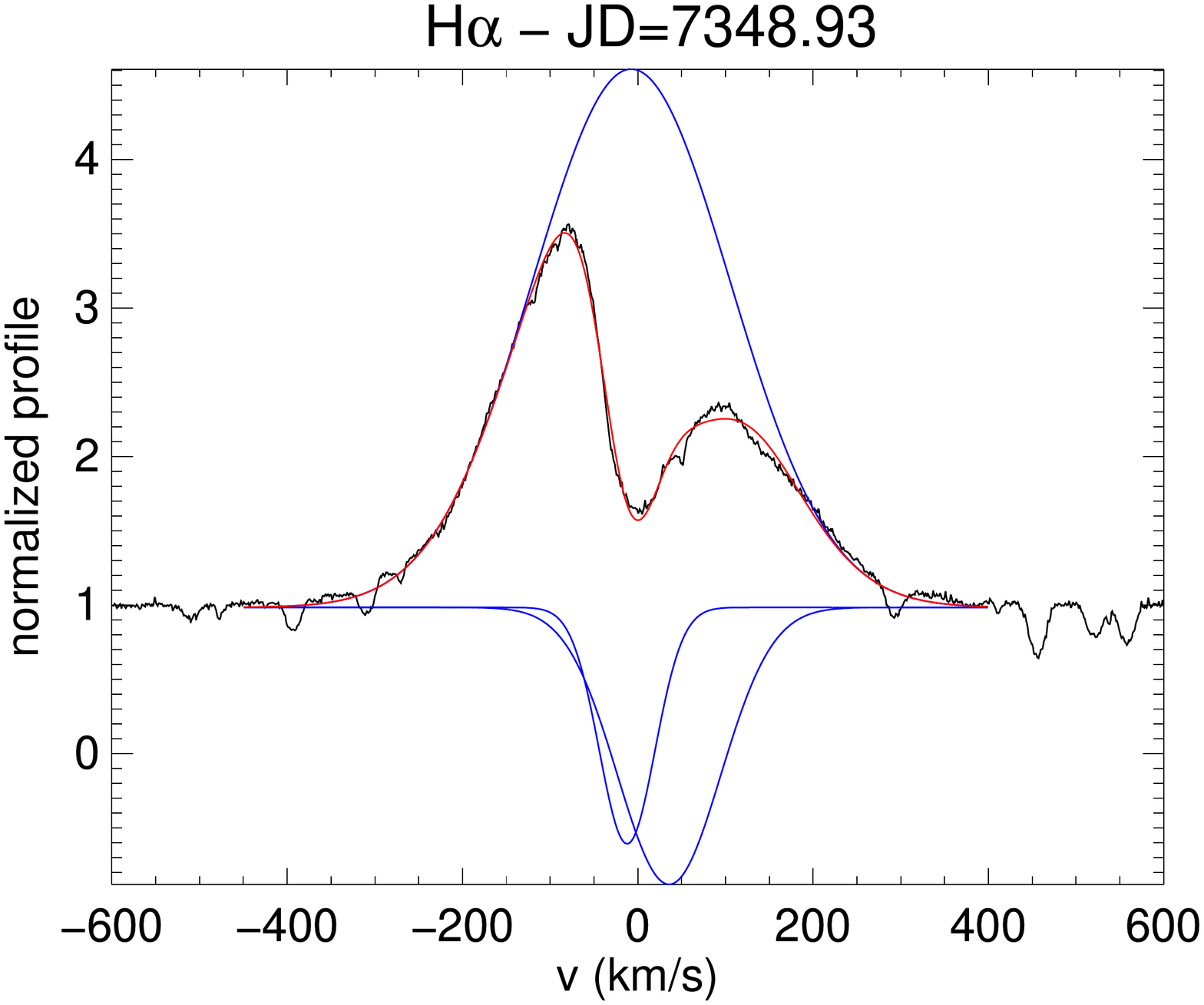}\includegraphics[width=3.2cm]{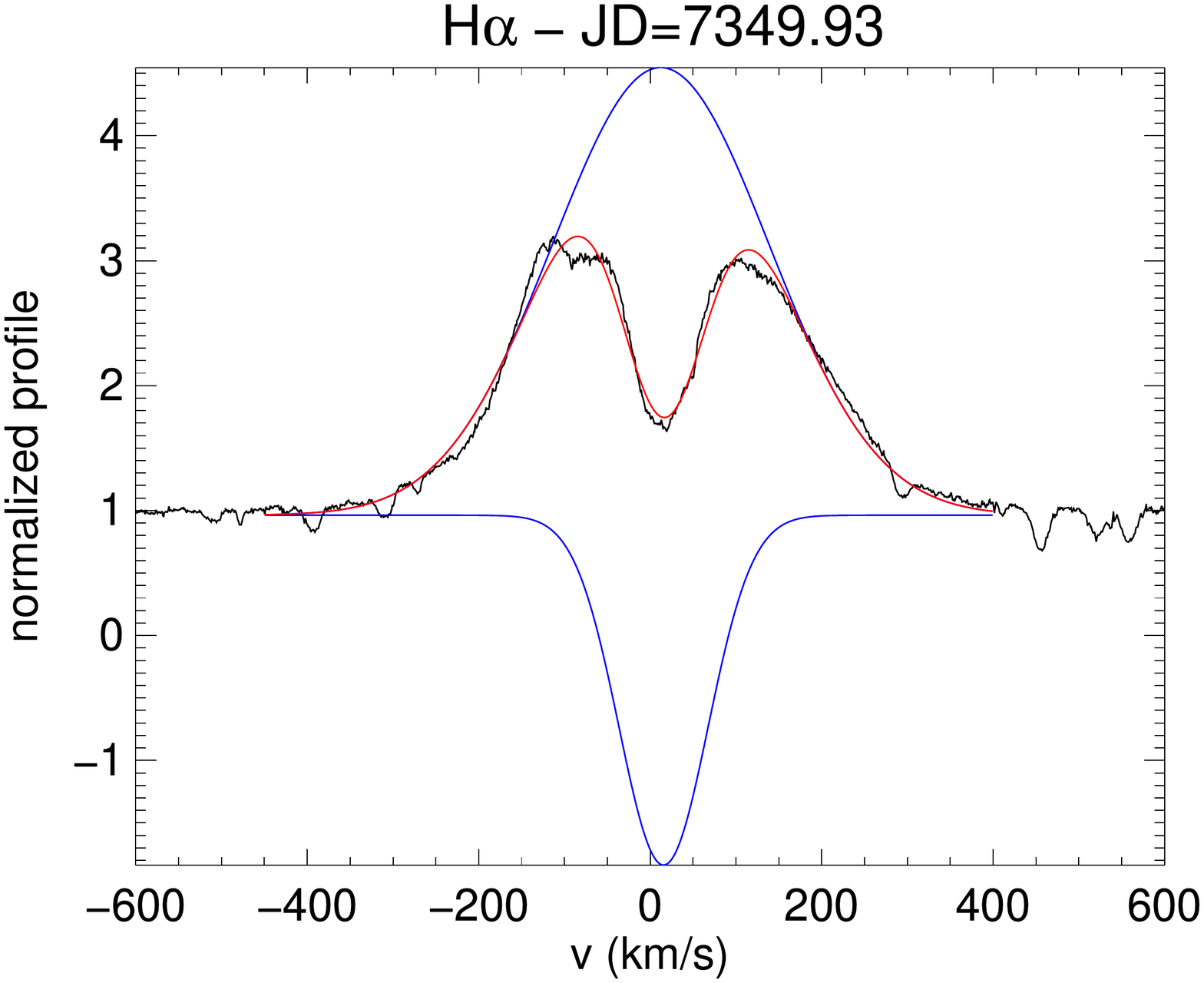}\includegraphics[width=3.2cm]{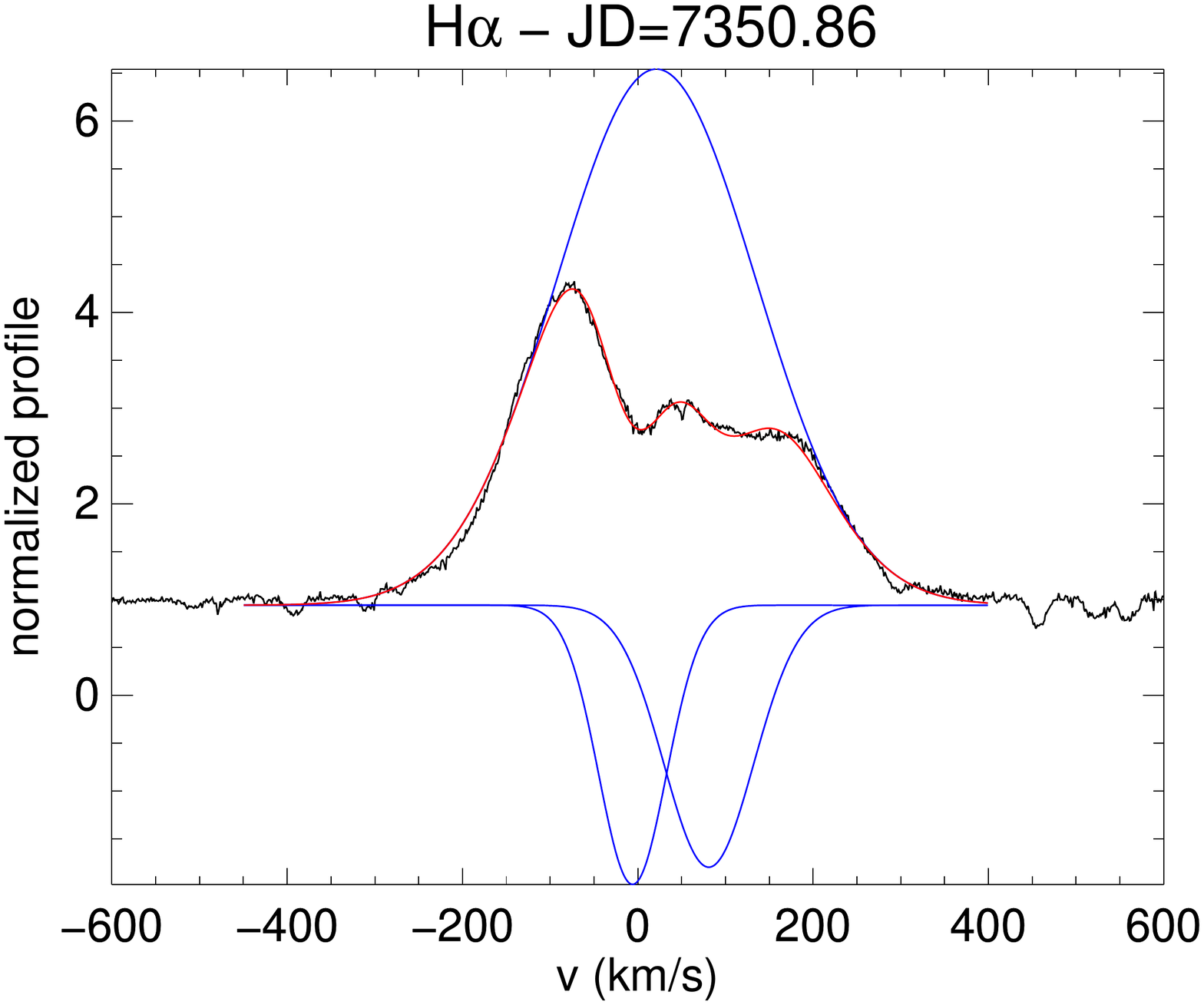}\includegraphics[width=3.2cm]{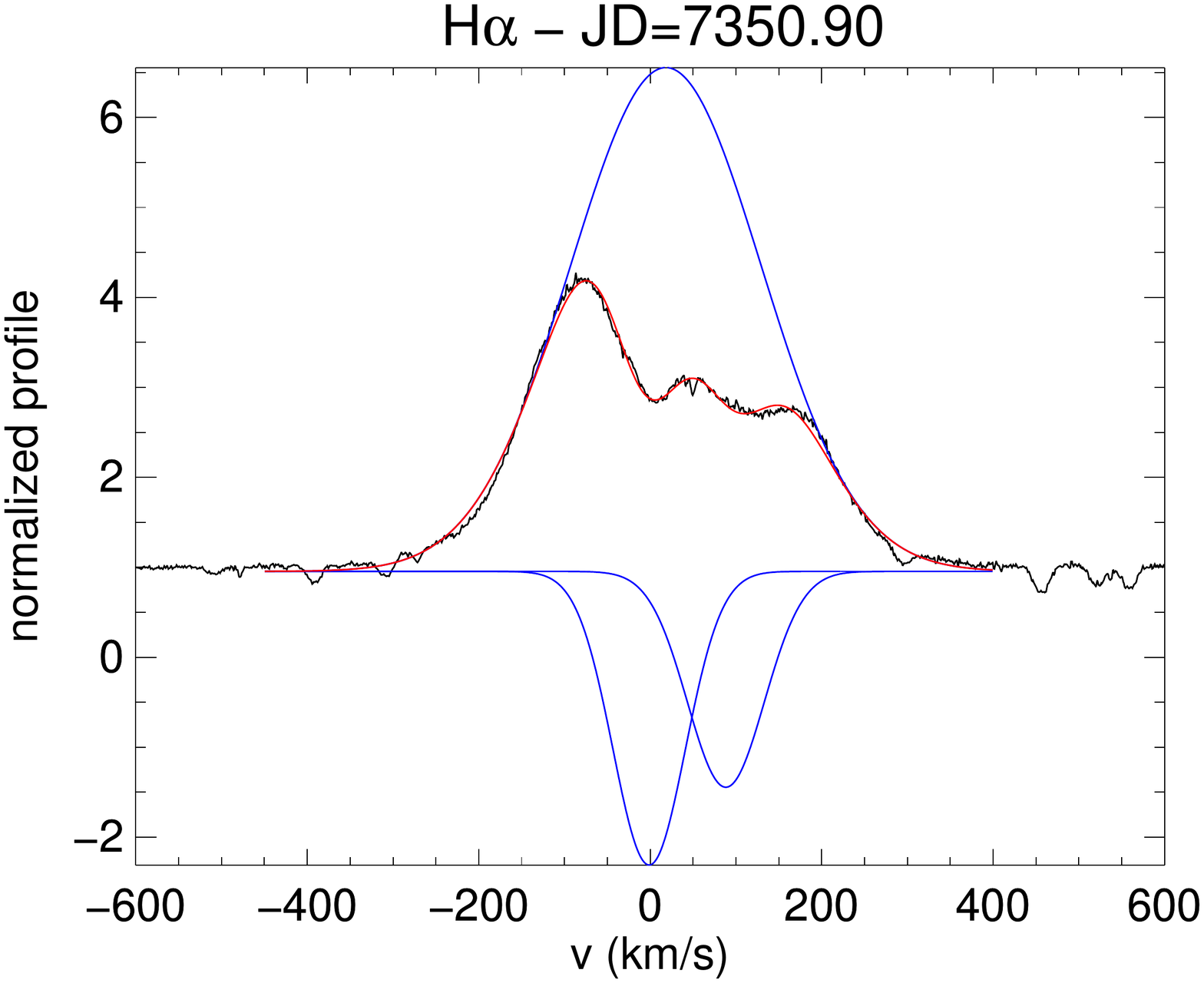}
\includegraphics[width=3.2cm]{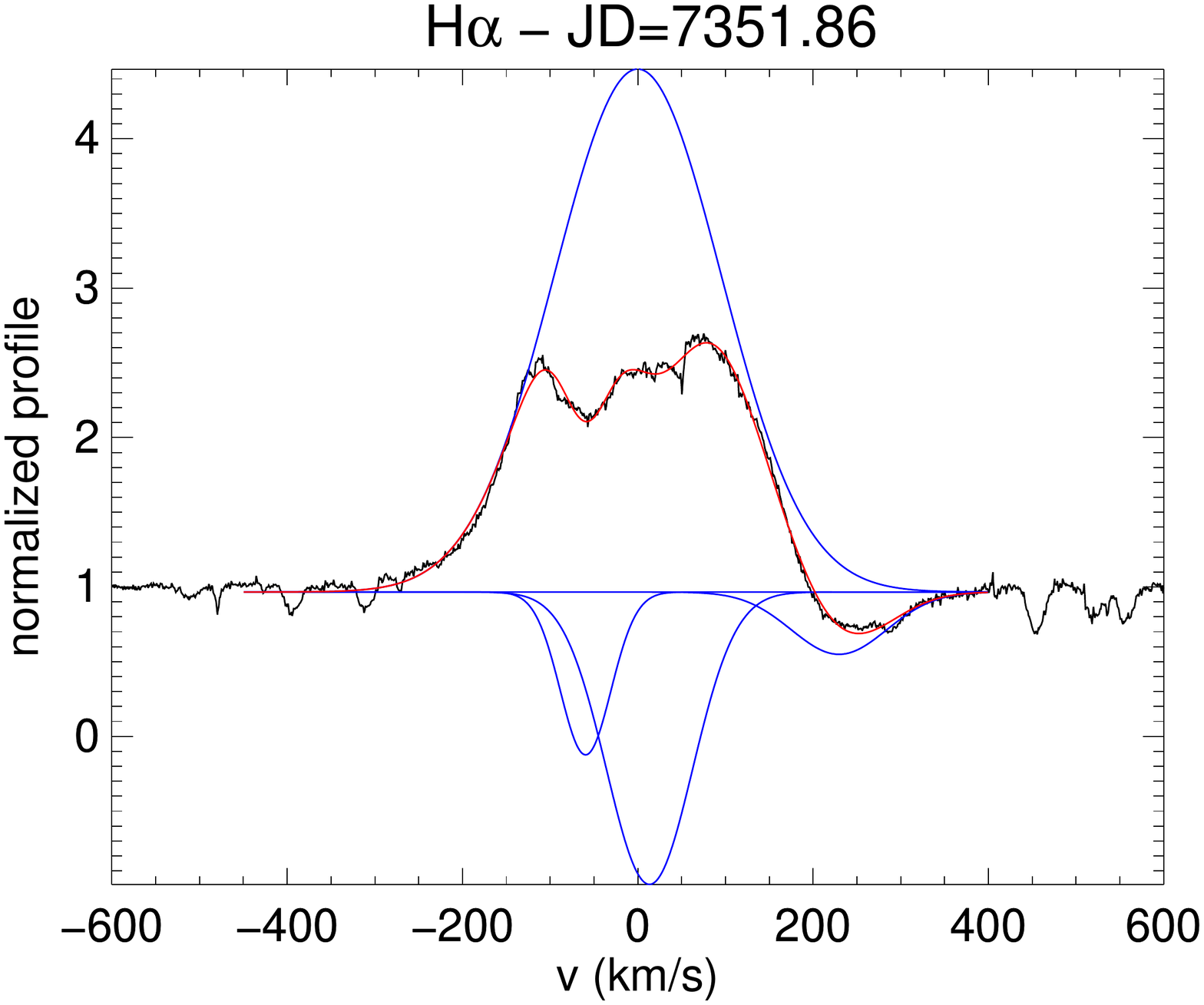}\includegraphics[width=3.2cm]{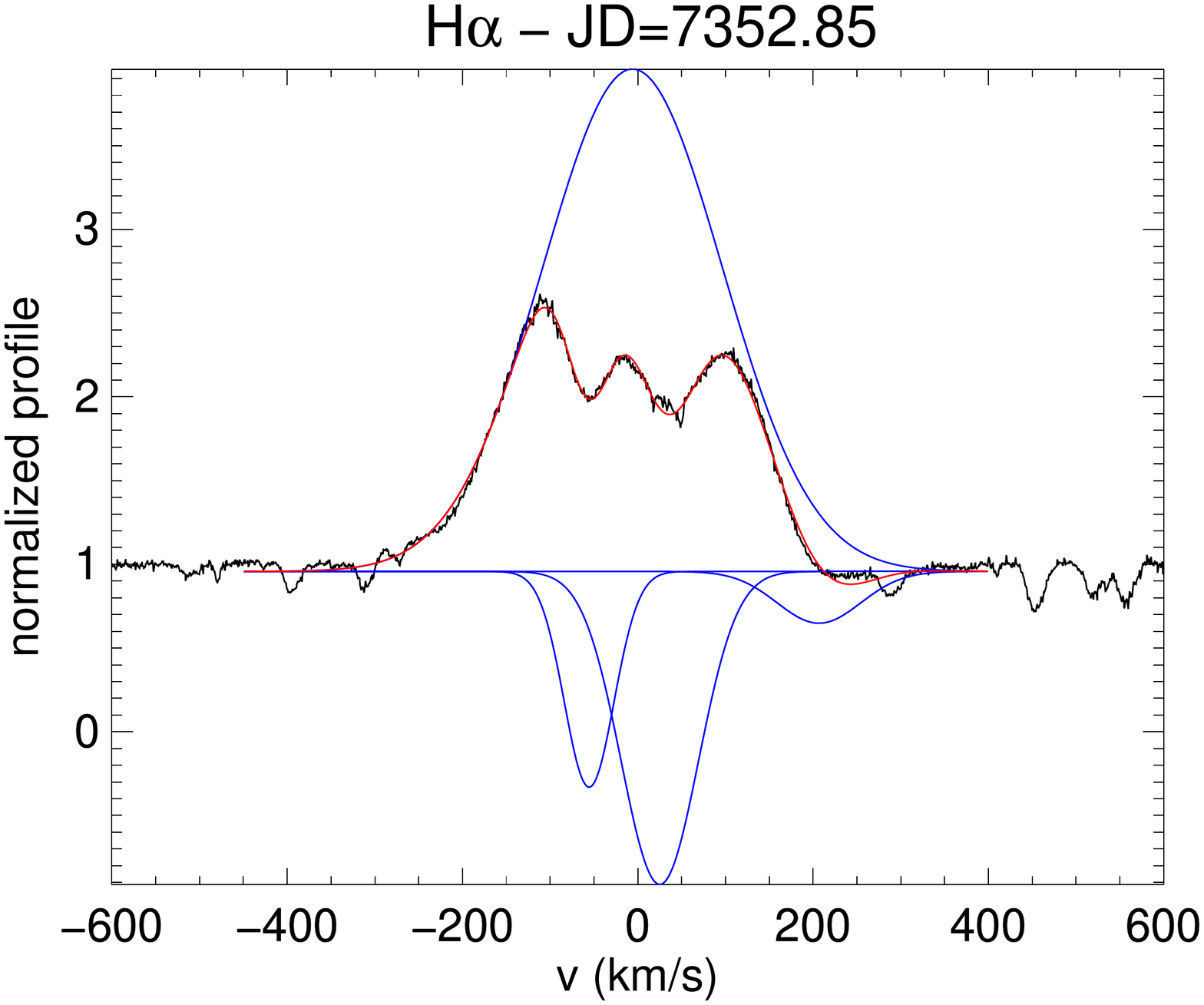}\includegraphics[width=3.2cm]{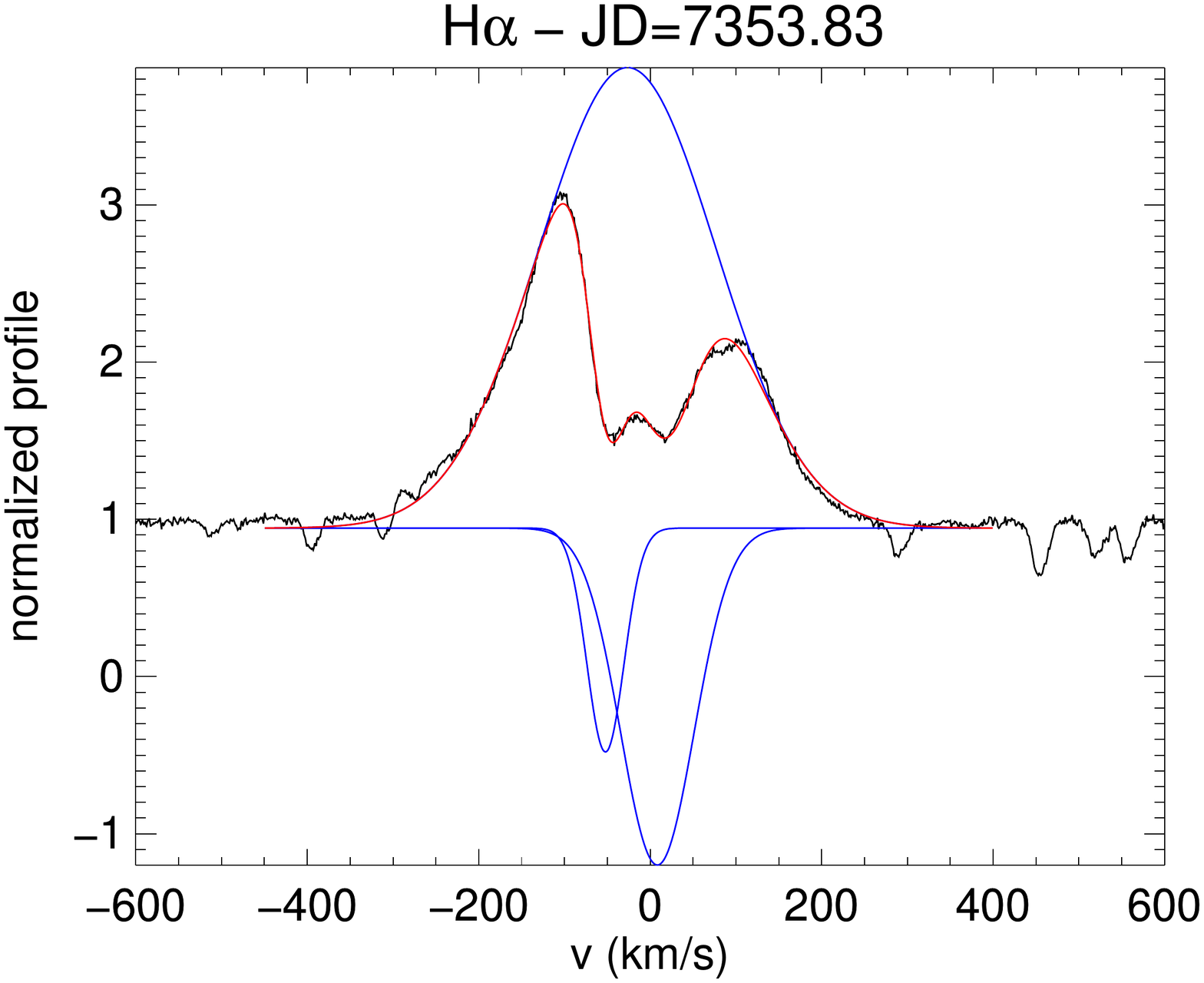}\includegraphics[width=3.2cm]{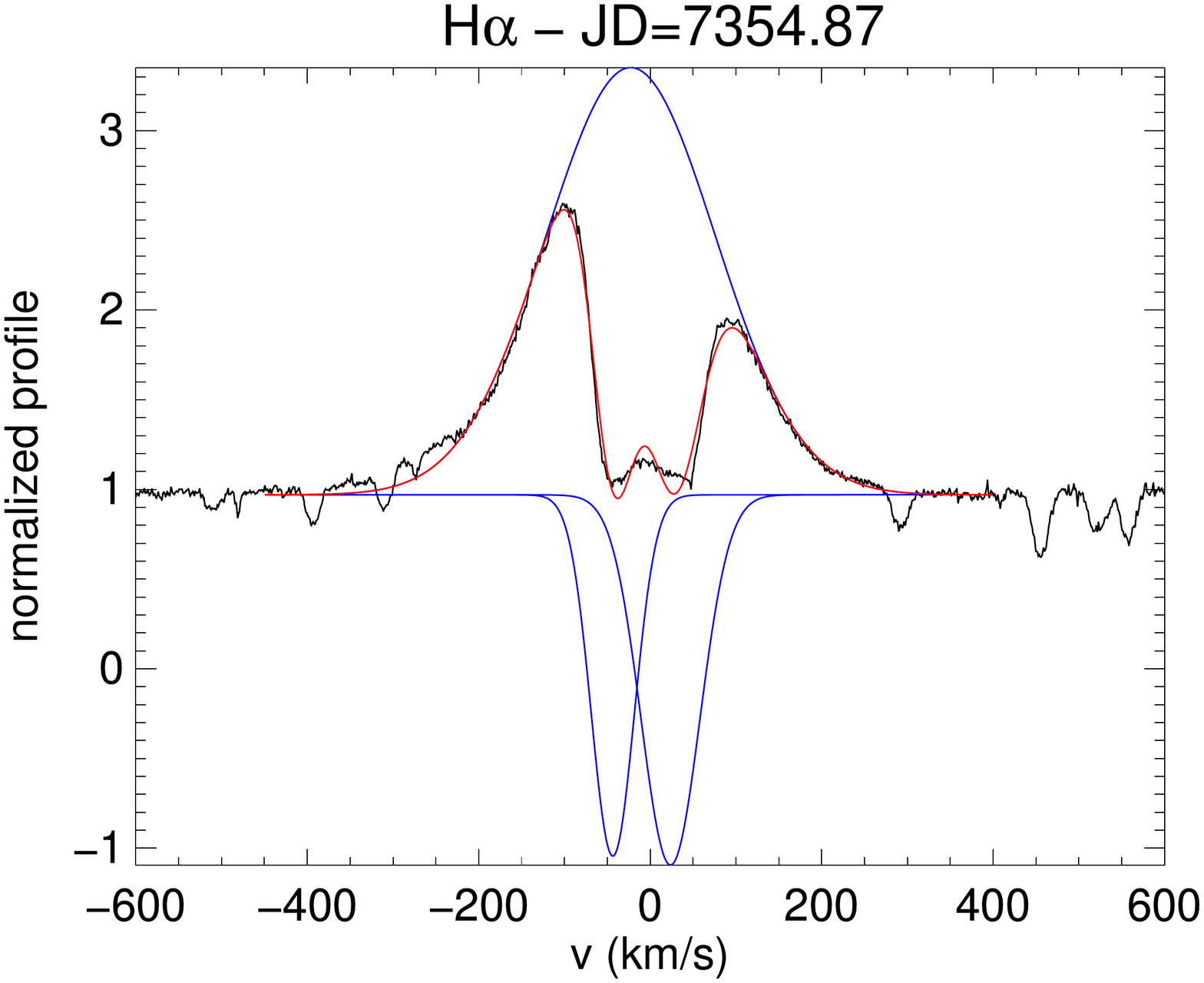}\includegraphics[width=3.2cm]{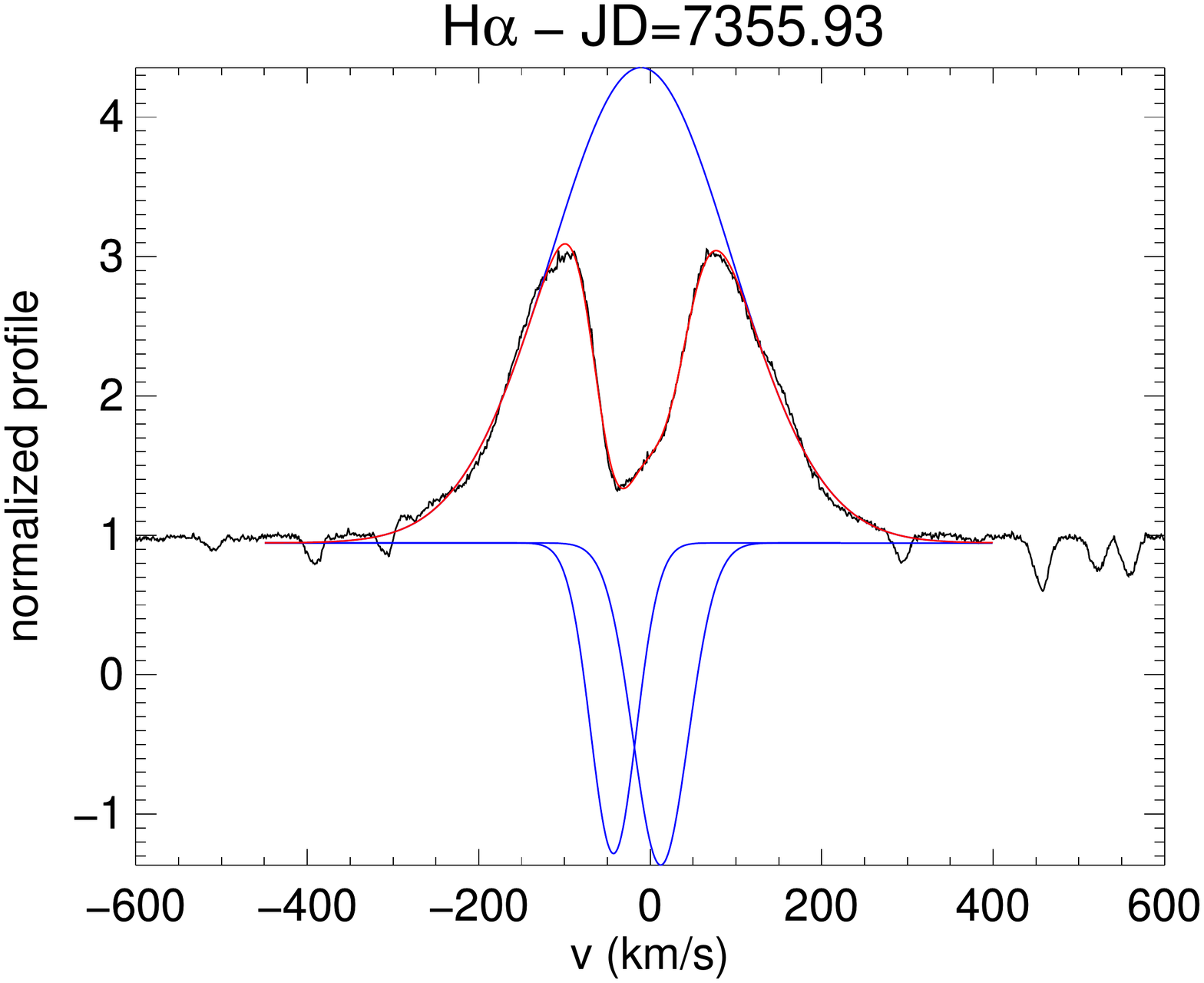}
\includegraphics[width=3.2cm]{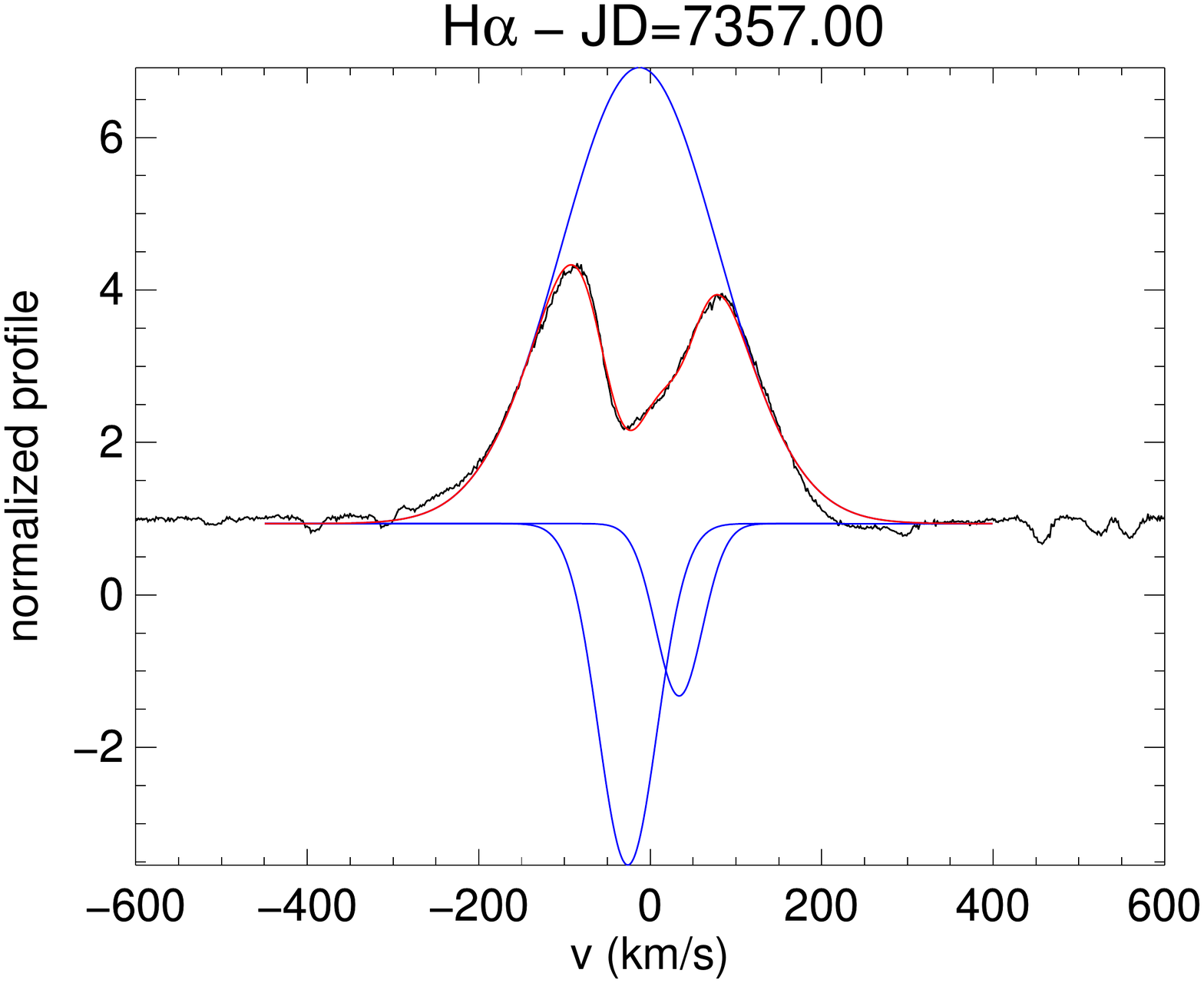}\includegraphics[width=3.2cm]{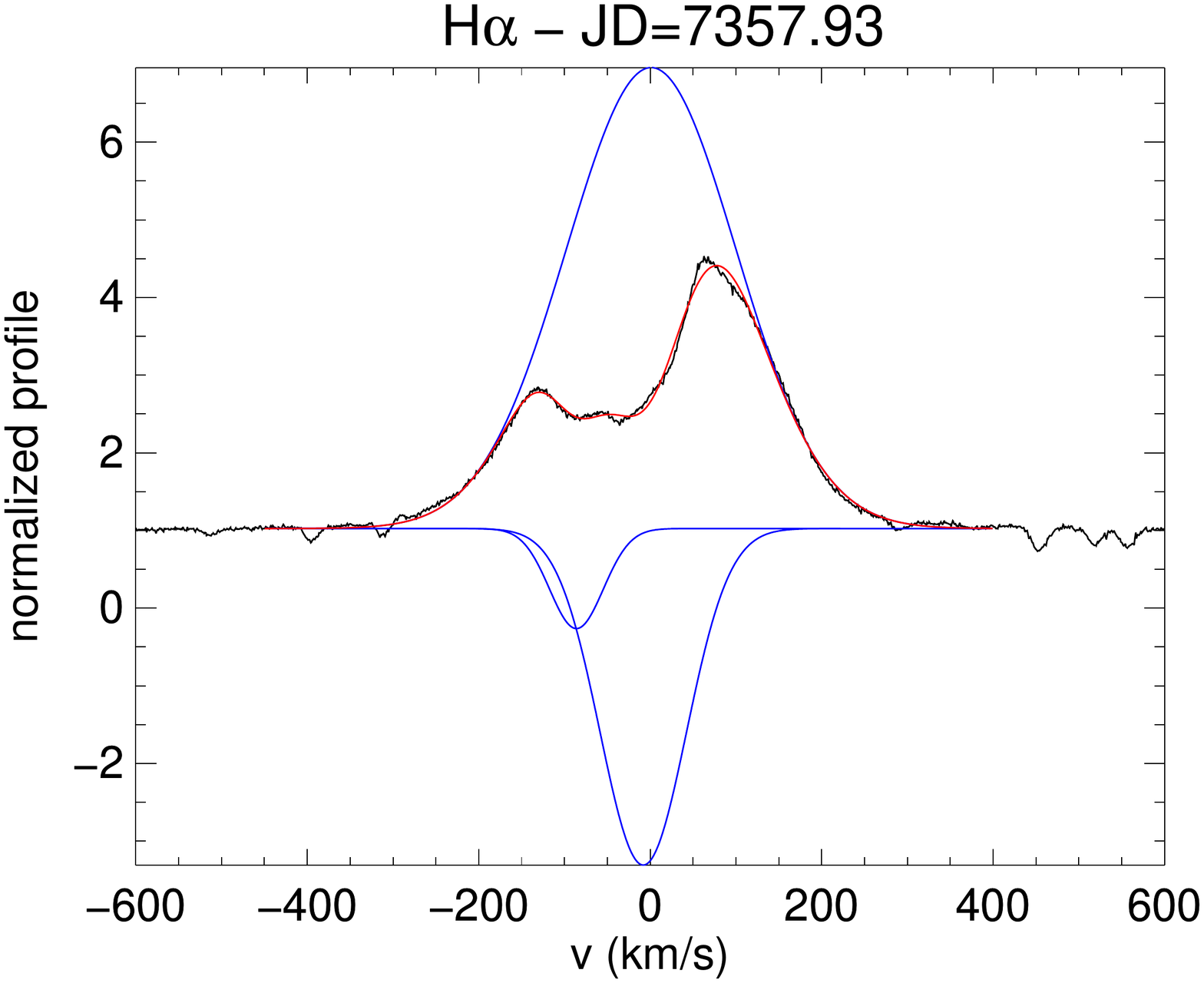}\includegraphics[width=3.2cm]{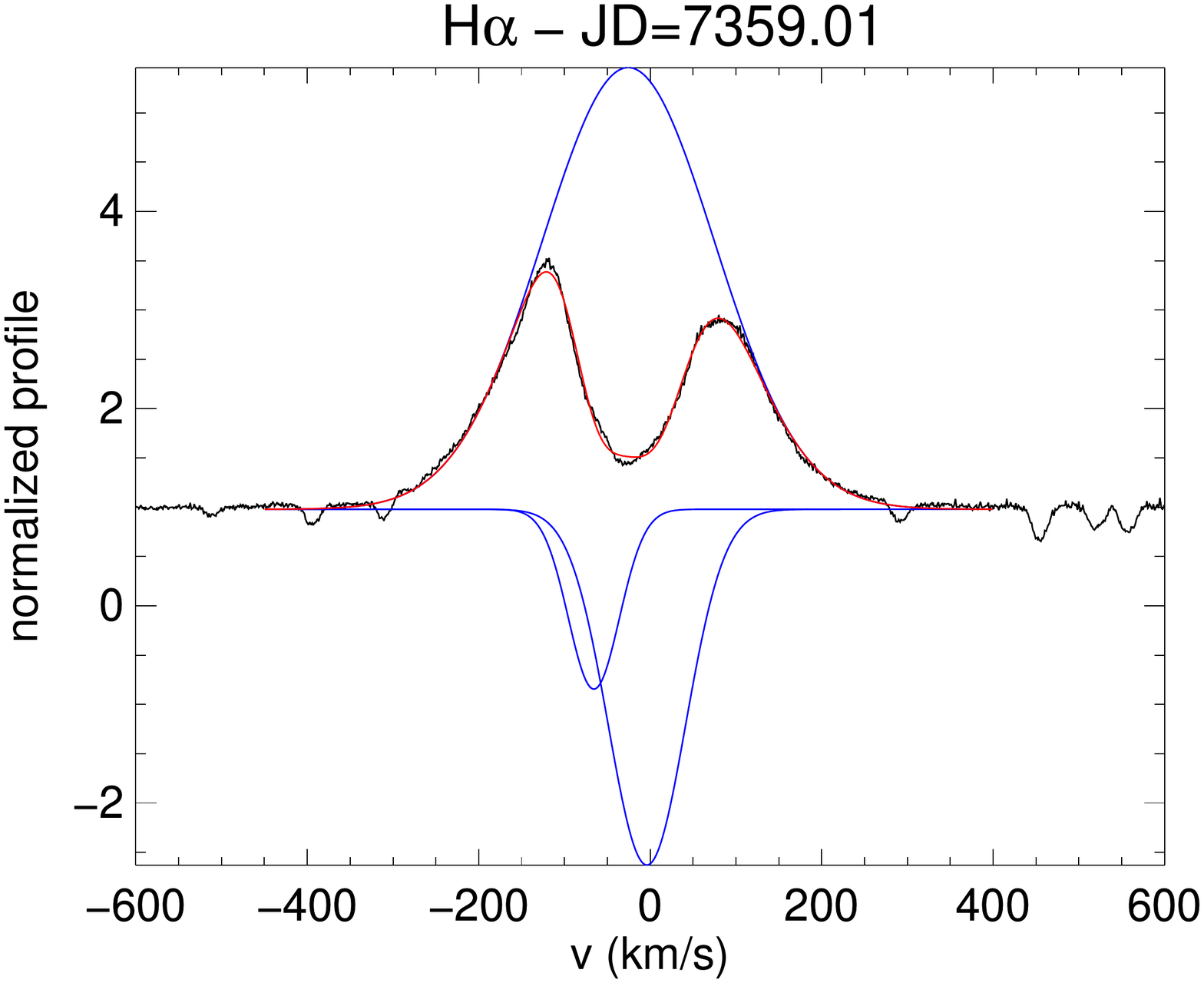}\includegraphics[width=3.2cm]{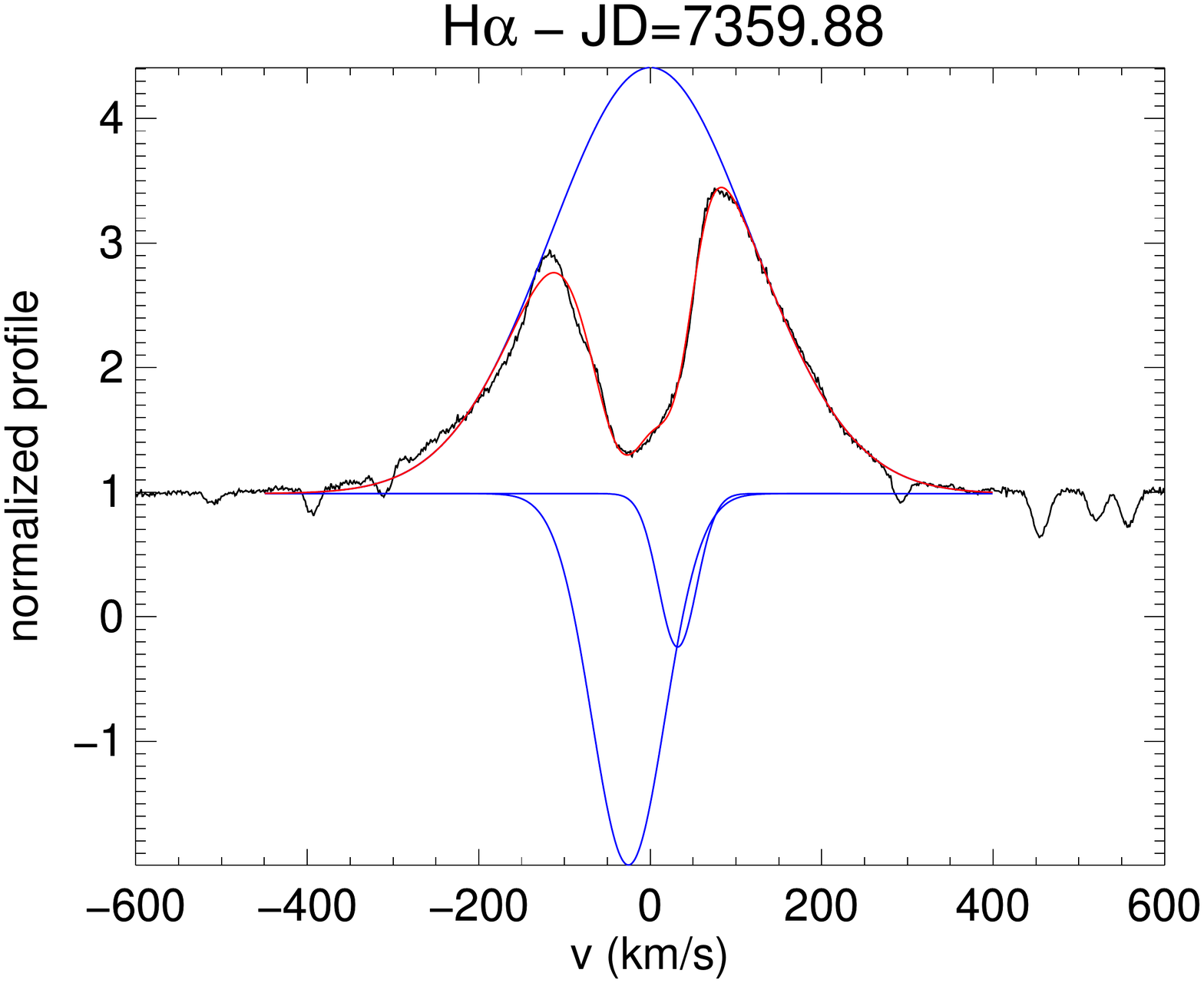}
\caption{H$\alpha$ profile decomposition. The observed profile is shown in black, the Gaussian
components in blue, and the sum of the blue components corresponds to the red lines.}
\label{ha_decomposition}
\end{figure*}

The H$\alpha$ emission line profiles of LkCa 15 were decomposed with emission and
absorption Gaussian components, as shown in Fig. \ref{ha_decomposition}.

\end{appendix}

\end{document}